\documentclass[english,nofootinbib]{revtex4-2}
\usepackage[LGR,T1]{fontenc}
\usepackage{textcomp}
\usepackage[latin9]{inputenc}
\usepackage{color}
\usepackage{array}
\usepackage{booktabs}
\usepackage{mathtools}
\usepackage{multirow}
\usepackage{dsfont}
\usepackage{amsmath}
\usepackage{stackrel}
\usepackage{graphicx}
\usepackage{esint}
\usepackage{booktabs}
\usepackage{array}

\makeatletter

\providecommand{\tabularnewline}{\\}

\usepackage{babel}

\usepackage{diagbox}

\makeatother

\usepackage{babel}
\begin{document}
\title{Satellite-based communication for phase-matching measurement-device-independent
quantum key distribution}
\author{Arindam Dutta$^{1,2}$}
\email{arindamsalt@gmail.com}

\email{https://orcid.org/0000-0003-3909-7519}

\author{Subhashish Banerjee$^{1}$}
\email{subhashish@iitj.ac.in}

\email{https://orcid.org/0000-0002-7739-4680}

\author{Anirban Pathak$^{2\,}$}
\email{anirban.pathak@gmail.com}

\email{https://orcid.org/0000-0003-4195-2588}

\affiliation{$^{1}$Department of Physics, Indian Institute of Technology Jodhpur,
Jodhpur 342030, Rajasthan, India~\\
$^{2}$Department of Physics and Materials Science \& Engineering,
Jaypee Institute of Information Technology, A 10, Sector 62, Noida,
UP-201309, India}
\begin{abstract}
This study investigates the feasibility of the phase-matching measurement-device-independent
quantum key distribution (PM-MDI QKD) protocol proposed by Lin and
L{\"u}tkenhaus for satellite-based quantum communication. The protocol's
key rate, known to exceed the repeaterless PLOB bound, is evaluated
in the asymptotic limit under noisy conditions typical of satellite
communications, including loss-only scenarios. The setup involves
two ground-based parties connected via fiber (loss-only or noisy)
and a space-based third party linked to one of these two ground-based
parties through free-space communication. Simulations using the elliptic-beam
approximation model the average key rate (AKR) and its probability
distribution (PDR) across varying zenith angles and fiber distances.
Down-link free-space communication is assessed under day and night
conditions, with intensity optimization for each graphical point.
Dynamic configurations of satellite and ground stations are also considered.
Results indicate that AKR decays more slowly under loss-only conditions,
while PDR analysis shows higher key rates produce more concentrated
distributions. These findings demonstrate the potential of PM-MDI
QKD protocols for achieving reliable key rates in satellite-based
quantum communication.
\end{abstract}
\maketitle

\section{introduction}

In today's information-driven world, protecting communications and
data is essential across various domains, such as financial transactions,
privacy, and the integrity of IoT systems. Classical cryptosystems
like RSA (Rivest-Shamir-Adleman) rely on computational complexity
\cite{RSA83}, but these defenses could be undermined by large-scale
quantum computers. Quantum key distribution (QKD) \cite{BB84,SPR17,E91,B92,BBM92,SAR+04,SPC+09}
offers a solution to this threat, providing security that remains
immune to advances in algorithms or computational power \cite{CJC+20}.
A major challenge in QKD, whether for individual links or within a
network, is also relevant to other schemes \cite{DP+24,DP+23}, is
how the key rate scales with channel loss, represented by the single-photon
transmissivity $\eta$. Traditional QKD protocols typically show a
key rate scaling in the asymptotic limit as $R^{\infty}=O(\eta)$.
Repeaterless optical channels have established bounds \cite{PGB+09,TGW14,PLOB17}.
The tight bound on QKD performance, in terms of the secret key rate
per optical mode, is the Pirandola-Ottaviani-Laurenza-Banchi (PLOB)
bound given by $R^{\infty}\le\log_{2}\frac{1}{1-\eta}$, which can
be reached \cite{PLOB17}. The PLOB repeaterless bound was later extended
to free-space \cite{P21} and satellite communications \cite{P+21}.
Quantum repeaters \cite{BDC+98}, aimed at improving performance by
inserting intermediate stations, have shown progress, but no quantum
repeater has yet outperformed direct optical channels or broken the
repeaterless bounds. Proposals have been introduced for minimal quantum
devices capable of demonstrating quantum repeater functionality by
surpassing the repeaterless bounds through a simple single-node configuration
\cite{LJK+16}. However, this potential quantum advantage has yet
to be experimentally verified. For instance, Refs. \cite{XLW+22,LZL+23,LLL+24} highlight some of
the most recent advancements in the field of QKD. In addition to these,
notable progress has been achieved in chip-based QKD technologies,
as reported in \cite{BPG+22,WHD+23}. These developments are paving
the way for more compact and scalable quantum communication systems.
Furthermore, efforts toward practical network deployment have also
gained momentum, as demonstrated in \cite{ZWM+22,HCL+24}. Recently, phase-matching measurement-device-independent
(PM-MDI) protocols have generated significant interest by successfully
surpassing the repeaterless bound with the use of appropriate test
states \cite{TLF+12,ferenczi2013security,LYD+18}. This discovery
has sparked considerable excitement within the quantum communication
community. In the original study \cite{LYD+18}, it was argued that,
in the infinite key limit, the secret key rate scales as $R^{\infty}=O(\sqrt{\eta})$,
where $\eta$ represents the single-photon transmissivity of the total
distance, rather than of individual segments. It is noteworthy that
such a performance can be achieved by an MDI protocol without relying
on quantum memory or other advanced components.

Most of the security analyses \cite{MZZ18,TLW18} of PM-MDI QKD protocols
have utilized a framework inspired by Shor-Preskill\textquoteright s
quantum error correction approach \cite{SP200}, which Koashi later
refined \cite{K09}. This approach was further extended to accommodate
a broader range of privacy amplification protocols \cite{TH13}. Because
of pessimistic estimates of the phase error rate, the secret key rate
bound in this approach may be generous. Additionally, variations of
the protocol suggested in prior studies introduce significant sifting
costs because of the need for phase randomization of signal states.
A modification proposed in Ref. \cite{LL18} of the PM-MDI QKD protocol
addresses this by distinguishing between test states, used for detecting
eavesdropping, and signal states, which are used to establish secret
keys without phase randomization. A security analysis for this modified
protocol is then conducted using Renner\textquoteright s framework
\cite{R08}, which is known for its flexibility in error correction
and privacy amplification methods. This framework is general enough
to adapt to any QKD protocol and other quantum schemes \cite{DP22,DP23}.
Interestingly, although these two security-proof frameworks have typically
been viewed as independent, recent efforts have been made to unify
them \cite{T20}. Authors in \cite{LL18} begin by evaluating the
security of the protocol within a framework that assumes an infinite
number of test states, akin to the initial decoy state analysis in
weak coherent pulse BB84 protocols \cite{LMC05}. Under this assumption,
a key rate formula is analytically derived for cases where Alice and
Bob detect correlations in a loss-only scenario. A broader framework
is also developed to account for noise, using numerical methods to
demonstrate the robustness of the proposed protocol. Unlike traditional
approaches that rely on phase-randomized decoy states, this protocol
uses non-phase-randomized coherent states, enabling a modified form
of tomography on the quantum channel and untrusted measurement devices.
This method extends the decoy state concept by employing general test
states to assess the channel and detect potential adversarial attacks. 

In QKD protocols that use optical fiber, the polarization state can
be altered due to random birefringence fluctuations within the fiber
\cite{VR99,GK2000}. Additionally, signal attenuation and interference
from environmental noise during QKD transmissions \cite{STP+16} over
optical fibers limit the ability to achieve significant key rates
beyond metropolitan-scale networks \cite{TTS+11,BAL17}. One possible
solution is to use optical satellite links, which can potentially
overcome the distance limitations of terrestrial photonic communication
systems \cite{UJK+09,PAT+19,DP24,DP+22,SWU13,SB19}. In open space
conditions, polarization is less affected by atmospheric turbulence,
but polarization variation occurs in the satellite\textquoteright s
reference frame due to its motion \cite{P21,P+21}. Although this
presents substantial technological challenges, various experimental
and theoretical studies \cite{SWU13,BTD09} have demonstrated the
feasibility of this approach using existing ground-based technologies
approved for space operations \cite{TCT+22,VVM19}. Over the past
decade, numerous experiments in free-space conditions have validated
the practicality of QKD setups on mobile platforms, including trucks
\cite{BHG+15}, hot-air balloons \cite{WYL+13}, aircraft \cite{NMR+13,PKB+17},
and drones \cite{LTG+20}. This quantum internet, integrated with
existing classical internet infrastructure, will connect quantum information
processors, making possible new capabilities beyond the reach of classical
information techniques \cite{WEH18}. Quantum technology is poised
to enhance security protocols---confidentiality, integrity, authenticity,
and non-repudiation---across various e-commerce transactions and
sensitive communications \cite{YFL+23,CLW+24}. Although this analysis
does not include field tests with prepare-and-measure schemes, prior
studies on both \emph{terrestrial} \cite{MWF+07,DBD+24} and \emph{satellite-based}
\cite{LCL+17,LCH+18} QKD have shown high-rate decoy-state key exchanges
over free-space links. Additionally, entanglement-based QKD protocols
eliminate the need to trust the satellite source in dual down-link
setups.

Based on the findings in Refs. \cite{LKB19,DMB+24}, down-link communication
involves transmitting from the satellite to the ground, where atmospheric
effects primarily influence the final stages of propagation. In contrast,
up-link communication, where signals are sent from the ground to the
satellite, is impacted by atmospheric effects during the initial phase.
These effects are significantly more pronounced in up-links than in
down-links. The associated phenomena, such as beam deflection and
broadening, result in angular distortions that affect the final beam
size and contribute to channel losses. The severity of these effects
depends on the distance the beam travels after experiencing what is
referred to as the \emph{kick-in effect}. In up-link configurations,
beam broadening occurs near the ground transmitter, and that effect
continues over long distances before reaching the satellite. Conversely,
in down-link scenarios, the beam mainly travels through a vacuum,
with atmospheric effects becoming significant only in the last $15-20$
kilometers before reaching the ground receiver. Building on previous
research, this study focuses on analyzing the PM-MDI QKD protocol
introduced in \cite{LL18} for secure long-distance free-space quantum
communication via satellite, with Bob in a down-link configuration
and Charlie at a ground station connected to Alice via a fiber channel.
While Ref. \cite{LL18} demonstrated the protocol's robustness using
numerical methods, our study derives analytical calculations for the
noisy scenario to obtain the key rate equation in the asymptotic limit.
We provide an intuitive analysis of the performance of the PM-MDI
QKD protocol with both the satellite and fiber-based systems, considering
both loss-only and noisy scenarios.

The structure of this paper is as follows. Section \ref{sec:II} delves
into the PM-MDI QKD protocol, providing an in-depth analysis of the
key rate equation for various fiber channels to assess the protocol\textquoteright s
performance in satellite-based systems. This section also discusses
how atmospheric conditions affect free-space communication link and
considers the elliptical-beam approximation at the receiver. Section
\ref{sec:III} offers a comprehensive evaluation of the MDI protocol's
performance, supported by simulation results. The conclusions and
key findings are discussed in Section \ref{sec:IV}. Appendix A briefly
outlines the key rate derivation for the loss-only scenario, while
Appendix B details the analytical derivation for the noisy scenario,
which underpins the simulation results for the satellite-based PM-MDI
QKD protocol.

\section{Theory on satellite based PM-MDI QKD protocol }\label{sec:II}

\begin{figure}[h]
\begin{centering}
\includegraphics[scale=0.5]{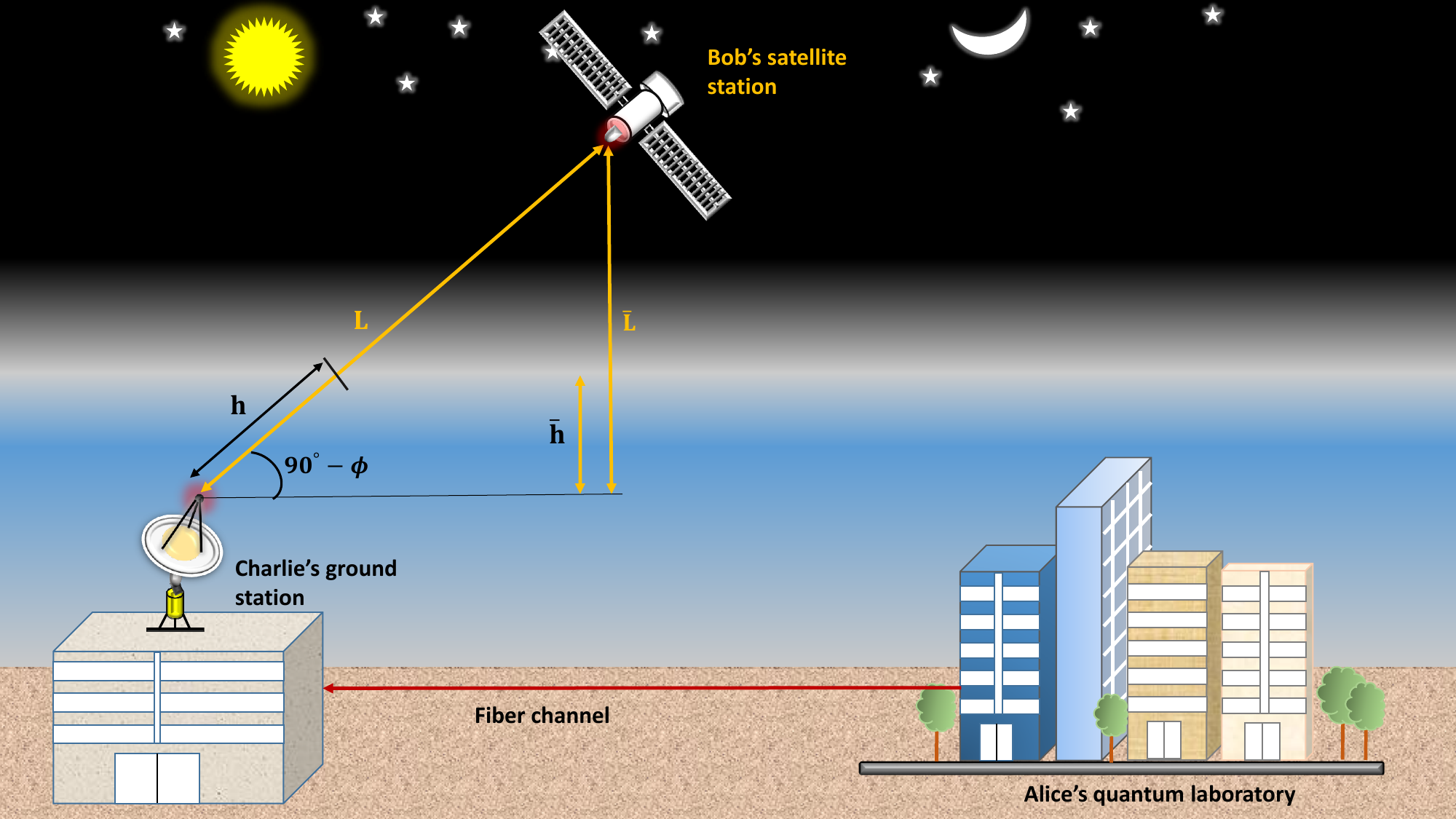}
\par\end{centering}
\caption{(Color online) Schematic of satellite-based PM-MDI--QKD.}\label{fig:PM_MDI_QKD_Satellite_Setup}
\end{figure}

In this section, we describe the PM-MDI QKD protocol using standard
notations as proposed in Ref. \cite{LL18} and analyze the key rate
equation under a collective attack scenario. To implement this PM-MDI
QKD scheme in LEO satellite communication, we discuss important aspects
of elliptic beam approximation for signal transmission in free-space
communications. For our proposed satellite-based PM-MDI setup, Alice
and Charlie are on the ground, while Bob is in space, as illustrated
in Figure \ref{fig:PM_MDI_QKD_Satellite_Setup}. Alice uses fiber
channels to send her states to Charlie, while Bob uses satellite link,
with transmittance modeled by the probability distribution of the
transmittance (PDT) \cite{LKB19}. Practically, it is more advantageous
for Alice to be inside the city and Charlie outside, as this configuration
fully utilizes fiber-based networks and minimizes additional light
pollution and scattering at the receiver end. Our simulation considers
both a loss-only scenario and a noisy scenario (which includes realistic
imperfections) for the fiber channel between Alice and Charlie. In
Ref. \cite{LL18}, the key rate for the loss-only scenario was analytically
derived for fiber-based communication, while the noisy scenario was
addressed numerically. In contrast, this wo\textcolor{black}{rk presents
an explicit analytical calculation of the key rate equation in a noisy
scenario, securing the key against collective attacks \cite{LJ20}.
We implement this analysis within a satellite-based MDI-QKD framework.}
In our simulation, we use optimized intensity values (average photon
number) to calculate each point on the average key rate plots, thereby
maximizing the key rate (see Figure \ref{fig:AKR_PLOT_Loss_Noisy}). 

\subsection{PM-MDI QKD protocol and security analysis}

\emph{Description of PM-MDI QKD protocol}: Alice and Bob each select
a random bit, $m_{\mathcal{A}}$ and $m_{B}$, according to a predefined
probability distribution, where $m_{\mathcal{A}},m_{\mathcal{B}}\in\{0,1\}$.
If $m_{\mathcal{A}}=0$, Alice designates the round as key-generation
mode; if $m_{\mathcal{A}}=1$, she designates it as test mode, and
Bob follows the same rule. In test mode, Alice (Bob) randomly selects
a phase $\theta_{\mathcal{A}}\left(\theta_{\mathcal{B}}\right)\in\left[0,2\pi\right)$
and an intensity $\mu_{\mathcal{A}}\left(\mu_{\mathcal{B}}\right)$.
She (he) then prepares a coherent state $|\sqrt{\mu_{\mathcal{A}}}e^{i\theta_{\mathcal{A}}}\rangle\left(|\sqrt{\mu_{\mathcal{B}}}e^{i\theta_{\mathcal{B}}}\rangle\right)$
and transmits it to the untrusted third party, Charlie, via a fiber
(satellite link) quantum channel. In key-generation mode, Alice (Bob)
randomly generates a bit value $k_{\mathcal{A}}\left(k_{\mathcal{B}}\right)\in\left\{ 0,1\right\} $
with a uniform probability distribution, selects the predefined intensity
$\mu$, and sends a coherent state $|\sqrt{\mu}e^{i\pi k_{\mathcal{A}}}\rangle\left(|\sqrt{\mu}e^{i\pi k_{\mathcal{B}}}\rangle\right)$
to Charlie.

In each round, Charlie conducts a joint measurement on the signals
obtained from Alice and Bob and announces the outcome \footnote{We use the abbreviations $+,-,?$ and ${\rm d}$ to correspond to
Charlie's outcome: detectors ${\rm D}_{+}$ clicks, detector ${\rm D}_{-}$
clicks, no-detector clicks and both the detectors clicks, respectively.} as one of $\left\{ +,-,?,{\rm d}\right\} $. Throughout this paper,
Charlie's announcement is referred to as $\gamma$. This process is
repeated multiple times, and after all announcements are made, Alice
and Bob proceed with \emph{key sifting}, \emph{parameter estimation}
and \emph{post-processing} steps. They use an authenticated classical
channel to talk with each other and classify all rounds into two disjoint
sets which are for key generation and parameter estimation. Alice
and Bob disclose their choices of $m_{\mathcal{A}}$ and $m_{\mathcal{B}}$
for each round, followed by Charlie's announcement $\gamma$. If both
$m_{\mathcal{A}}$ and $m_{\mathcal{B}}$ equal $0$, it would indicate
that both of them have selected the key-generation mode for that round.
If in such a round, Charlie's announcement happens to be $\gamma\in\left\{ +,-\right\} $,
they retain the data from that round for key generation. Data from
all other rounds is used for parameter estimation. During parameter
estimation, Alice and Bob reveal the values of $\mu_{\mathcal{A}},\mu_{\mathcal{B}},\theta_{\mathcal{A}},\theta_{\mathcal{B}}$
and $k_{\mathcal{A}},k_{\mathcal{B}}$ for rounds where either both
chose the test mode or one chose the test mode while the other chose
the key mode. If their analysis suggests that Eve has acquired excessive
information about the signals, which would prevent secure key generation,
they abort the protocol. Otherwise, they continue, with Alice generating
a raw key from her bit value $k_{\mathcal{A}}$ in each key-generation
round. To align his key with that of Alice, Bob may find it convenient
to flip his bit value when the announcement is $\gamma=$ ``$-$''.
Following this, Alice and Bob perform error correction and privacy
amplification, as done in standard QKD protocols, to produce a secret
key.

\emph{Security analysis}: This analysis provides a brief interpretation
of security against collective attacks in the limit of an infinite
key. To establish the protocol's security against such attacks, the
source-replacement scheme \cite{CLL04,FL12} is applied to both Alice's
and Bob's sources, converting the protocol to an equivalent entanglement-based
version. The security of this entanglement-based protocol is then
demonstrated by evaluating the secret key generation rate. In each
round, Alice selects a state from the set of possible signal states
$\left\{ |\phi_{m}\rangle_{A^{\prime}}\right\} $ based on a predetermined
probability distribution $\left\{ p_{m}\right\} $. Here $\{|m\rangle_{\mathcal{A}}\}$
is the orthogonal basis element for Alice's register system $\mathcal{A}$
that is used to record the state selection prepared in register $A^{\prime}.$
Further, Bob also selects a state from the same set $\left\{ |\phi_{n}\rangle_{B^{\prime}}\right\} $
according to a probability distribution $\left\{ q_{n}\right\} $.
In the \emph{source-replacement scheme} \cite{LL18}, Alice's and
Bob's sources effectively generate the state,

\begin{equation}
\begin{array}{lcl}
|\Psi\rangle_{\mathcal{AB}A^{\prime}B^{\prime}} & = & \left(\underset{m}{\sum}\sqrt{p_{m}}|m\rangle_{\mathcal{A}}|\phi_{m}\rangle_{A^{\prime}}\right)\otimes\left(\underset{m}{\sum}\sqrt{q_{n}}|n\rangle_{\mathcal{B}}|\phi_{n}\rangle_{B^{\prime}}\right)\\
\\ & = & \underset{m,n}{\sum}\sqrt{p_{m}q_{n}}|m,n\rangle_{\mathcal{AB}}|\phi_{m},\phi_{n}\rangle_{A^{\prime}B^{\prime}}
\end{array}.\label{eq:Source_Replacement_State1}
\end{equation}
The register $\mathcal{A}$ is used to record the state selections
prepared in register $A^{\prime}$ by Alice, while the register $\mathcal{B}$
records the state selections in register $B^{\prime}$ by Bob. An
orthonormal basis $\left\{ |m\rangle_{\mathcal{A}}\right\} $ exists
for Alice's register system $\mathcal{A}$, corresponding to the states
$\left\{ |\phi_{m}\rangle_{A^{\prime}}\right\} $, and another orthonormal
basis $\left\{ |n\rangle_{\mathcal{B}}\right\} $ exists for Bob's
register system $\mathcal{B}$, corresponding to the states $\left\{ |\phi_{n}\rangle_{B^{\prime}}\right\} $.
Importantly, Eve does not have access to registers $\mathcal{A}$
and $\mathcal{B}$. Alice keeps her register $\mathcal{A}$ and sends
the system $A^{\prime}$ to Charlie, while Bob keeps $\mathcal{B}$
and sends $B^{\prime}$. To communicate their state choices to Charlie
in each round, Alice performs a local measurement on her register
$\mathcal{A}$ using a positive-operator valued measure (POVM) $M_{\mathcal{A}}=\left\{ |m\rangle\langle m|\right\} $,
and Bob applies his POVM $M_{\mathcal{B}}=\left\{ |n\rangle\langle n|\right\} $
to his register $\mathcal{B}$. The source-replacement scheme in the
key-generation mode is considered for the signal states, while test
states in the test mode are used to constrain Eve's actions within
the subspace spanned by the signal states. The set of signal states
in the key-generation mode, denoted by $\mathcal{S}$, includes $\left\{ |+\sqrt{\mu},+\sqrt{\mu}\rangle,|+\sqrt{\mu},-\sqrt{\mu}\rangle,|-\sqrt{\mu},-\sqrt{\mu}\rangle,|-\sqrt{\mu},+\sqrt{\mu}\rangle\right\} $\footnote{Two coherent states $\left\{ |+\sqrt{\mu}\rangle,|-\sqrt{\mu}\rangle\right\} $
span a two-dimensional space and can be expressed as $|\pm\sqrt{\mu}\rangle=c_{0}|e_{0}\rangle\pm c_{1}|e_{1}\rangle$,
where $\left\{ |e_{0}\rangle,|e_{1}\rangle\right\} $ is an orthogonal
basis. The coefficients satisfy $\left|c_{0}\right|^{2}+\left|c_{1}\right|^{2}=1$
and $\left|c_{0}\right|^{2}-\left|c_{1}\right|^{2}=\langle+\sqrt{\mu}|-\sqrt{\mu}\rangle$
\cite{LL18}.}. Each state is a two-mode coherent state generated by both Alice
and Bob\footnote{Here, $\mathcal{S}$ is used as the basis spanning set, and we omit
subscripts $A^{\prime}B^{\prime}$ for the elements of $\mathcal{S}$.}, and the signal states in the set $\mathcal{S}$ are represented
as column vectors in the basis $\mathds{B}$ \cite{LL18}:

\[
\begin{array}{lcl}
|+\mu,+\mu\rangle=\left(\begin{array}{c}
c_{0}^{2}\\
c_{1}^{2}\\
c_{0}c_{1}\\
c_{0}c_{1}
\end{array}\right), &  & |-\mu,-\mu\rangle=\left(\begin{array}{c}
c_{0}^{2}\\
c_{1}^{2}\\
-c_{0}c_{1}\\
-c_{0}c_{1}
\end{array}\right)\\
\\|+\mu,-\mu\rangle=\left(\begin{array}{c}
c_{0}^{2}\\
-c_{1}^{2}\\
-c_{0}c_{1}\\
c_{0}c_{1}
\end{array}\right), &  & |-\mu,+\mu\rangle=\left(\begin{array}{c}
c_{0}^{2}\\
-c_{1}^{2}\\
c_{0}c_{1}\\
-c_{0}c_{1}
\end{array}\right)
\end{array}.
\]

In the MDI QKD protocol, the eavesdropper (Eve) has full control over
the quantum channels connecting Alice, Bob, and the intermediate node,
Charlie, as well as the measurement devices at Charlie\textquoteright s
location. Since these devices are untrusted and not characterized,
it is assumed that Eve can mimic Charlie to perform the measurements.
Eve performs measurements, described by a POVM ${\rm E}$, on the
states from Alice and Bob in registers $A^{\prime}$ and $B^{\prime}$
. For simplicity, we assume that ${\rm E}$ consists of four elements
similar to Charlie's measurements: $\left\{ +,-,?,{\rm d}\right\} $.
Alice and Bob may only retain the $\left\{ +,-\right\} $ outcomes
for key distillation, but the $\left\{ ?,{\rm d}\right\} $ outcomes
can be used for parameter estimation. We denote the outcomes of the
POVM ${\rm E}$ as $\left\{ {\rm E}^{\gamma}\right\} $ for $\gamma\in\left\{ +,-,?,{\rm d}\right\} $.
Eve applies a ``completely positive trace-preserving'' (CPTP) map
\cite{LL18}, denoted as $\mathcal{E}_{A^{\prime}B^{\prime}\longrightarrow EC}$,
on the quantum states in registers $A^{\prime}$ and $B^{\prime}$.
The measurement results are recorded in the classical register $C$,
while the post-measurement quantum state is stored in register $E$.
Generally, $\mathcal{E}_{A^{\prime}B^{\prime}\longrightarrow EC}$
can be expressed as:

\begin{equation}
\begin{array}{lcl}
\mathcal{E}_{A^{\prime}B^{\prime}\longrightarrow EC}\left(Y\right) & = & \underset{\gamma}{\sum}\mathcal{E}_{\gamma}\left(Y\right)\otimes|\gamma\rangle\langle\gamma|_{C}.\end{array}\label{eq:CPTP=000020map}
\end{equation}
Here, $\mathcal{E}_{\gamma}$ denotes a ``completely positive trace-nonincreasing''
map \cite{LL18}, $Y$ is a linear operator on systems $A^{\prime}B^{\prime}$,
and $\left\{ |\gamma\rangle\right\} $ represents an orthonormal basis
for register $C$. Eve cannot alter the quantum states in registers
$\mathcal{A}$ and $\mathcal{B}$. Consequently, when Eve directly
interacts with the state $|\Psi\rangle_{\mathcal{AB}A^{\prime}B^{\prime}}$
from the source-replacement scheme, the resulting joint system $\rho_{\mathcal{AB}EC}$
shared by Alice, Bob, and Eve, along with the classical register $C$
containing the measurement outcomes, is described as follows:

\begin{equation}
\begin{array}{lcl}
\rho_{\mathcal{AB}EC} & = & \mathds{1}\otimes\mathcal{E}_{A^{\prime}B^{\prime}\longrightarrow EC}\left(|\Psi\rangle\langle\Psi|_{\mathcal{AB}A^{\prime}B^{\prime}}\right)\\
\\ & = & \underset{m,n,m^{\prime},n^{\prime}}{\sum}\sqrt{p_{m}p_{m^{\prime}}q_{n}q_{n^{\prime}}}|m,n\rangle\langle m^{\prime},n^{\prime}|_{\mathcal{AB}}\otimes\underset{\gamma}{\sum}\left(\sqrt{{\rm E}^{\gamma}}|\phi_{m},\phi_{n}\rangle\langle\phi_{m^{\prime}},\phi_{n^{\prime}}|\left(\sqrt{{\rm E}^{\gamma}}\right)^{\dagger}\right)_{E}\otimes|\gamma\rangle\langle\gamma|_{C}
\end{array}.\label{eq:After=000020CPTP=000020map}
\end{equation}
Alice and Bob carry out the POVM $M_{\mathcal{A}}$ and $M_{\mathcal{B}}$
on their respective registers $\mathcal{A}$ and $\mathcal{B}$. After
performing these measurements, Alice records her results in a classical
register $A$, and Bob records his in a classical register $B$. Without
loss of generality, the state $\rho_{ABEC}$ can be expressed from
the state $\rho_{\mathcal{AB}EC}$ via a CPTP map as follows:

\[
\begin{array}{lcl}
\rho_{ABEC} & = & \underset{a,b,\gamma}{\sum}p\left(\gamma\right)p\left(a,b|\gamma\right)|a\rangle\langle a|_{A}\otimes|b\rangle\langle b|_{B}\otimes\rho_{E}^{a,b,\gamma}\otimes|\gamma\rangle\langle\gamma|_{C}\\
\\ & = & \underset{\gamma}{\sum}p\left(\gamma\right)\rho_{ABE}^{\gamma}\otimes|\gamma\rangle\langle\gamma|_{C}
\end{array},
\]
and

\[
\begin{array}{lcl}
\rho_{ABE}^{\gamma} & = & \underset{a,b}{\sum}p\left(a,b|\gamma\right)|a\rangle\langle a|_{A}\otimes|b\rangle\langle b|_{B}\otimes\rho_{E}^{a,b,\gamma}\end{array},
\]
where $p(\gamma)$ represents the marginal probability derived from
the joint probability distribution $p\left(a,b,\gamma\right)$, and
$p\left(a,b|\gamma\right)=\frac{p\left(a,b,\gamma\right)}{p\left(\gamma\right)}$
denotes the conditional probability. In this context, $\rho_{E}^{a,b,\gamma}$
refers to Eve\textquoteright s conditional state, given that Alice
possesses $a$ in register $A$, Bob has $b$ in register $B$, and
the central node declares $\gamma$. This can be expressed as

\[
\begin{array}{lcl}
\rho_{E}^{a,b,\gamma} & = & \frac{\sqrt{{\rm E}^{\gamma}}|\phi_{a},\phi_{b}\rangle\langle\phi_{a},\phi_{b}|\sqrt{\left({\rm E}^{\gamma}\right)^{\dagger}}}{\boldsymbol{{\rm Tr}}\left(\sqrt{{\rm E}^{\gamma}}|\phi_{a},\phi_{b}\rangle\langle\phi_{a},\phi_{b}|\sqrt{\left({\rm E}^{\gamma}\right)^{\dagger}}\right)}\end{array}.
\]
The Devetak-Winter formula applies in scenarios where error correction
is not necessarily optimal (at the Shannon limit), providing the number
of secret bits that can be extracted from the state $\rho_{ABE}^{\gamma}$
as $r\left(\rho_{ABE}^{\gamma}\right)$, defined as:

\begin{equation}
\begin{array}{lcl}
r\left(\rho_{ABE}^{\gamma}\right) & = & {\rm max}\left[1-\delta_{{\rm EC}}^{\gamma}-\chi\left(A:E\right)_{\rho_{ABE}^{\gamma}},0\right]\end{array},\label{eq:Key-Rate=000020Equation}
\end{equation}
where $\delta_{{\rm EC}}^{\gamma}$ indicates the information leakage
per signal during error correction for rounds corresponding to the
outcome $\gamma$, and

\begin{equation}
\begin{array}{lcl}
\chi\left(A:E\right)_{\rho_{ABE}^{\gamma}} & = & S\left(\rho_{E}^{\gamma}\right)-\underset{a}{\sum}p\left(a|\gamma\right)S\left(\rho_{E}^{a,\gamma}\right)\end{array}\label{eq:Holevo=000020Information}
\end{equation}
is the Holevo information, where $S\left(\rho\right)=-\boldsymbol{{\rm Tr}}\left(\rho{\rm log}_{2}\rho\right)$
represents the von Neumann entropy. Ultimately, we derive the final
key rate equation for collective attacks in the asymptotic key limit
(refer to Ref. \cite{LL18} for more details), 

\begin{equation}
\begin{array}{lcl}
R^{\infty} & =\underset{\rho_{ABEC}\in\zeta}{{\rm min}} & \underset{\gamma}{\sum}p\left(\gamma\right)r\left(\rho_{ABE}^{\gamma}\right)\end{array}\label{eq:Final_Key_Rate_Equantion}
\end{equation}
\emph{Fiber channel description}:\emph{ }Here, we provide a brief
description of the fiber link between Alice and Charlie (Eve) under
conditions of loss and noise, as discussed in \cite{LL18}, and the
satellite link between Bob and Charlie (Eve), as described in \cite{LKB19}.
Specifically, we present the explicit analytical derivation of the
key rate for noisy scenario (see Appendix B). For the sake of simplicity,
we assume the symmetric transitivity of these two links as proposed
in \cite{LL18}. In the scenario considering only loss, the coherent
states prepared by Alice and Bob are $|\alpha_{{\rm A}}\rangle_{{\rm A^{\prime}}}$
and $|\alpha_{{\rm B}}\rangle_{{\rm B^{\prime}}}$, respectively,
which they transmit to Charlie through a lossy channel. After transmission\footnote{We take transmittance in fiber channel defined as $\eta=10^{-\frac{0.2L}{10}}$
\cite{MMA+23,LL18}. The choice of $\text{\ensuremath{\eta}}$ is
consistent with the transmittance of the commercially available telecom-grade
optical fibers which usually have a loss of 0.2 dB/km. }, the state evolves to $\left|\eta^{\frac{1}{4}}\alpha_{{\rm A}},\eta^{\frac{1}{4}}\alpha_{{\rm B}}\right\rangle _{I_{{\rm A}}I_{{\rm B}}}$,
while Eve's state is represented by $\left|\sqrt{1-\eta^{\frac{1}{2}}}\alpha_{{\rm A}},\sqrt{1-\eta^{\frac{1}{2}}}\alpha_{{\rm B}}\right\rangle _{E_{{\rm A}}E_{{\rm B}}}$.
After passing through the beam splitter, the state transforms into
$\left|\frac{\eta^{\frac{1}{4}}\left(\alpha_{{\rm A}}+\alpha_{{\rm B}}\right)}{\sqrt{2}},\frac{\eta^{\frac{1}{4}}\left(\alpha_{{\rm A}}-\alpha_{{\rm B}}\right)}{\sqrt{2}}\right\rangle _{{\rm O_{A}}{\rm O_{B}}}$
as the output state. Ultimately, the asymptotic key generation rate
as a function of $\eta$ and intensity $\mu$ in this loss-only scenario
is derived (see Appendix A for details),

\begin{equation}
R_{{\rm loss}}^{\infty}=\left(1-e^{-2\mu\sqrt{\eta}}\right)\left[1-h\left(\frac{1-e^{-4\mu\left(1-\sqrt{\eta}\right)}e^{-2\mu\sqrt{\eta}}}{2}\right)\right].\label{eq:Key-rate=000020of=000020Loss-Only=000020Scenario-1}
\end{equation}
To account for realistic imperfections, we consider factors such as
mode mismatch, phase mismatch, detector dark counts, detector inefficiency,
and error correction inefficiency. In \cite{LL18} numerical methods
were used to analyze noisy scenarios. In contrast, our research employs
analytical methods to derive the key rate equation, which we then
apply to LEO satellite quantum communication. As in the loss-only
scenario, the initial states of Alice and Bob are $\left|\alpha_{{\rm A}},\alpha_{{\rm B}}\right\rangle $.
After accounting for all imperfections, the final state can be expressed
as follows: 

\[
\begin{array}{lcl}
|\alpha_{{\rm final}}\rangle & = & \left|\frac{\eta^{\frac{1}{4}}\left(\alpha_{{\rm A}}+\sqrt{{\rm M}}\,\alpha_{{\rm B}}e^{i\delta}\right)}{\sqrt{2}},\frac{\eta^{\frac{1}{4}}\sqrt{1-{\rm M}}\,\alpha_{{\rm B}}e^{i\delta}}{\sqrt{2}},\frac{\eta^{\frac{1}{4}}\left(\alpha_{{\rm A}}-\sqrt{{\rm M}}\,\alpha_{{\rm B}}e^{i\delta}\right)}{\sqrt{2}},-\frac{\eta^{\frac{1}{4}}\sqrt{1-{\rm M}}\,\alpha_{{\rm B}}e^{i\delta}}{\sqrt{2}}\right\rangle _{{\rm O_{A1}O_{A2}}{\rm O_{B1}O_{B2}}},\\
|\beta_{{\rm final}}\rangle & = & \left|\frac{\eta^{\frac{1}{4}}\left(\beta_{{\rm A}}+\sqrt{{\rm M}}\,\beta_{{\rm B}}e^{i\delta}\right)}{\sqrt{2}},\frac{\eta^{\frac{1}{4}}\sqrt{1-{\rm M}}\,\beta_{{\rm B}}e^{i\delta}}{\sqrt{2}},\frac{\eta^{\frac{1}{4}}\left(\beta_{{\rm A}}-\sqrt{{\rm M}}\,\beta_{{\rm B}}e^{i\delta}\right)}{\sqrt{2}},-\frac{\eta^{\frac{1}{4}}\sqrt{1-{\rm M}}\,\beta_{{\rm B}}e^{i\delta}}{\sqrt{2}}\right\rangle _{{\rm O_{A1}O_{A2}}{\rm O_{B1}O_{B2}}}.
\end{array}
\]
In Appendix B, we analytically derive the necessary elements for determining
the key rate in a noisy scenario. Using Equations (\ref{eq:Holevo_Information_Noisy_Scenario}
- \ref{eq:Key_Rate_Plus_Minus_Noisy}) from Appendix B, we can obtain
the secret key generation rate under realistic imperfections as follows:

\begin{equation}
\begin{array}{lcl}
R_{{\rm noisy}}^{\infty} & = & \underset{\gamma}{\sum}p\left(\gamma\right)r\left(\rho_{ABE}^{\gamma}\right)\\
 & = & p\left(+\right)r\left(\rho_{ABE}^{+}\right)+p\left(-\right)r\left(\rho_{ABE}^{-}\right)
\end{array}.\label{eq:Final_Key_Rate_Noisy_Scenario-1}
\end{equation}

\subsection{Satellite-based optical links with the elliptic beam approximation}

\emph{Free space link description}: We employ the PDT model using the elliptic-beam approximation \cite{VSV16}
to account for the impact of random scatterers, such as haze particles
and raindrops, on optical beam propagation. These scatterers influence
the beam in two primary ways: (i) distortion of beam shape, including
stochastic deflection of the centroid, and (ii) random losses due
to scattering. Theoretical modeling is used to describe the first
effect, while the derived PDTs are validated against experimental
probability distributions \cite{VSV+17}. This model is compared with
field measurements conducted in Erlangen over a 1.6-km atmospheric
link under both daytime and nighttime conditions. The nighttime experiments
coincided with the development of a hazy turbulent atmosphere, whereas
the daytime links were influenced by light rainfall. The strong correlation
between theoretical predictions and experimental results highlights
the effectiveness of the generalized elliptic-beam model in accurately
describing quantum light propagation through turbulent atmospheric
conditions \cite{VSV+17}. In the context of satellite-based optical
links, atmospheric properties vary significantly based on parameters
such as air density, pressure, temperature, and ionized particle concentration.
These variations lead to different atmospheric layer thicknesses depending
on location. To simplify the analysis, we adopt an approximate atmospheric
model \cite{LKB19}, where the atmosphere is assumed to be uniform
up to a specific altitude $\overline{{\rm h}}$, beyond which vacuum
extends to the satellite altitude $\overline{{\rm L}}$, as illustrated
in Figure \ref{fig:PM_MDI_QKD_Satellite_Setup}. Instead of a continuously
varying atmospheric profile, this model is characterized by only two
parameters: the physical quantity\textquoteright s value within the
uniform atmospheric layer and the effective thickness$\overline{{\rm h}}$,
providing a computationally efficient framework for analyzing satellite-based
quantum communication.

 Our objective is to evaluate the performance of key rates under different weather conditions of the
PM-MDI QKD protocol. We will utilize the free space channel between
Bob and Charlie, employing the channel transmission $\eta$ for light
propagation through atmospheric links. This analysis will use the
elliptic-beam approximation with a generalized method \cite{VSV16,VSV+17,LKB19}.
This approach affects the channel transmittance, influenced by beam
parameters and the radius of the receiving aperture. Variations in
temperature and pressure due to atmospheric turbulence cause fluctuations
in the air's refractive index, leading to losses that impact the transmitted
photons. These photons are detected by a receiver with a limited aperture.
The transmitted signal can be affected by various degradation factors
such as beam wandering, deformation, and broadening. To analyze this,
we consider a Gaussian beam propagating along the $z$-axis and reaching
the aperture plane at a distance $z={\rm L}$. In this analysis, we
recognize that the assumption of ideal Gaussian beams emitted by the
transmitter is not entirely accurate. Standard telescopes typically
produce beams with intensity distributions that approximate a circular
Gaussian profile, albeit with some deviations often due to truncation
effects at the edges of optical elements. A significant consequence
of these imperfections is the inherent broadening of the beam due
to diffraction. Our model addresses this by adjusting the parameter
representing the initial beam width $\left(\mathcal{W}_{0}\right)$,
which accounts for the increased divergence in the far-field caused
by the imperfect quasi-Gaussian beam. To capture this effect, we model
the transmission of the elliptical beam through a circular aperture
and consider the statistical characteristics of the elliptical beam
as it propagates through turbulence using a Gaussian approximation.
However, our approach simplifies certain aspects, particularly the
assumption of isotropic atmospheric turbulence. For more detailed
formulations, readers are referred to the Supplemental Material of
Ref. \cite{VSV16}. This ``quasi-Gaussian'' beam \cite{VSV12,VSV16}
travels through a channel that includes both atmospheric and vacuum
regions, originating either from a satellite-based transmitter or
a ground station. The fluctuating intensity transmittance of a signal
through a circular aperture, with a telescope's receiving radius $r$,
can be expressed as follows:

\begin{equation}
\begin{array}{lcl}
\eta & = & \int_{\left|\rho\right|^{2}=r^{2}}{\rm d^{2}\boldsymbol{\rho}\left|u\left(\mathbf{\boldsymbol{\rho}},L\right)\right|^{2}.}\end{array}\label{eq:Transmittance}
\end{equation}
The term $u\left(\mathbf{\boldsymbol{\rho}},{\rm L}\right)$ represents
the beam envelope at the receiver plane, positioned at a distance
${\rm L}$ from the transmitter. The expression $\left|u\left(\mathbf{\boldsymbol{\rho}},{\rm L}\right)\right|^{2}$
signifies the normalized intensity across the transverse plane, where
$\boldsymbol{\rho}$ is the position vector in this plane. The vector
parameter ${\rm \boldsymbol{v}}$ describes the state of the beam
at the receiver plane as

\begin{equation}
{\rm \boldsymbol{v}}=\left(x_{0},y_{0},\mathcal{W}_{1},\mathcal{W}_{2},\varphi\right),\label{eq:Vector=000020of=000020beam-parameters}
\end{equation}
where the symbols $x_{0}$, $y_{0}$ denote the beam's centroid coordinates,
and $\mathcal{W}_{1(2)}$, $\varphi$ the major (minor) semi-axes
of the elliptical beam profile and the orientation angle of the elliptical
beam, respectively. The beam parameters, along with the radius of
the receiving aperture $r$, determine the transmittance. The atmosphere
is typically modeled as consisting of distinct layers, each characterized
by various physical properties such as air density, temperature, pressure,
and ionized particles, with the structure and thickness of these layers
varying based on geographic location. For simplicity, we utilize a
model for a satellite-based optical link where the atmosphere is taken
to be uniform up to a specified altitude $\overline{{\rm h}}$, beyond
which it transitions into a vacuum extending to the satellite at altitude
$\overline{{\rm L}}$, as depicted in the Figure \ref{fig:PM_MDI_QKD_Satellite_Setup}.
Instead of varying physical properties continuously with altitude,
this model focuses on two primary parameters, the physical property
value within the uniform atmospheric layer and the effective altitude
range, $\overline{{\rm h}}$. This approach is justified since atmospheric
effects are most significant within the first $15$ to $20$ km above
the Earth's surface, particularly as LEO satellites typically operate
at altitudes above 400 km. For this analysis, $\overline{{\rm L}}$
is set to $500$ km, with the zenith angle considered within $\left[0^{0},75^{0}\right]$.
Based on these assumptions, the relevant altitude range for satellite
orbits for key distribution is approximately ${\rm L}\in\left[500,1900\right]$
km\footnote{The relationship between the total free-space link length and the
zenith angle is given by ${\rm L}=\overline{{\rm L}}\sec\phi$.}. The effective atmospheric thickness $\overline{{\rm h}}$ is maintained
at $20$ km. It is also assumed that atmospheric parameters are constant
(with values greater than $0$) within this layer and drop to $0$
beyond it \cite{LKB19}.

Now, we consider the transmittance, as expressed in Eq. (\ref{eq:Transmittance}),
for an elliptical beam incident on a circular aperture of radius $r$.
The transmittance is given by the following expression \cite{VSV16}:

\begin{equation}
\begin{array}{lcl}
\eta\left(x_{0},y_{0},\mathcal{W}_{1},\mathcal{W}_{2},\varphi\right) & = & \frac{2\,\chi_{{\rm ex}t}}{\pi\mathcal{W}_{1}\mathcal{W}_{2}}\int_{0}^{r}\rho\,{\rm d}\rho\int_{0}^{2\pi}{\rm d}\theta{\rm e^{-2A_{1}\left(\rho cos\theta-\rho_{0}\right)^{2}}}{\rm e^{-2A_{2}\rho^{2}sin^{2}\theta}}e^{-2{\rm A}_{3}\left(\rho{\rm cos}\theta-\rho_{0}\right)\rho{\rm sin}\theta}.\end{array}\label{eq:PDT=000020Equation}
\end{equation}
In this scenario, $r$ is the aperture's radius, while $\rho$ and
$\theta$ are the polar coordinates of the vector $\boldsymbol{\rho}$.
Similarly, $\rho_{0}$ and $\theta_{0}$ are the polar coordinates
corresponding to the vector $\boldsymbol{\rho}_{0}$, where $x=\rho{\rm \,cos}\theta,$
$y=\rho{\rm \,sin}\theta,$$x_{0}=\rho_{0}{\rm \,sin}\theta_{0}$
and $y_{0}=\rho_{0}{\rm \,sin}\theta_{0}$, and

\[
\begin{array}{lcl}
{\rm A}_{1} & = & \left(\frac{{\rm cos}^{2}\left(\varphi-\theta_{0}\right)}{\mathcal{W}_{1}^{2}}+\frac{{\rm sin}^{2}\left(\varphi-\theta_{0}\right)}{\mathcal{W}_{2}^{2}}\right),\\
{\rm A}_{2} & = & \left(\frac{{\rm sin}^{2}\left(\varphi-\theta_{0}\right)}{\mathcal{W}_{1}^{2}}+\frac{{\rm cos}^{2}\left(\varphi-\theta_{0}\right)}{\mathcal{W}_{2}^{2}}\right),\\
{\rm A}_{3} & = & \left(\frac{1}{\mathcal{W}_{1}^{2}}-\frac{1}{\mathcal{W}_{2}^{2}}\right){\rm sin\,2\left(\varphi-\theta_{0}\right).}
\end{array}
\]
These expressions can be used for numerical integration, as shown
in Eq. (\ref{eq:PDT=000020Equation}), using the Monte Carlo method
or another appropriate technique. To facilitate integration with the
Monte Carlo method, $N$ sets of values for the vector ${\rm \boldsymbol{v}}$
need to be generated (see Eq. (\ref{eq:Vector=000020of=000020beam-parameters})).
It is assumed that the angle $\left(\varphi-\theta_{0}\right)$ is
uniformly distributed over the interval $[0,\frac{\pi}{2}]$, along
with other parameters\footnote{To calculate transmittance, $\mathcal{W}_{{\rm i}}$ must first be
derived from $\Theta_{{\rm i}}$ using the relation $\begin{array}{lcl}
\Theta_{{\rm i}} & = & \ln\left(\frac{\mathcal{W}_{{\rm i}}^{2}}{\mathcal{W}_{{\rm 0}}^{2}}\right),\end{array}$ where ${\rm i}=1,2$. Here, $\mathcal{W}_{0}$ represents the beam
spot radius at the transmitter.}. The variables $(x_{0},y_{0},\Theta_{{\rm 1}},\Theta_{{\rm 2}})$
are assumed to follow a normal distribution \cite{WHW+18}. By substituting
the simulated values of ${\rm \boldsymbol{v}}$ into Eq. (\ref{eq:PDT=000020Equation}),
numerical integration can be performed. This process also incorporates
the \emph{extinction factor}\footnote{The parameter $\chi_{{\rm ext}}(\phi)$ represents the extinction
losses due to atmospheric back-scattering and absorption. Its value
changes based on the elevation angle $\left(90^{\circ}-\phi\right)$
or zenith angle $(\phi)$ \cite{BSH+13,VBB2000}.}, $\chi_{{\rm ext}}$, resulting in $N$ atmospheric transmittance
values, denoted as $\eta\left({\rm \boldsymbol{v}_{i}}\right)$, where
$i$ ranges from $1$ to $N$.

In the following section, we will assess the performance of the PM-MDI
QKD protocol when applied to both satellite and fiber optic links.
This evaluation requires the calculation of average key rates (AKR)
based on the PDT for various link lengths and configurations. This
can be represented as,

\begin{equation}
\begin{array}{lclcl}
\bar{R} & = & \intop_{0}^{1}R(\eta)\,P(\eta)\,{\rm d}\eta & = & \stackrel[{\rm i}=1]{N_{bins}}{\sum}R(\eta_{{\rm i}})\,P(\eta_{{\rm i}}).\end{array}\label{eq:Average=000020key-rate}
\end{equation}
Here, $\bar{R}$ represents the average key rate, while $R(\eta)$
denotes the key rate for a specific transmittance value and the PDT
is $P(\eta)$. To compute the integral average, the interval $[0,1]$
is divided into $N_{bins}$ bins, each centered at $\eta_{{\rm i}}$
for $i$ ranging from $1$ to $N_{bins}$, with the rates summed according
to their respective weights. The value of $P(\eta_{{\rm i}})$ is
determined using random sampling as described in the preceding paragraph.
The key rate formulations for different scenarios, $R(\eta)$, can
be found in Eqs. (\ref{eq:Key-rate=000020of=000020Loss-Only=000020Scenario-1})
and (\ref{eq:Final_Key_Rate_Noisy_Scenario-1}).

\section{Performance analysis of satellite-based PM-MDI QKD protocol }\label{sec:III}

In this section, we thoroughly examine the effect of PDT\footnote{Refer to Figures 3 and 4 in Ref. \cite{LKB19} for PDT following the
random sampling with beam parameters ${\rm \boldsymbol{v}}$ in a
down-link configuration.} on the key rate following the weighted sum, and the probability distribution
of the average key rate (PDR) for both loss-only and noisy scenarios
(fiber channel) for the PM-MDI QKD protocol. The minimum distance
between Bob and Charlie (the satellite's altitude) is kept constant
at $\overline{{\rm L}}=500$ km, focusing on scenarios involving LEO
satellites, like the Chinese satellite \emph{Micius} \cite{LCL+17,YCL+17,RXY+17,YCLL+17}.
However, the fiber channel link distance between Alice and Charlie
may vary to make this study more general. We present the results of
numerical simulations for the satellite-based PM-MDI QKD scheme under
asymptotic conditions, using the experimental parameters detailed
in Table I\footnote{Here, we use the wavelength $\lambda=1550$ nm.}
in Ref. \cite{DMB+24}. Satellite-based optical communication links are highly susceptible
to weather conditions. Atmospheric turbulence and scattering particles,
such as haze and fog, introduce random fluctuations in the air\textquoteright s
relative permittivity across various spatial and temporal scales.
These fluctuations complicate light propagation by causing random
beam deviations and distortions. Consequently, this leads to reduced
transmittance due to geometric losses associated with the finite aperture
of the receiver, along with random phase front variations. Therefore,
a comprehensive model of these atmospheric effects is essential for
accurately assessing the performance of such links in quantum communication
protocols \cite{VSV+17,LKB19}. In this context, $C_{n}^{2}$ represents
the \emph{refractive index structure constant},
which remains valid only for the specific geographical location and
atmospheric conditions from which it was experimentally derived (see
\cite{LKB19} and references therein). Additional losses caused by
back-scattering and absorption in the atmosphere are accounted for
using the extinction parameter $\chi_{{\rm ext}}$. In this model,
the atmospheric conditions are fully characterized by the values of
$C_{n}^{2}$ and $n_{0}$ (the density of scattering particles), along
with the atmospheric layer thickness $\overline{{\rm h}}$ (where
$h=\overline{{\rm h}}\sec\phi$) and the extinction factor $\chi_{{\rm ext}}$.
Since weather conditions vary, the values of $C_{n}^{2}$ and $n_{0}$
will also differ accordingly (see Table \ref{tab:Table1}). The parameters
$C_{n}^{2}$, $n_{0}$, and $h$ are typically determined through
experimental data fitting. However, in this study, these values are
parameterized systematically to develop a predictive model. Simulations
are performed under various atmospheric conditions, including clear,
slightly foggy, and moderately foggy nights, as well as non-windy,
moderately windy, and windy days \cite{DMB+24}. A key aspect of this
analysis is the comparison between nighttime and daytime operations.
Daytime conditions are characterized by higher temperatures, which
result in stronger winds and increased mixing between atmospheric
layers, leading to greater turbulence and higher values of $C_{n}^{2}$
compared to nighttime. On clear days, the lower atmosphere generally
contains less moisture than at night in the same location, reducing
beam spreading due to scattering particles. At night, lower temperatures
create a less turbulent atmosphere but also contribute to haze and
mist formation. The presence of haze and mist at night increases the
value of $n_{0}$ compared to daytime. In these conditions, the effects
of scattering over particulate matter can outweigh those caused by
turbulence. Crucial parameters in this context include atmospheric
effects, the radii of the transmitting and receiving telescopes, and
the signal wavelength. For the satellite, a transmitting telescope
with a radius of $r_{{\rm sat}}=0.15$ m $\left(\mathcal{W}_{0}\right)$
is chosen, while the ground station telescope features a radius of
$r_{{\rm grnd}}=0.5$ m, with the signal wavelength set at $\lambda=1550$
nm.

\begin{table}
\centering
\begin{tabular*}{12cm}{@{\extracolsep{\fill}}ccc}
\hline 
Parameter & Value & Short description\tabularnewline
\hline 
$n_{0}$ & 0.61 ${\rm m^{-3}}$ & Night-time condition 1\tabularnewline
$n_{0}$ & 0.01 ${\rm m^{-3}}$ & Day-time condition 1\tabularnewline
$n_{0}$ & 3.00 ${\rm m^{-3}}$ & Night-time condition 2\tabularnewline
$n_{0}$ & 0.05 ${\rm m^{-3}}$ & Day-time condition 2\tabularnewline
$n_{0}$ & 6.10 ${\rm m^{-3}}$ & Night-time condition 3\tabularnewline
$n_{0}$ & 0.10 ${\rm m^{-3}}$ & Day-time condition 3\tabularnewline
$C_{n}^{2}$ & $1.12\times10^{-16}$ ${\rm m^{-\frac{2}{3}}}$ & Night-time condition 1\tabularnewline
$C_{n}^{2}$ & $1.64\times10^{-16}$ ${\rm m^{-\frac{2}{3}}}$ & Day-time condition 1\tabularnewline
$C_{n}^{2}$ & $5.50\times10^{-16}$ ${\rm m^{-\frac{2}{3}}}$ & Night-time condition 2\tabularnewline
$C_{n}^{2}$ & $8.00\times10^{-16}$ ${\rm m^{-\frac{2}{3}}}$ & Day-time condition 2\tabularnewline
$C_{n}^{2}$ & $1.10\times10^{-15}$ ${\rm m^{-\frac{2}{3}}}$ & Night-time condition 3\tabularnewline
$C_{n}^{2}$ & $1.60\times10^{-15}$ ${\rm m^{-\frac{2}{3}}}$ & Day-time condition 3\tabularnewline
\hline 
\end{tabular*}\caption{ Parameters correspond to the different atmospheric weather conditions
\cite{DMB+24}.}\label{tab:Table1}
\end{table}

According to the findings of Refs. \cite{LKB19,DMB+24}, in down-links
that refer to communications from satellite to the ground, atmospheric
effects become significant only during the final phase of propagation.
Specifically, when $z$ surpasses $({\rm L-h})$. Conversely, in up-link
communication, these effects are relevant only when $z$ is less than
${\rm h}$, with $z$ representing the longitudinal coordinate. The
atmospheric effects are significantly more severe for up-links than
for down-links. These effects, such as beam deflection and broadening,
involve angular influences that alter the final beam diameter, thus
affecting channel losses. The magnitude of these effects is directly
related to the distance traveled after the onset of what is known
as the \emph{kick in effect}. In up-links, these phenomena arise near
the transmitter, causing the beam to broaden over several hundred
kilometers before reaching the satellite. Conversely, in down-link
transmissions, the majority of the beam's path is in vacuum, with
atmospheric interference becoming significant only within the final
15 to 20 km before it reaches the receiver. Additionally, up-links
and down-links differ in terms of the origin of fluctuations in the
beam centroid position, denoted by $(x_{0},y_{0})$. In up-links,
atmospheric deflections are much more significant than pointing errors
($\varphi$), which can be disregarded. In down-link scenarios, the
beam size is already significantly larger than the atmospheric turbulence
near the upper atmosphere, which minimizes beam wandering due to atmospheric
disturbances. As a result, pointing errors become the dominant factor
influencing performance. 

\begin{figure}[h]
\begin{centering}
\includegraphics[scale=0.4]{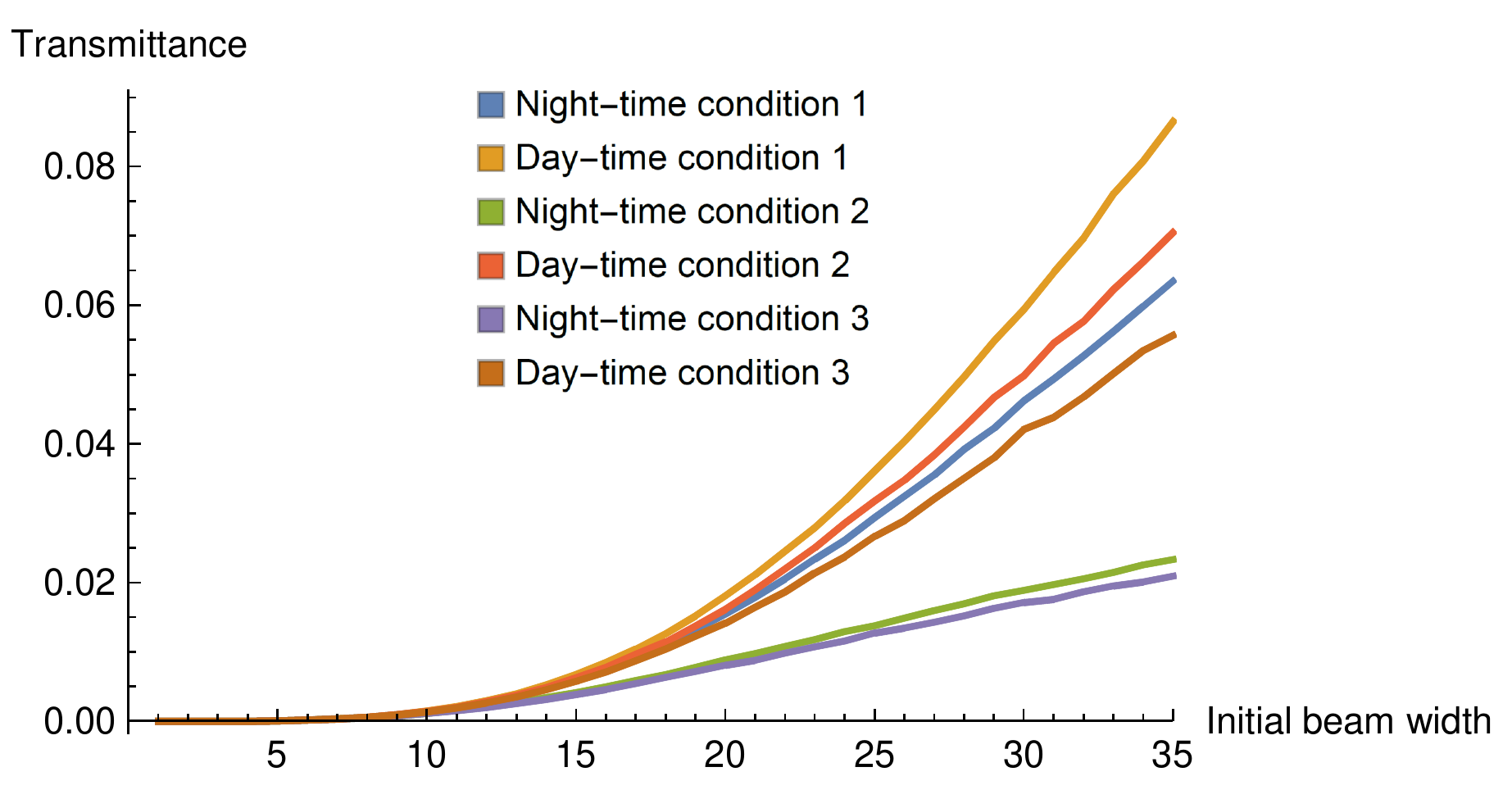}
\par\end{centering}
\caption{(Color online) Plot illustrating the variation in transmittance of
the PM-MDI QKD protocol as a function of the initial beam width $\mathcal{W}_{0}$
at the zenith position.}\label{fig:Transmittance_InitialBeamWidth}
\end{figure}

 We also simulate the
impact of the initial beam width on transmittance. For this analysis,
we assume the satellite is positioned at the zenith angle $\left(\phi=0^{\circ}\right)$
and vary the initial beam width $\left(\mathcal{W}_{0}\right)$ in
the range $r_{{\rm sat}}=0$ to $0.35$ m. As shown in Figure \ref{fig:Transmittance_InitialBeamWidth},
transmittance increases significantly with the increase in initial
beam width. Notably, under day 1 condition, the transmittance reaches
its maximum value at $r_{{\rm sat}}=0.35$, outperforming all other
weather scenarios. For the same beam width, the transmittance values
for day 2, night 1, and day 3 are nearly identical. However, under
night 2 and night 3 conditions, the transmittance is substantially
lower compared to the other weather conditions. At $r_{{\rm sat}}=0.15$
m, the transmittance appears nearly uniform across all conditions.
A larger beam width helps reduce beam divergence, resulting in better
overlap with the ground station\textquoteright s receiving aperture---an
advantage in terms of signal collection. However, a wider beam also
becomes more susceptible to atmospheric turbulence, leading to increased
beam wandering, scintillation, and phase distortions, all of which
can adversely affect the quantum bit error rate (QBER). For our AKR
and PDR simulations, we select $r_{{\rm sat}}=0.15$ m as the initial
beam width to balance these competing effects.

\begin{figure}[h]
\begin{centering}
\includegraphics[scale=0.5]{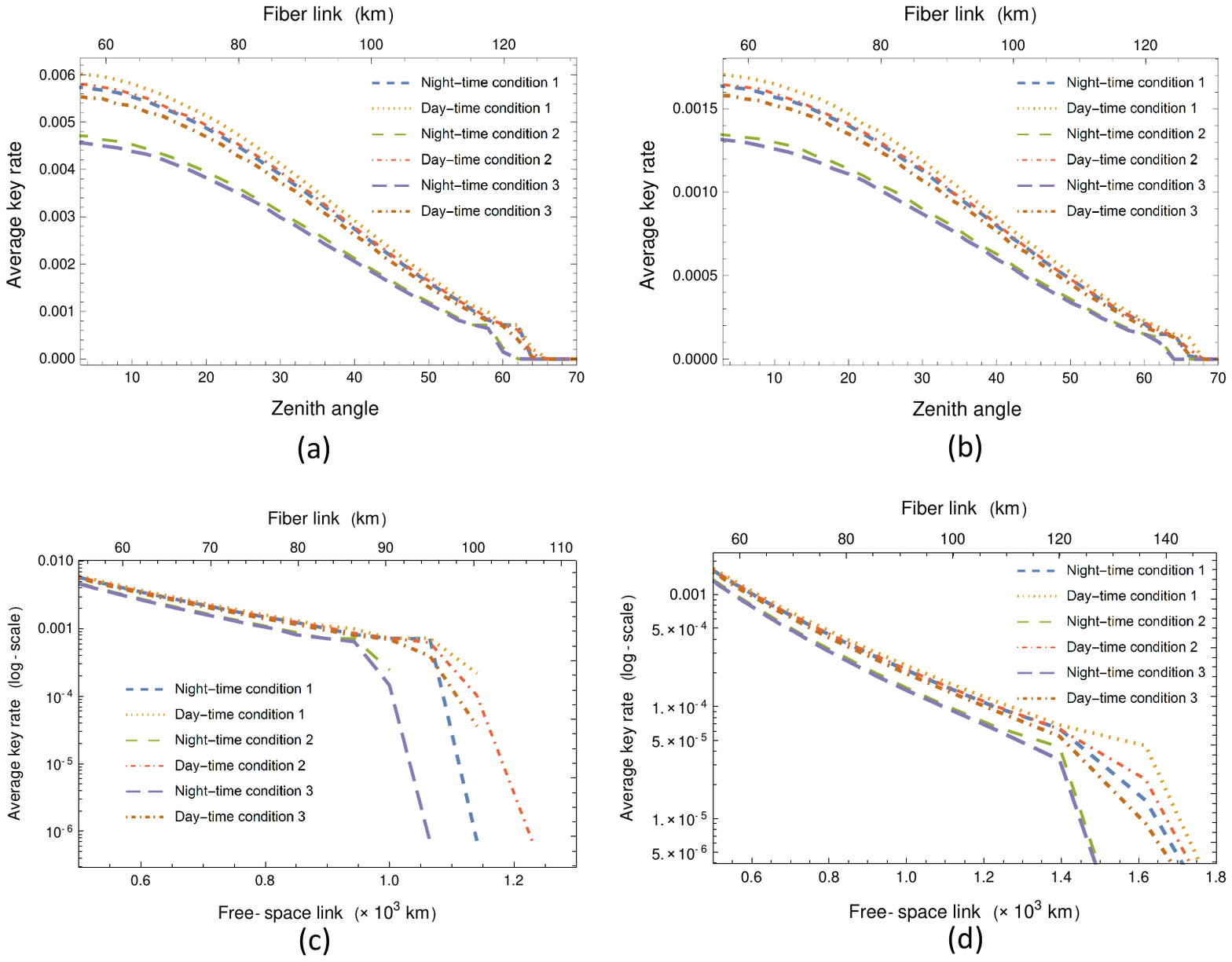}
\par\end{centering}
\caption{(Color online) Plot depicting the variation of AKR (per pulse) of
the PM-MDI QKD protocol as a function of zenith angle (free-space
link) and corresponding fiber link in different weather conditions
i.e., day-time scenarios 1, 2 and 3, characterized by no wind, moderate
wind and windy conditions, respectively (referred to as Day 1/2/3).
Night-time scenarios 1, 2, 3 are represented by clear, slightly foggy,
and moderate foggy conditions, respectively (referred to as Night
1/2/3): (a) AKR as a function of zenith angle and corresponding fiber
link for down-link configuration under different weather conditions
and loss-only scenario, respectively, (b) Average key rate as a function
of zenith angle and corresponding fiber link for down-link configuration
under different weather conditions and noisy scenario, respectively,
 (c) and (d) present
the same results as shown in (a) and (b), respectively; however, for
improved visualization of the impact of link length on the average
key rate, a logarithmic scale is applied to the y-axis. On the x-axis,
the free-space link is represented in terms of the zenith angle, as
in (a) and (b).}\label{fig:AKR_PLOT_Loss_Noisy}
\end{figure}

Based on the above facts, we simulate the PM-MDI QKD setup using a
down-link configuration in the free-space link between Charlie and
Bob. We employ Eqs. (\ref{eq:Key-rate=000020of=000020Loss-Only=000020Scenario-1})
and (\ref{eq:Final_Key_Rate_Noisy_Scenario-1}), along with the PDT
of the free-space link (see Eq. (\ref{eq:PDT=000020Equation})), to
generate the simulation results of the AKR for a lossy and noisy fiber
link between Alice and Charlie, and the free-space-link between Charlie
and Bob. The simulation results are shown in Figure \ref{fig:AKR_PLOT_Loss_Noisy}.
To generalize our findings, we evaluate the AKR for various zenith
angles in the free-space link and different distances in the fiber
link. Since this protocol uses non-phase-randomized coherent states
$|\sqrt{\mu}e^{i\theta}\rangle$, we optimize the intensity ($\mu$)
for each point on the plot. Additionally, each data point on the graph
is derived from $1000$ parameter samples from Eq. (\ref{eq:Vector=000020of=000020beam-parameters})
with suitable distribution and calculated using Eq. (\ref{eq:PDT=000020Equation}). It should be noted that
Figure \ref{fig:AKR_PLOT_Loss_Noisy} (a) and (b) present the same
results as Figure \ref{fig:AKR_PLOT_Loss_Noisy} (c) and (d), respectively.
The latter figures are formatted for improved visualization of the
X-axis. Specifically, the zenith angles in (a) and (b) are equivalent
to the free-space link distances in (c) and (d), using the transformation
${\rm L}=\overline{{\rm L}}\sec\phi$. Additionally, the Y-axis in
Figures (c) and (d) employs a logarithmic scale to correspond with
the linear Y-axis used in Figures (a) and (b).
Figures \ref{fig:AKR_PLOT_Loss_Noisy} (a) and (b) show that under
day-time condition 1, the highest AKR at the zenith position for the
loss-only and noisy scenarios are approximately $0.006$ and $0.0017$,
respectively. Notably, the AKR is higher for the loss-only scenario
in comparison to the noisy scenario that accounts for realistic imperfections.
This result is logically expected. The AKR for both scenarios under
night-time condition 1 is nearly the same as in day-time condition
2. At the zenith position, the AKR for the loss-only and noisy scenarios
in day-time condition 2 are approximately $0.00580$ and $0.00165$,
respectively. The graph lines for night-time conditions 2 and 3 nearly
overlap in both loss-only and noisy scenarios, as shown in Figures
\ref{fig:AKR_PLOT_Loss_Noisy} (a) and (b)\textcolor{black}{.} The
key rates in these cases are notably lower than in the other scenarios.
At the zenith position, the AKR values are at their lowest under night-time
condition 3, at approximately $0.0046$ and $0.0013$ for loss-only
and noisy scenarios, respectively. The order of different weather
conditions that yield higher key rate values is as follows, day 1,
night 1 (or day 2), day 3, night 2, and night 3. The ratios of the
AKR between loss and noisy scenarios at zenith for day condition 1,
day condition 2, day condition 3, and night condition 3 are $3.53,3.51,3.48$,
and $3.53$, respectively\footnote{The ratio for night condition 1 and night condition 2 is disregarded
as their graph lines are close to day condition 2 and night condition
3, respectively.}. These ratios suggest that the lossy fiber-link configuration is
more advantageous than the noisy fiber-link configuration. In Figures
\ref{fig:AKR_PLOT_Loss_Noisy} (c) and (d), the AKR remains significant,
up to the order of $10^{-6}\left(0.5\times10^{-6}\right)$ with $1250$
km ($1750$ km) and $110$ km ($140$ km) for free-space and fiber-link,
respectively, in the loss-only (noisy) scenario. In the noisy channel,
the AKR is lower than that in the loss-only channel but sustains longer
distances for both fiber and satellite links. Moreover, the AKR decreases
more rapidly with distance in the noisy channel than in the loss-only
channel. This is because, in addition to photon loss---which characterizes
the loss-only channel---the noisy channel introduces additional quantum
noise sources such as mode mismatch, phase fluctuations, error correction
inefficiency, and detector inefficiencies. These factors degrade the
fidelity of the transmitted quantum states, leading to a more pronounced
reduction in the AKR compared to the loss-only case. Since MDI-QKD
is inherently robust to certain types of noise, it can still enable
secure key extraction under moderate noise levels. Techniques like
post-selection can help extract keys from noisy signals over extended
distances, although at the cost of reduced key rates. Therefore, while
the PM-MDI QKD protocol supports long-distance key generation even
in noisy conditions, it exhibits a steeper decline in AKR as both
noise and loss accumulate with distance. As a result, when comparing
AKR performance over distance, the loss-only channel consistently
outperforms the noisy channel. Daytime conditions generally provide better channel
transmission (free-space-link) compared to nighttime. This trend is
consistent in both scenarios. A key focus is the comparison between
day-time and night-time operations. During the day, higher temperatures
lead to stronger winds and increased mixing across different atmospheric
layers, which results in more noticeable turbulence. However, clear
days typically have lower moisture content in the lower atmosphere
than at night, causing less beam spreading due to particle scattering.
At night, cooler temperatures reduce turbulence but lead to the formation
of mist and haze. In these conditions, scattering has a greater impact
during night-time than turbulence does during the day. Here, it is
important to note that we have visualized the situation in such a
way that the wavelength of the signal in both the free space channel
and the fiber channel used in the PM-MDI QKD scheme is 1550 nm. The
choice of the wavelength for communication through optical fiber is
justified as the loss is minimal at this wavelength, but the choice
for the free space part needs a bit of discussion as the free space
communication is often done at 800 nm due to the fact that the single-photon
detectors are more efficient at that wavelength. The present scenario
is different from the investigated scheme of the PM-MDI QKD scheme,
we need to use the same wavelength in both the channels. Here, we
select 1550 nm. As detectors are less efficient, free space channel
will have a relatively lesser number of clicks, but there are certain
advantages of using 1550 nm for free-space communication. Specifically,
there is an atmospheric window at 1550 nm and the transmission efficiency
at 1550 nm is slightly higher than that at 800 nm \cite{LYL+17}.
Further, the sunlight contains a considerably large amount of 800
nm light compared to 1550 nm (in fact the intensity of 800 nm in sunlight
is about 5 times that of 1550 nm). This reduces the possibility of
false detection and allows us to simulate a situation where free-space
quantum communication is performed in the daytime, too. Finally, Rayleigh
scattering at 1550 nm can be computed to be $\sim$ 7\% of its value
at 800 nm \cite{LYL+17}. In fact, in Ref. \cite{LYL+17}, the noise
count rate of 1550 nm was measured in the daylight scenario to simulate
satellite-to-Earth communication, and the result was found to be smaller
by a factor of 22.5 in comparison to the same for 850 nm light. The
above rationale is used for the choice of wavelength in the present
study.

\begin{figure}[h]
\begin{centering}
\includegraphics[scale=0.5]{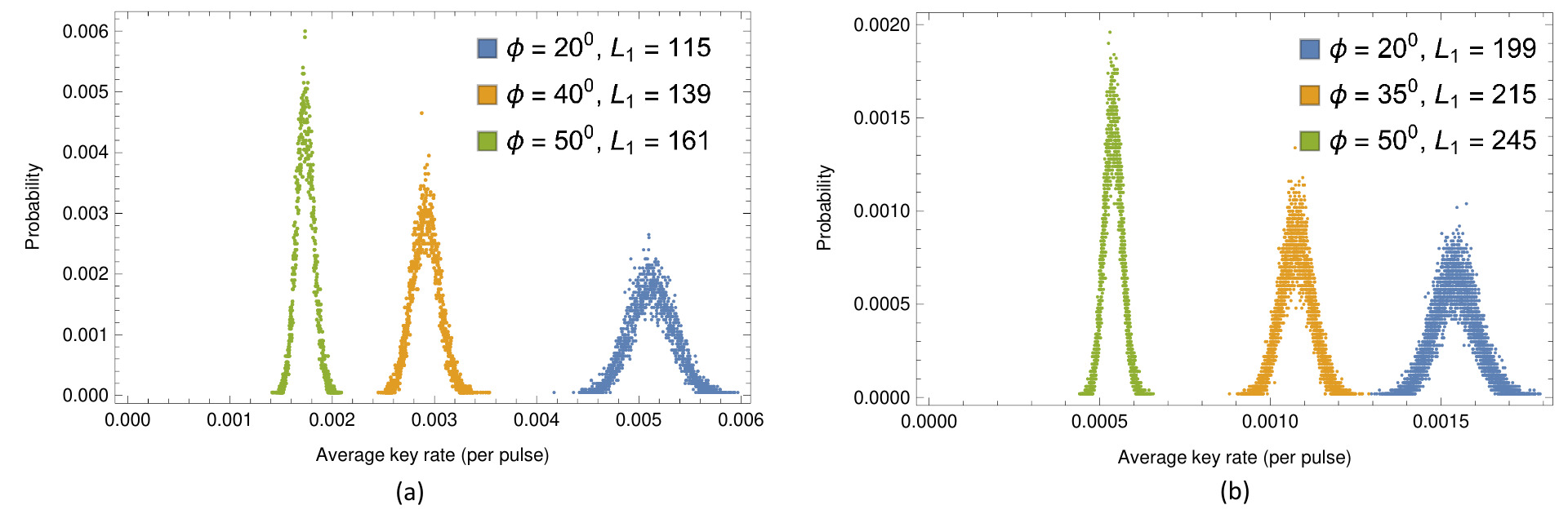}
\par\end{centering}
\caption{(Color online) Plot for distribution of average key-rate (AKR) of PM-MDI
QKD protocol variation with different zenith angles $\left(\phi\right)$
and corresponding fiber link length ($L_{1}$) under condition Day-1
in down-link configuration: (a) Probability distribution of key-rate
for loss-only scenario in fiber link, (b) Probability distribution
of key-rate for noisy scenario in fiber link under down-link configuration.}\label{fig:PDR_Loss_Noisy}
\end{figure}

We now discuss the PDR in both loss-only and noisy scenarios, as shown
in Figure \ref{fig:PDR_Loss_Noisy}. In this satellite-based PM-MDI
QKD scheme, we take the down-link configuration for a free-space link.
To achieve optimal performance, we simulate the PDR under day-time
condition 1, with appropriate zenith angles $\left(\phi\right)$ and
fiber-link distances $\left(L_{1}\right)$, for both loss-only and
noisy scenarios. A dataset of $10^{5}$ beam parameters is utilized
for simulating the AKR and approximating the results to six decimal
places for loss-only scenarios and seven decimal places for noisy
scenarios \footnote{This approximation is well-chosen and highly suitable for representing
PDR effectively.}. In the loss-only scenario, as shown in Figure \ref{fig:PDR_Loss_Noisy}
(a), we compare different scenarios with $\phi=20^{\circ},L_{1}=115\,{\rm km}$;
$\phi=40^{\circ},L_{1}=139\,{\rm km}$; and $\phi=50^{\circ},L_{1}=161\,{\rm km}$.
Here, it may be noted that the selected values $\phi$ and $L_{1}$
are nor arbitrary. In fact, they are obtained systematically, as in
our plots, the key rates are calculated only for those values where
the transmittance of fiber and satellite are the same. This constraint
leads to specific distant locations in the ground station corresponding
to specific values of zenith angles in the free space channel. The
highest AKR is observed for $\phi=20^{\circ},L_{1}=115\,{\rm km}$,
although the maximum value of the probability of AKR is higher for
$\phi=50^{\circ},L_{1}=161\,{\rm km}$. For $\phi=40^{\circ},L_{1}=139\,{\rm km}$,
the AKR value is higher (lower), and the maximum probability of AKR
is lower (higher) compared to the cases of $\phi=50^{\circ},L_{1}=161\,{\rm km}$
$\left(\phi=20^{\circ},L_{1}=115\,{\rm km}\right)$. Therefore,
greater zenith angles and fiber-link distances generally yield higher
maximum values of probability. Importantly, a higher key rate is associated
with a lower probability of occurrence. In Figure \ref{fig:PDR_Loss_Noisy}
(b), the PDR for noisy conditions is plotted with different values
for zenith angles and fiber-link distances. The PDR exhibits similar
characteristics in noisy (considering realistic imperfections of the
fiber channel) scenarios too, but both the AKR and the maximum probability
are lower than in the loss-only scenario, as expected. Additionally,
the data point density is higher for lower zenith angles and shorter
fiber-link distances in both scenarios (see Figure \ref{fig:PDR_Loss_Noisy}
(a) and (b)). In the loss-only scenario, there are notably higher
key rate values and probabilities compared to the noisy scenario.
The spread of the PDR along the AKR axis is significantly greater
in the loss-only scenario than in the noisy scenario. For instance,
at $\phi=20^{\circ}$, the spreads of the PDR along the AKR axis
are approximately $0.0015$ and $0.0005$ for the loss-only and noisy
scenarios, respectively. This indicates that the fiber performs better
in the loss-only scenario compared to the noisy one. The specific
shape of the PDT suggests that the PDR shape would remain consistent
across varying zenith angles (or equivalently, different free-space
link distances) and fiber-link distances.

\section{Conclusion}\label{sec:IV}

In this study, the possibility of implementing the PM-MDI QKD scheme
\cite{LL18} in a satellite-based quantum communication scenario is
investigated. Here, Bob's setup is located on the satellite, while
Charlie is positioned at a ground station where atmospheric noise
is minimal. Bob transmits quantum signals to Charlie via a down-link
configuration in free-space communication. Alice, who may be situated
in a more noisy atmospheric environment, such as a city, is connected
to Charlie through a fiber channel that may experience both loss and
noise. We derive an analytic expression for the key rate in noisy
conditions and use it alongside the key rate for a loss-only scenario
to simulate the performance of the PM-MDI QKD scheme in a satellite-based
setup with an elliptic-beam approximation. We conduct a detailed analysis
of the AKR under varying zenith angles and fiber-link distances, considering
different weather and day-night conditions. Each point in the graphical
representation is obtained using the optimized intensity value ($\mu$).
The results show that the AKR remains in the order of $10^{-6}$ for
the loss-only scenario and $0.5\times10^{-6}$ for the noisy scenario,
with a more pronounced decay in the noisy environment. \textcolor{black}{In
the noisy channel, the AKR is lower than that in the loss-only channel,
but a noisy channel can support communication over relatively longer
distances for both fiber and satellite links. }

\textcolor{black}{Moreover, the AKR decreases faster with distance
in the noisy channel than in the loss-only channel but maintains a
lower key rate over longer distances compared to the loss-only scenario.
Consequently, when evaluating the AKR in relation to distance, the
loss-only channel is found to demonstrate better performance than
the noisy channel.} We also plot the probability distribution of the
AKR, observing that the density of data points and the key rate is
higher for lower zenith angles and shorter fiber-link distances, whereas
both decrease for higher zenith angles and longer distances. The maximum
probability of AKR is found to be higher for lower key rates in both
scenarios. The probability distribution of the AKR maintains a consistent
shape across all scenarios, as we used a specific probability distribution
for the beam parameters. It should be noted that different probability
distributions, based on specific atmospheric conditions and altitudes,
can be employed for more accurate simulations, and empirical data
can further enhance accuracy.

Using a low-loss or less-noisy fiber is the most effective approach
for implementing the PM-MDI QKD protocol. Our work can be extended
by analyzing transmittance in satellite and fiber-based scenarios
with independent variations. Finite key analysis can also be conducted
for both fiber and satellite-based scenarios involving free-space
and fiber transmittance. Our findings suggest that achieving a significant
key rate between distant parties (Alice and Bob) with variable positioning
is feasible using a satellite-based implementation of the PM-MDI QKD
protocol. The implementation is achievable with current technology,
as it uses a non-phase randomized coherent state source. Our analytical
calculations also provide a foundation for further investigation of
key rates under finite key conditions and different attack scenarios.

\subsection*{Acknowledgment: }

The authors acknowledge support from the Indian Space Research Organisation
(ISRO) project no: ISRO/RES/3/906/22-23.

\section*{Availability of data and materials}

No additional data is needed for this work.

\section*{Competing interests}

The authors declare that they have no competing interests.

\bibliographystyle{apsrev4-2}
\bibliography{SatelliteSecond}

\begin{thebibliography}{81}%
\makeatletter
\providecommand \@ifxundefined [1]{%
 \@ifx{#1\undefined}
}%
\providecommand \@ifnum [1]{%
 \ifnum #1\expandafter \@firstoftwo
 \else \expandafter \@secondoftwo
 \fi
}%
\providecommand \@ifx [1]{%
 \ifx #1\expandafter \@firstoftwo
 \else \expandafter \@secondoftwo
 \fi
}%
\providecommand \natexlab [1]{#1}%
\providecommand \enquote  [1]{``#1''}%
\providecommand \bibnamefont  [1]{#1}%
\providecommand \bibfnamefont [1]{#1}%
\providecommand \citenamefont [1]{#1}%
\providecommand \href@noop [0]{\@secondoftwo}%
\providecommand \href [0]{\begingroup \@sanitize@url \@href}%
\providecommand \@href[1]{\@@startlink{#1}\@@href}%
\providecommand \@@href[1]{\endgroup#1\@@endlink}%
\providecommand \@sanitize@url [0]{\catcode `\\12\catcode `\$12\catcode
  `\&12\catcode `\#12\catcode `\^12\catcode `\_12\catcode `\%12\relax}%
\providecommand \@@startlink[1]{}%
\providecommand \@@endlink[0]{}%
\providecommand \url  [0]{\begingroup\@sanitize@url \@url }%
\providecommand \@url [1]{\endgroup\@href {#1}{\urlprefix }}%
\providecommand \urlprefix  [0]{URL }%
\providecommand \Eprint [0]{\href }%
\providecommand \doibase [0]{https://doi.org/}%
\providecommand \selectlanguage [0]{\@gobble}%
\providecommand \bibinfo  [0]{\@secondoftwo}%
\providecommand \bibfield  [0]{\@secondoftwo}%
\providecommand \translation [1]{[#1]}%
\providecommand \BibitemOpen [0]{}%
\providecommand \bibitemStop [0]{}%
\providecommand \bibitemNoStop [0]{.\EOS\space}%
\providecommand \EOS [0]{\spacefactor3000\relax}%
\providecommand \BibitemShut  [1]{\csname bibitem#1\endcsname}%
\let\auto@bib@innerbib\@empty
%</preamble>
\bibitem [{\citenamefont {Rivest}\ \emph {et~al.}(1983)\citenamefont {Rivest},
  \citenamefont {Shamir},\ and\ \citenamefont {Adleman}}]{RSA83}%
  \BibitemOpen
  \bibfield  {author} {\bibinfo {author} {\bibfnamefont {R.~L.}\ \bibnamefont
  {Rivest}}, \bibinfo {author} {\bibfnamefont {A.}~\bibnamefont {Shamir}},\
  and\ \bibinfo {author} {\bibfnamefont {L.~M.}\ \bibnamefont {Adleman}},\
  }\href@noop {} {\bibinfo {title} {Cryptographic communications system and
  method}} (\bibinfo {year} {1983}),\ \bibinfo {note} {{US} Patent
  4,405,829}\BibitemShut {NoStop}%
\bibitem [{\citenamefont {Bennett}\ and\ \citenamefont
  {Brassard}(1984)}]{BB84}%
  \BibitemOpen
  \bibfield  {author} {\bibinfo {author} {\bibfnamefont {C.~H.}\ \bibnamefont
  {Bennett}}\ and\ \bibinfo {author} {\bibfnamefont {G.}~\bibnamefont
  {Brassard}},\ }\href@noop {} {\bibinfo {title} {Quantum cryptography:
  Public-key distribution and coin tossing, {in Proc. IEEE Int. Conf. on
  Computers, Systems, and Signal Processing (Bangalore, India, 1984), pp.
  175-179.}}} (\bibinfo {year} {1984})\BibitemShut {NoStop}%
\bibitem [{\citenamefont {Shenoy-Hejamadi}\ \emph {et~al.}(2017)\citenamefont
  {Shenoy-Hejamadi}, \citenamefont {Pathak},\ and\ \citenamefont
  {Radhakrishna}}]{SPR17}%
  \BibitemOpen
  \bibfield  {author} {\bibinfo {author} {\bibfnamefont {A.}~\bibnamefont
  {Shenoy-Hejamadi}}, \bibinfo {author} {\bibfnamefont {A.}~\bibnamefont
  {Pathak}},\ and\ \bibinfo {author} {\bibfnamefont {S.}~\bibnamefont
  {Radhakrishna}},\ }\href@noop {} {\bibfield  {journal} {\bibinfo  {journal}
  {Quanta}\ }\textbf {\bibinfo {volume} {6}},\ \bibinfo {pages} {1} (\bibinfo
  {year} {2017})}\BibitemShut {NoStop}%
\bibitem [{\citenamefont {Ekert}(1991)}]{E91}%
  \BibitemOpen
  \bibfield  {author} {\bibinfo {author} {\bibfnamefont {A.~K.}\ \bibnamefont
  {Ekert}},\ }\href@noop {} {\bibfield  {journal} {\bibinfo  {journal}
  {Physical Review Letters}\ }\textbf {\bibinfo {volume} {67}},\ \bibinfo
  {pages} {661} (\bibinfo {year} {1991})}\BibitemShut {NoStop}%
\bibitem [{\citenamefont {Bennett}(1992)}]{B92}%
  \BibitemOpen
  \bibfield  {author} {\bibinfo {author} {\bibfnamefont {C.~H.}\ \bibnamefont
  {Bennett}},\ }\href@noop {} {\bibfield  {journal} {\bibinfo  {journal}
  {Physical Review Letters}\ }\textbf {\bibinfo {volume} {68}},\ \bibinfo
  {pages} {3121} (\bibinfo {year} {1992})}\BibitemShut {NoStop}%
\bibitem [{\citenamefont {Bennett}\ \emph {et~al.}(1992)\citenamefont
  {Bennett}, \citenamefont {Brassard},\ and\ \citenamefont {Mermin}}]{BBM92}%
  \BibitemOpen
  \bibfield  {author} {\bibinfo {author} {\bibfnamefont {C.~H.}\ \bibnamefont
  {Bennett}}, \bibinfo {author} {\bibfnamefont {G.}~\bibnamefont {Brassard}},\
  and\ \bibinfo {author} {\bibfnamefont {N.~D.}\ \bibnamefont {Mermin}},\
  }\href@noop {} {\bibfield  {journal} {\bibinfo  {journal} {Physical Review
  Letters}\ }\textbf {\bibinfo {volume} {68}},\ \bibinfo {pages} {557}
  (\bibinfo {year} {1992})}\BibitemShut {NoStop}%
\bibitem [{\citenamefont {Scarani}\ \emph {et~al.}(2004)\citenamefont
  {Scarani}, \citenamefont {Acin}, \citenamefont {Ribordy},\ and\ \citenamefont
  {Gisin}}]{SAR+04}%
  \BibitemOpen
  \bibfield  {author} {\bibinfo {author} {\bibfnamefont {V.}~\bibnamefont
  {Scarani}}, \bibinfo {author} {\bibfnamefont {A.}~\bibnamefont {Acin}},
  \bibinfo {author} {\bibfnamefont {G.}~\bibnamefont {Ribordy}},\ and\ \bibinfo
  {author} {\bibfnamefont {N.}~\bibnamefont {Gisin}},\ }\href@noop {}
  {\bibfield  {journal} {\bibinfo  {journal} {Physical Review Letters}\
  }\textbf {\bibinfo {volume} {92}},\ \bibinfo {pages} {057901} (\bibinfo
  {year} {2004})}\BibitemShut {NoStop}%
\bibitem [{\citenamefont {Scarani}\ \emph {et~al.}(2009)\citenamefont
  {Scarani}, \citenamefont {Bechmann-Pasquinucci}, \citenamefont {Cerf},
  \citenamefont {Du{\v{s}}ek}, \citenamefont {L{\"u}tkenhaus},\ and\
  \citenamefont {Peev}}]{SPC+09}%
  \BibitemOpen
  \bibfield  {author} {\bibinfo {author} {\bibfnamefont {V.}~\bibnamefont
  {Scarani}}, \bibinfo {author} {\bibfnamefont {H.}~\bibnamefont
  {Bechmann-Pasquinucci}}, \bibinfo {author} {\bibfnamefont {N.~J.}\
  \bibnamefont {Cerf}}, \bibinfo {author} {\bibfnamefont {M.}~\bibnamefont
  {Du{\v{s}}ek}}, \bibinfo {author} {\bibfnamefont {N.}~\bibnamefont
  {L{\"u}tkenhaus}},\ and\ \bibinfo {author} {\bibfnamefont {M.}~\bibnamefont
  {Peev}},\ }\href@noop {} {\bibfield  {journal} {\bibinfo  {journal} {Reviews
  of Modern Physics}\ }\textbf {\bibinfo {volume} {81}},\ \bibinfo {pages}
  {1301} (\bibinfo {year} {2009})}\BibitemShut {NoStop}%
\bibitem [{\citenamefont {Chatterjee}\ \emph {et~al.}(2020)\citenamefont
  {Chatterjee}, \citenamefont {Joarder}, \citenamefont {Chatterjee},
  \citenamefont {Sanders},\ and\ \citenamefont {Sinha}}]{CJC+20}%
  \BibitemOpen
  \bibfield  {author} {\bibinfo {author} {\bibfnamefont {R.}~\bibnamefont
  {Chatterjee}}, \bibinfo {author} {\bibfnamefont {K.}~\bibnamefont {Joarder}},
  \bibinfo {author} {\bibfnamefont {S.}~\bibnamefont {Chatterjee}}, \bibinfo
  {author} {\bibfnamefont {B.~C.}\ \bibnamefont {Sanders}},\ and\ \bibinfo
  {author} {\bibfnamefont {U.}~\bibnamefont {Sinha}},\ }\href@noop {}
  {\bibfield  {journal} {\bibinfo  {journal} {Physical Review Applied}\
  }\textbf {\bibinfo {volume} {14}},\ \bibinfo {pages} {024036} (\bibinfo
  {year} {2020})}\BibitemShut {NoStop}%
\bibitem [{\citenamefont {Dutta}\ and\ \citenamefont
  {Pathak}(2025{\natexlab{a}})}]{DP+24}%
  \BibitemOpen
  \bibfield  {author} {\bibinfo {author} {\bibfnamefont {A.}~\bibnamefont
  {Dutta}}\ and\ \bibinfo {author} {\bibfnamefont {A.}~\bibnamefont {Pathak}},\
  }\href@noop {} {\bibfield  {journal} {\bibinfo  {journal} {Modern Physics
  Letters A}\ ,\ \bibinfo {pages} {2450196}} (\bibinfo {year}
  {2025}{\natexlab{a}})}\BibitemShut {NoStop}%
\bibitem [{\citenamefont {Dutta}\ and\ \citenamefont
  {Pathak}(2025{\natexlab{b}})}]{DP+23}%
  \BibitemOpen
  \bibfield  {author} {\bibinfo {author} {\bibfnamefont {A.}~\bibnamefont
  {Dutta}}\ and\ \bibinfo {author} {\bibfnamefont {A.}~\bibnamefont {Pathak}},\
  }\href@noop {} {\bibfield  {journal} {\bibinfo  {journal} {Physica Scripta}\
  }\textbf {\bibinfo {volume} {100}},\ \bibinfo {pages} {035101} (\bibinfo
  {year} {2025}{\natexlab{b}})}\BibitemShut {NoStop}%
\bibitem [{\citenamefont {Pirandola}\ \emph {et~al.}(2009)\citenamefont
  {Pirandola}, \citenamefont {Garc{\'\i}a-Patr{\'o}n}, \citenamefont
  {Braunstein},\ and\ \citenamefont {Lloyd}}]{PGB+09}%
  \BibitemOpen
  \bibfield  {author} {\bibinfo {author} {\bibfnamefont {S.}~\bibnamefont
  {Pirandola}}, \bibinfo {author} {\bibfnamefont {R.}~\bibnamefont
  {Garc{\'\i}a-Patr{\'o}n}}, \bibinfo {author} {\bibfnamefont {S.~L.}\
  \bibnamefont {Braunstein}},\ and\ \bibinfo {author} {\bibfnamefont
  {S.}~\bibnamefont {Lloyd}},\ }\href@noop {} {\bibfield  {journal} {\bibinfo
  {journal} {Physical Review Letters}\ }\textbf {\bibinfo {volume} {102}},\
  \bibinfo {pages} {050503} (\bibinfo {year} {2009})}\BibitemShut {NoStop}%
\bibitem [{\citenamefont {Takeoka}\ \emph {et~al.}(2014)\citenamefont
  {Takeoka}, \citenamefont {Guha},\ and\ \citenamefont {Wilde}}]{TGW14}%
  \BibitemOpen
  \bibfield  {author} {\bibinfo {author} {\bibfnamefont {M.}~\bibnamefont
  {Takeoka}}, \bibinfo {author} {\bibfnamefont {S.}~\bibnamefont {Guha}},\ and\
  \bibinfo {author} {\bibfnamefont {M.~M.}\ \bibnamefont {Wilde}},\ }\href@noop
  {} {\bibfield  {journal} {\bibinfo  {journal} {Nature Communications}\
  }\textbf {\bibinfo {volume} {5}},\ \bibinfo {pages} {5235} (\bibinfo {year}
  {2014})}\BibitemShut {NoStop}%
\bibitem [{\citenamefont {Pirandola}\ \emph {et~al.}(2017)\citenamefont
  {Pirandola}, \citenamefont {Laurenza}, \citenamefont {Ottaviani},\ and\
  \citenamefont {Banchi}}]{PLOB17}%
  \BibitemOpen
  \bibfield  {author} {\bibinfo {author} {\bibfnamefont {S.}~\bibnamefont
  {Pirandola}}, \bibinfo {author} {\bibfnamefont {R.}~\bibnamefont {Laurenza}},
  \bibinfo {author} {\bibfnamefont {C.}~\bibnamefont {Ottaviani}},\ and\
  \bibinfo {author} {\bibfnamefont {L.}~\bibnamefont {Banchi}},\ }\href@noop {}
  {\bibfield  {journal} {\bibinfo  {journal} {Nature communications}\ }\textbf
  {\bibinfo {volume} {8}},\ \bibinfo {pages} {15043} (\bibinfo {year}
  {2017})}\BibitemShut {NoStop}%
\bibitem [{\citenamefont {Pirandola}(2021{\natexlab{a}})}]{P21}%
  \BibitemOpen
  \bibfield  {author} {\bibinfo {author} {\bibfnamefont {S.}~\bibnamefont
  {Pirandola}},\ }\href@noop {} {\bibfield  {journal} {\bibinfo  {journal}
  {Physical Review Research}\ }\textbf {\bibinfo {volume} {3}},\ \bibinfo
  {pages} {013279} (\bibinfo {year} {2021}{\natexlab{a}})}\BibitemShut
  {NoStop}%
\bibitem [{\citenamefont {Pirandola}(2021{\natexlab{b}})}]{P+21}%
  \BibitemOpen
  \bibfield  {author} {\bibinfo {author} {\bibfnamefont {S.}~\bibnamefont
  {Pirandola}},\ }\href@noop {} {\bibfield  {journal} {\bibinfo  {journal}
  {Physical Review Research}\ }\textbf {\bibinfo {volume} {3}},\ \bibinfo
  {pages} {023130} (\bibinfo {year} {2021}{\natexlab{b}})}\BibitemShut
  {NoStop}%
\bibitem [{\citenamefont {Briegel}\ \emph {et~al.}(1998)\citenamefont
  {Briegel}, \citenamefont {D{\"u}r}, \citenamefont {Cirac},\ and\
  \citenamefont {Zoller}}]{BDC+98}%
  \BibitemOpen
  \bibfield  {author} {\bibinfo {author} {\bibfnamefont {H.-J.}\ \bibnamefont
  {Briegel}}, \bibinfo {author} {\bibfnamefont {W.}~\bibnamefont {D{\"u}r}},
  \bibinfo {author} {\bibfnamefont {J.~I.}\ \bibnamefont {Cirac}},\ and\
  \bibinfo {author} {\bibfnamefont {P.}~\bibnamefont {Zoller}},\ }\href@noop {}
  {\bibfield  {journal} {\bibinfo  {journal} {Physical Review Letters}\
  }\textbf {\bibinfo {volume} {81}},\ \bibinfo {pages} {5932} (\bibinfo {year}
  {1998})}\BibitemShut {NoStop}%
\bibitem [{\citenamefont {Luong}\ \emph {et~al.}(2016)\citenamefont {Luong},
  \citenamefont {Jiang}, \citenamefont {Kim},\ and\ \citenamefont
  {L{\"u}tkenhaus}}]{LJK+16}%
  \BibitemOpen
  \bibfield  {author} {\bibinfo {author} {\bibfnamefont {D.}~\bibnamefont
  {Luong}}, \bibinfo {author} {\bibfnamefont {L.}~\bibnamefont {Jiang}},
  \bibinfo {author} {\bibfnamefont {J.}~\bibnamefont {Kim}},\ and\ \bibinfo
  {author} {\bibfnamefont {N.}~\bibnamefont {L{\"u}tkenhaus}},\ }\href@noop {}
  {\bibfield  {journal} {\bibinfo  {journal} {Applied Physics B}\ }\textbf
  {\bibinfo {volume} {122}},\ \bibinfo {pages} {1} (\bibinfo {year}
  {2016})}\BibitemShut {NoStop}%
\bibitem [{\citenamefont {Xie}\ \emph {et~al.}(2022)\citenamefont {Xie},
  \citenamefont {Lu}, \citenamefont {Weng}, \citenamefont {Cao}, \citenamefont
  {Jia}, \citenamefont {Bao}, \citenamefont {Wang}, \citenamefont {Fu},
  \citenamefont {Yin},\ and\ \citenamefont {Chen}}]{XLW+22}%
  \BibitemOpen
  \bibfield  {author} {\bibinfo {author} {\bibfnamefont {Y.-M.}\ \bibnamefont
  {Xie}}, \bibinfo {author} {\bibfnamefont {Y.-S.}\ \bibnamefont {Lu}},
  \bibinfo {author} {\bibfnamefont {C.-X.}\ \bibnamefont {Weng}}, \bibinfo
  {author} {\bibfnamefont {X.-Y.}\ \bibnamefont {Cao}}, \bibinfo {author}
  {\bibfnamefont {Z.-Y.}\ \bibnamefont {Jia}}, \bibinfo {author} {\bibfnamefont
  {Y.}~\bibnamefont {Bao}}, \bibinfo {author} {\bibfnamefont {Y.}~\bibnamefont
  {Wang}}, \bibinfo {author} {\bibfnamefont {Y.}~\bibnamefont {Fu}}, \bibinfo
  {author} {\bibfnamefont {H.-L.}\ \bibnamefont {Yin}},\ and\ \bibinfo {author}
  {\bibfnamefont {Z.-B.}\ \bibnamefont {Chen}},\ }\href@noop {} {\bibfield
  {journal} {\bibinfo  {journal} {PRX Quantum}\ }\textbf {\bibinfo {volume}
  {3}},\ \bibinfo {pages} {020315} (\bibinfo {year} {2022})}\BibitemShut
  {NoStop}%
\bibitem [{\citenamefont {Li}\ \emph {et~al.}(2023)\citenamefont {Li},
  \citenamefont {Zhang}, \citenamefont {Lu}, \citenamefont {Li}, \citenamefont
  {Jiang}, \citenamefont {Liu}, \citenamefont {Huang}, \citenamefont {Li},
  \citenamefont {Wang}, \citenamefont {Wang} \emph {et~al.}}]{LZL+23}%
  \BibitemOpen
  \bibfield  {author} {\bibinfo {author} {\bibfnamefont {W.}~\bibnamefont
  {Li}}, \bibinfo {author} {\bibfnamefont {L.}~\bibnamefont {Zhang}}, \bibinfo
  {author} {\bibfnamefont {Y.}~\bibnamefont {Lu}}, \bibinfo {author}
  {\bibfnamefont {Z.-P.}\ \bibnamefont {Li}}, \bibinfo {author} {\bibfnamefont
  {C.}~\bibnamefont {Jiang}}, \bibinfo {author} {\bibfnamefont
  {Y.}~\bibnamefont {Liu}}, \bibinfo {author} {\bibfnamefont {J.}~\bibnamefont
  {Huang}}, \bibinfo {author} {\bibfnamefont {H.}~\bibnamefont {Li}}, \bibinfo
  {author} {\bibfnamefont {Z.}~\bibnamefont {Wang}}, \bibinfo {author}
  {\bibfnamefont {X.-B.}\ \bibnamefont {Wang}}, \emph {et~al.},\ }\href@noop {}
  {\bibfield  {journal} {\bibinfo  {journal} {Physical Review Letters}\
  }\textbf {\bibinfo {volume} {130}},\ \bibinfo {pages} {250802} (\bibinfo
  {year} {2023})}\BibitemShut {NoStop}%
\bibitem [{\citenamefont {Liu}\ \emph {et~al.}(2024)\citenamefont {Liu},
  \citenamefont {Luo}, \citenamefont {Luo}, \citenamefont {Li}, \citenamefont
  {Zhang},\ and\ \citenamefont {Wei}}]{LLL+24}%
  \BibitemOpen
  \bibfield  {author} {\bibinfo {author} {\bibfnamefont {X.}~\bibnamefont
  {Liu}}, \bibinfo {author} {\bibfnamefont {D.}~\bibnamefont {Luo}}, \bibinfo
  {author} {\bibfnamefont {Z.}~\bibnamefont {Luo}}, \bibinfo {author}
  {\bibfnamefont {S.}~\bibnamefont {Li}}, \bibinfo {author} {\bibfnamefont
  {Z.}~\bibnamefont {Zhang}},\ and\ \bibinfo {author} {\bibfnamefont
  {K.}~\bibnamefont {Wei}},\ }\href@noop {} {\bibfield  {journal} {\bibinfo
  {journal} {Physical Review Applied}\ }\textbf {\bibinfo {volume} {22}},\
  \bibinfo {pages} {064018} (\bibinfo {year} {2024})}\BibitemShut {NoStop}%
\bibitem [{\citenamefont {Beutel}\ \emph {et~al.}(2022)\citenamefont {Beutel},
  \citenamefont {Br{\"u}ckerhoff-Pl{\"u}ckelmann}, \citenamefont {Gehring},
  \citenamefont {Kovalyuk}, \citenamefont {Zolotov}, \citenamefont {Goltsman},\
  and\ \citenamefont {Pernice}}]{BPG+22}%
  \BibitemOpen
  \bibfield  {author} {\bibinfo {author} {\bibfnamefont {F.}~\bibnamefont
  {Beutel}}, \bibinfo {author} {\bibfnamefont {F.}~\bibnamefont
  {Br{\"u}ckerhoff-Pl{\"u}ckelmann}}, \bibinfo {author} {\bibfnamefont
  {H.}~\bibnamefont {Gehring}}, \bibinfo {author} {\bibfnamefont
  {V.}~\bibnamefont {Kovalyuk}}, \bibinfo {author} {\bibfnamefont
  {P.}~\bibnamefont {Zolotov}}, \bibinfo {author} {\bibfnamefont
  {G.}~\bibnamefont {Goltsman}},\ and\ \bibinfo {author} {\bibfnamefont
  {W.~H.}\ \bibnamefont {Pernice}},\ }\href@noop {} {\bibfield  {journal}
  {\bibinfo  {journal} {Optica}\ }\textbf {\bibinfo {volume} {9}},\ \bibinfo
  {pages} {1121} (\bibinfo {year} {2022})}\BibitemShut {NoStop}%
\bibitem [{\citenamefont {Wei}\ \emph {et~al.}(2023)\citenamefont {Wei},
  \citenamefont {Hu}, \citenamefont {Du}, \citenamefont {Hua}, \citenamefont
  {Zhao}, \citenamefont {Chen}, \citenamefont {Huang},\ and\ \citenamefont
  {Xiao}}]{WHD+23}%
  \BibitemOpen
  \bibfield  {author} {\bibinfo {author} {\bibfnamefont {K.}~\bibnamefont
  {Wei}}, \bibinfo {author} {\bibfnamefont {X.}~\bibnamefont {Hu}}, \bibinfo
  {author} {\bibfnamefont {Y.}~\bibnamefont {Du}}, \bibinfo {author}
  {\bibfnamefont {X.}~\bibnamefont {Hua}}, \bibinfo {author} {\bibfnamefont
  {Z.}~\bibnamefont {Zhao}}, \bibinfo {author} {\bibfnamefont {Y.}~\bibnamefont
  {Chen}}, \bibinfo {author} {\bibfnamefont {C.}~\bibnamefont {Huang}},\ and\
  \bibinfo {author} {\bibfnamefont {X.}~\bibnamefont {Xiao}},\ }\href@noop {}
  {\bibfield  {journal} {\bibinfo  {journal} {Photonics Research}\ }\textbf
  {\bibinfo {volume} {11}},\ \bibinfo {pages} {1364} (\bibinfo {year}
  {2023})}\BibitemShut {NoStop}%
\bibitem [{\citenamefont {Zhong}\ \emph {et~al.}(2022)\citenamefont {Zhong},
  \citenamefont {Wang}, \citenamefont {Mandil}, \citenamefont {Lo},\ and\
  \citenamefont {Qian}}]{ZWM+22}%
  \BibitemOpen
  \bibfield  {author} {\bibinfo {author} {\bibfnamefont {X.}~\bibnamefont
  {Zhong}}, \bibinfo {author} {\bibfnamefont {W.}~\bibnamefont {Wang}},
  \bibinfo {author} {\bibfnamefont {R.}~\bibnamefont {Mandil}}, \bibinfo
  {author} {\bibfnamefont {H.-K.}\ \bibnamefont {Lo}},\ and\ \bibinfo {author}
  {\bibfnamefont {L.}~\bibnamefont {Qian}},\ }\href@noop {} {\bibfield
  {journal} {\bibinfo  {journal} {Physical Review Applied}\ }\textbf {\bibinfo
  {volume} {17}},\ \bibinfo {pages} {014025} (\bibinfo {year}
  {2022})}\BibitemShut {NoStop}%
\bibitem [{\citenamefont {Huang}\ \emph {et~al.}(2024)\citenamefont {Huang},
  \citenamefont {Chen}, \citenamefont {Luo}, \citenamefont {He}, \citenamefont
  {Liu}, \citenamefont {Zhang},\ and\ \citenamefont {Wei}}]{HCL+24}%
  \BibitemOpen
  \bibfield  {author} {\bibinfo {author} {\bibfnamefont {C.}~\bibnamefont
  {Huang}}, \bibinfo {author} {\bibfnamefont {Y.}~\bibnamefont {Chen}},
  \bibinfo {author} {\bibfnamefont {T.}~\bibnamefont {Luo}}, \bibinfo {author}
  {\bibfnamefont {W.}~\bibnamefont {He}}, \bibinfo {author} {\bibfnamefont
  {X.}~\bibnamefont {Liu}}, \bibinfo {author} {\bibfnamefont {Z.}~\bibnamefont
  {Zhang}},\ and\ \bibinfo {author} {\bibfnamefont {K.}~\bibnamefont {Wei}},\
  }\href@noop {} {\bibfield  {journal} {\bibinfo  {journal} {Science China
  Physics, Mechanics \& Astronomy}\ }\textbf {\bibinfo {volume} {67}},\
  \bibinfo {pages} {240312} (\bibinfo {year} {2024})}\BibitemShut {NoStop}%
\bibitem [{\citenamefont {Tamaki}\ \emph {et~al.}(2012)\citenamefont {Tamaki},
  \citenamefont {Lo}, \citenamefont {Fung},\ and\ \citenamefont {Qi}}]{TLF+12}%
  \BibitemOpen
  \bibfield  {author} {\bibinfo {author} {\bibfnamefont {K.}~\bibnamefont
  {Tamaki}}, \bibinfo {author} {\bibfnamefont {H.-K.}\ \bibnamefont {Lo}},
  \bibinfo {author} {\bibfnamefont {C.-H.~F.}\ \bibnamefont {Fung}},\ and\
  \bibinfo {author} {\bibfnamefont {B.}~\bibnamefont {Qi}},\ }\href@noop {}
  {\bibfield  {journal} {\bibinfo  {journal} {Physical Review A}\ }\textbf
  {\bibinfo {volume} {85}},\ \bibinfo {pages} {042307} (\bibinfo {year}
  {2012})}\BibitemShut {NoStop}%
\bibitem [{\citenamefont {Ferenczi}(2013)}]{ferenczi2013security}%
  \BibitemOpen
  \bibfield  {author} {\bibinfo {author} {\bibfnamefont {A.}~\bibnamefont
  {Ferenczi}},\ }\href@noop {} {\bibfield  {journal} {\bibinfo  {journal}
  {Ph.D. thesis, University of Waterloo}\ } (\bibinfo {year}
  {2013})}\BibitemShut {NoStop}%
\bibitem [{\citenamefont {Lucamarini}\ \emph {et~al.}(2018)\citenamefont
  {Lucamarini}, \citenamefont {Yuan}, \citenamefont {Dynes},\ and\
  \citenamefont {Shields}}]{LYD+18}%
  \BibitemOpen
  \bibfield  {author} {\bibinfo {author} {\bibfnamefont {M.}~\bibnamefont
  {Lucamarini}}, \bibinfo {author} {\bibfnamefont {Z.~L.}\ \bibnamefont
  {Yuan}}, \bibinfo {author} {\bibfnamefont {J.~F.}\ \bibnamefont {Dynes}},\
  and\ \bibinfo {author} {\bibfnamefont {A.~J.}\ \bibnamefont {Shields}},\
  }\href@noop {} {\bibfield  {journal} {\bibinfo  {journal} {Nature}\ }\textbf
  {\bibinfo {volume} {557}},\ \bibinfo {pages} {400} (\bibinfo {year}
  {2018})}\BibitemShut {NoStop}%
\bibitem [{\citenamefont {Ma}\ \emph {et~al.}(2018)\citenamefont {Ma},
  \citenamefont {Zeng},\ and\ \citenamefont {Zhou}}]{MZZ18}%
  \BibitemOpen
  \bibfield  {author} {\bibinfo {author} {\bibfnamefont {X.}~\bibnamefont
  {Ma}}, \bibinfo {author} {\bibfnamefont {P.}~\bibnamefont {Zeng}},\ and\
  \bibinfo {author} {\bibfnamefont {H.}~\bibnamefont {Zhou}},\ }\href@noop {}
  {\bibfield  {journal} {\bibinfo  {journal} {Physical Review X}\ }\textbf
  {\bibinfo {volume} {8}},\ \bibinfo {pages} {031043} (\bibinfo {year}
  {2018})}\BibitemShut {NoStop}%
\bibitem [{\citenamefont {Tamaki}\ \emph {et~al.}(2018)\citenamefont {Tamaki},
  \citenamefont {Lo}, \citenamefont {Wang},\ and\ \citenamefont
  {Lucamarini}}]{TLW18}%
  \BibitemOpen
  \bibfield  {author} {\bibinfo {author} {\bibfnamefont {K.}~\bibnamefont
  {Tamaki}}, \bibinfo {author} {\bibfnamefont {H.-K.}\ \bibnamefont {Lo}},
  \bibinfo {author} {\bibfnamefont {W.}~\bibnamefont {Wang}},\ and\ \bibinfo
  {author} {\bibfnamefont {M.}~\bibnamefont {Lucamarini}},\ }\href@noop {}
  {\bibfield  {journal} {\bibinfo  {journal} {arXiv preprint arXiv:1805.05511}\
  } (\bibinfo {year} {2018})}\BibitemShut {NoStop}%
\bibitem [{\citenamefont {Shor}\ and\ \citenamefont {Preskill}(2000)}]{SP200}%
  \BibitemOpen
  \bibfield  {author} {\bibinfo {author} {\bibfnamefont {P.~W.}\ \bibnamefont
  {Shor}}\ and\ \bibinfo {author} {\bibfnamefont {J.}~\bibnamefont
  {Preskill}},\ }\href@noop {} {\bibfield  {journal} {\bibinfo  {journal}
  {Physical Review Letters}\ }\textbf {\bibinfo {volume} {85}},\ \bibinfo
  {pages} {441} (\bibinfo {year} {2000})}\BibitemShut {NoStop}%
\bibitem [{\citenamefont {Koashi}(2009)}]{K09}%
  \BibitemOpen
  \bibfield  {author} {\bibinfo {author} {\bibfnamefont {M.}~\bibnamefont
  {Koashi}},\ }\href@noop {} {\bibfield  {journal} {\bibinfo  {journal} {New
  Journal of Physics}\ }\textbf {\bibinfo {volume} {11}},\ \bibinfo {pages}
  {045018} (\bibinfo {year} {2009})}\BibitemShut {NoStop}%
\bibitem [{\citenamefont {Tsurumaru}\ and\ \citenamefont
  {Hayashi}(2013)}]{TH13}%
  \BibitemOpen
  \bibfield  {author} {\bibinfo {author} {\bibfnamefont {T.}~\bibnamefont
  {Tsurumaru}}\ and\ \bibinfo {author} {\bibfnamefont {M.}~\bibnamefont
  {Hayashi}},\ }\href@noop {} {\bibfield  {journal} {\bibinfo  {journal} {IEEE
  transactions on information theory}\ }\textbf {\bibinfo {volume} {59}},\
  \bibinfo {pages} {4700} (\bibinfo {year} {2013})}\BibitemShut {NoStop}%
\bibitem [{\citenamefont {Lin}\ and\ \citenamefont
  {L{\"u}tkenhaus}(2018)}]{LL18}%
  \BibitemOpen
  \bibfield  {author} {\bibinfo {author} {\bibfnamefont {J.}~\bibnamefont
  {Lin}}\ and\ \bibinfo {author} {\bibfnamefont {N.}~\bibnamefont
  {L{\"u}tkenhaus}},\ }\href@noop {} {\bibfield  {journal} {\bibinfo  {journal}
  {Physical Review A}\ }\textbf {\bibinfo {volume} {98}},\ \bibinfo {pages}
  {042332} (\bibinfo {year} {2018})}\BibitemShut {NoStop}%
\bibitem [{\citenamefont {Renner}(2008)}]{R08}%
  \BibitemOpen
  \bibfield  {author} {\bibinfo {author} {\bibfnamefont {R.}~\bibnamefont
  {Renner}},\ }\href@noop {} {\bibfield  {journal} {\bibinfo  {journal}
  {International Journal of Quantum Information}\ }\textbf {\bibinfo {volume}
  {6}},\ \bibinfo {pages} {1} (\bibinfo {year} {2008})}\BibitemShut {NoStop}%
\bibitem [{\citenamefont {Dutta}\ and\ \citenamefont
  {Pathak}(2022{\natexlab{a}})}]{DP22}%
  \BibitemOpen
  \bibfield  {author} {\bibinfo {author} {\bibfnamefont {A.}~\bibnamefont
  {Dutta}}\ and\ \bibinfo {author} {\bibfnamefont {A.}~\bibnamefont {Pathak}},\
  }\href@noop {} {\bibfield  {journal} {\bibinfo  {journal} {Quantum
  Information Processing}\ }\textbf {\bibinfo {volume} {21}},\ \bibinfo {pages}
  {369} (\bibinfo {year} {2022}{\natexlab{a}})}\BibitemShut {NoStop}%
\bibitem [{\citenamefont {Dutta}\ and\ \citenamefont {Pathak}(2023)}]{DP23}%
  \BibitemOpen
  \bibfield  {author} {\bibinfo {author} {\bibfnamefont {A.}~\bibnamefont
  {Dutta}}\ and\ \bibinfo {author} {\bibfnamefont {A.}~\bibnamefont {Pathak}},\
  }\href@noop {} {\bibfield  {journal} {\bibinfo  {journal} {Quantum
  Information Processing}\ }\textbf {\bibinfo {volume} {22}},\ \bibinfo {pages}
  {13} (\bibinfo {year} {2023})}\BibitemShut {NoStop}%
\bibitem [{\citenamefont {Tsurumaru}(2020)}]{T20}%
  \BibitemOpen
  \bibfield  {author} {\bibinfo {author} {\bibfnamefont {T.}~\bibnamefont
  {Tsurumaru}},\ }\href@noop {} {\bibfield  {journal} {\bibinfo  {journal}
  {IEEE Transactions on Information Theory}\ }\textbf {\bibinfo {volume}
  {66}},\ \bibinfo {pages} {3465} (\bibinfo {year} {2020})}\BibitemShut
  {NoStop}%
\bibitem [{\citenamefont {Lo}\ \emph {et~al.}(2005)\citenamefont {Lo},
  \citenamefont {Ma},\ and\ \citenamefont {Chen}}]{LMC05}%
  \BibitemOpen
  \bibfield  {author} {\bibinfo {author} {\bibfnamefont {H.-K.}\ \bibnamefont
  {Lo}}, \bibinfo {author} {\bibfnamefont {X.}~\bibnamefont {Ma}},\ and\
  \bibinfo {author} {\bibfnamefont {K.}~\bibnamefont {Chen}},\ }\href@noop {}
  {\bibfield  {journal} {\bibinfo  {journal} {Physical Review Letters}\
  }\textbf {\bibinfo {volume} {94}},\ \bibinfo {pages} {230504} (\bibinfo
  {year} {2005})}\BibitemShut {NoStop}%
\bibitem [{\citenamefont {VanWiggeren}\ and\ \citenamefont {Roy}(1999)}]{VR99}%
  \BibitemOpen
  \bibfield  {author} {\bibinfo {author} {\bibfnamefont {G.~D.}\ \bibnamefont
  {VanWiggeren}}\ and\ \bibinfo {author} {\bibfnamefont {R.}~\bibnamefont
  {Roy}},\ }\href@noop {} {\bibfield  {journal} {\bibinfo  {journal} {Applied
  Optics}\ }\textbf {\bibinfo {volume} {38}},\ \bibinfo {pages} {3888}
  (\bibinfo {year} {1999})}\BibitemShut {NoStop}%
\bibitem [{\citenamefont {Gordon}\ and\ \citenamefont
  {Kogelnik}(2000)}]{GK2000}%
  \BibitemOpen
  \bibfield  {author} {\bibinfo {author} {\bibfnamefont {J.}~\bibnamefont
  {Gordon}}\ and\ \bibinfo {author} {\bibfnamefont {H.}~\bibnamefont
  {Kogelnik}},\ }\href@noop {} {\bibfield  {journal} {\bibinfo  {journal}
  {Proceedings of the National Academy of Sciences}\ }\textbf {\bibinfo
  {volume} {97}},\ \bibinfo {pages} {4541} (\bibinfo {year}
  {2000})}\BibitemShut {NoStop}%
\bibitem [{\citenamefont {Sharma}\ \emph {et~al.}(2016)\citenamefont {Sharma},
  \citenamefont {Thapliyal}, \citenamefont {Pathak},\ and\ \citenamefont
  {Banerjee}}]{STP+16}%
  \BibitemOpen
  \bibfield  {author} {\bibinfo {author} {\bibfnamefont {V.}~\bibnamefont
  {Sharma}}, \bibinfo {author} {\bibfnamefont {K.}~\bibnamefont {Thapliyal}},
  \bibinfo {author} {\bibfnamefont {A.}~\bibnamefont {Pathak}},\ and\ \bibinfo
  {author} {\bibfnamefont {S.}~\bibnamefont {Banerjee}},\ }\href@noop {}
  {\bibfield  {journal} {\bibinfo  {journal} {Quantum Information Processing}\
  }\textbf {\bibinfo {volume} {15}},\ \bibinfo {pages} {4681} (\bibinfo {year}
  {2016})}\BibitemShut {NoStop}%
\bibitem [{\citenamefont {Toyoshima}\ \emph {et~al.}(2011)\citenamefont
  {Toyoshima}, \citenamefont {Takenaka}, \citenamefont {Shoji}, \citenamefont
  {Takayama}, \citenamefont {Takeoka}, \citenamefont {Fujiwara}, \citenamefont
  {Sasaki} \emph {et~al.}}]{TTS+11}%
  \BibitemOpen
  \bibfield  {author} {\bibinfo {author} {\bibfnamefont {M.}~\bibnamefont
  {Toyoshima}}, \bibinfo {author} {\bibfnamefont {H.}~\bibnamefont {Takenaka}},
  \bibinfo {author} {\bibfnamefont {Y.}~\bibnamefont {Shoji}}, \bibinfo
  {author} {\bibfnamefont {Y.}~\bibnamefont {Takayama}}, \bibinfo {author}
  {\bibfnamefont {M.}~\bibnamefont {Takeoka}}, \bibinfo {author} {\bibfnamefont
  {M.}~\bibnamefont {Fujiwara}}, \bibinfo {author} {\bibfnamefont
  {M.}~\bibnamefont {Sasaki}}, \emph {et~al.},\ }\href@noop {} {\bibfield
  {journal} {\bibinfo  {journal} {International Journal of Optics}\ }\textbf
  {\bibinfo {volume} {2011}} (\bibinfo {year} {2011})}\BibitemShut {NoStop}%
\bibitem [{\citenamefont {Bedington}\ \emph {et~al.}(2017)\citenamefont
  {Bedington}, \citenamefont {Arrazola},\ and\ \citenamefont {Ling}}]{BAL17}%
  \BibitemOpen
  \bibfield  {author} {\bibinfo {author} {\bibfnamefont {R.}~\bibnamefont
  {Bedington}}, \bibinfo {author} {\bibfnamefont {J.~M.}\ \bibnamefont
  {Arrazola}},\ and\ \bibinfo {author} {\bibfnamefont {A.}~\bibnamefont
  {Ling}},\ }\href@noop {} {\bibfield  {journal} {\bibinfo  {journal} {npj
  Quantum Information}\ }\textbf {\bibinfo {volume} {3}},\ \bibinfo {pages}
  {30} (\bibinfo {year} {2017})}\BibitemShut {NoStop}%
\bibitem [{\citenamefont {Ursin}\ \emph {et~al.}(2009)\citenamefont {Ursin},
  \citenamefont {Jennewein}, \citenamefont {Kofler}, \citenamefont {Perdigues},
  \citenamefont {Cacciapuoti}, \citenamefont {de~Matos}, \citenamefont
  {Aspelmeyer}, \citenamefont {Valencia}, \citenamefont {Scheidl},
  \citenamefont {Acin} \emph {et~al.}}]{UJK+09}%
  \BibitemOpen
  \bibfield  {author} {\bibinfo {author} {\bibfnamefont {R.}~\bibnamefont
  {Ursin}}, \bibinfo {author} {\bibfnamefont {T.}~\bibnamefont {Jennewein}},
  \bibinfo {author} {\bibfnamefont {J.}~\bibnamefont {Kofler}}, \bibinfo
  {author} {\bibfnamefont {J.~M.}\ \bibnamefont {Perdigues}}, \bibinfo {author}
  {\bibfnamefont {L.}~\bibnamefont {Cacciapuoti}}, \bibinfo {author}
  {\bibfnamefont {C.~J.}\ \bibnamefont {de~Matos}}, \bibinfo {author}
  {\bibfnamefont {M.}~\bibnamefont {Aspelmeyer}}, \bibinfo {author}
  {\bibfnamefont {A.}~\bibnamefont {Valencia}}, \bibinfo {author}
  {\bibfnamefont {T.}~\bibnamefont {Scheidl}}, \bibinfo {author} {\bibfnamefont
  {A.}~\bibnamefont {Acin}}, \emph {et~al.},\ }\href@noop {} {\bibfield
  {journal} {\bibinfo  {journal} {Europhysics News}\ }\textbf {\bibinfo
  {volume} {40}},\ \bibinfo {pages} {26} (\bibinfo {year} {2009})}\BibitemShut
  {NoStop}%
\bibitem [{\citenamefont {Malpani}\ \emph {et~al.}(2019)\citenamefont
  {Malpani}, \citenamefont {Alam}, \citenamefont {Thapliyal}, \citenamefont
  {Pathak}, \citenamefont {Narayanan},\ and\ \citenamefont
  {Banerjee}}]{PAT+19}%
  \BibitemOpen
  \bibfield  {author} {\bibinfo {author} {\bibfnamefont {P.}~\bibnamefont
  {Malpani}}, \bibinfo {author} {\bibfnamefont {N.}~\bibnamefont {Alam}},
  \bibinfo {author} {\bibfnamefont {K.}~\bibnamefont {Thapliyal}}, \bibinfo
  {author} {\bibfnamefont {A.}~\bibnamefont {Pathak}}, \bibinfo {author}
  {\bibfnamefont {V.}~\bibnamefont {Narayanan}},\ and\ \bibinfo {author}
  {\bibfnamefont {S.}~\bibnamefont {Banerjee}},\ }\href@noop {} {\bibfield
  {journal} {\bibinfo  {journal} {Annalen der Physik}\ }\textbf {\bibinfo
  {volume} {531}},\ \bibinfo {pages} {1800318} (\bibinfo {year}
  {2019})}\BibitemShut {NoStop}%
\bibitem [{\citenamefont {Dutta}\ and\ \citenamefont {Pathak}(2024)}]{DP24}%
  \BibitemOpen
  \bibfield  {author} {\bibinfo {author} {\bibfnamefont {A.}~\bibnamefont
  {Dutta}}\ and\ \bibinfo {author} {\bibfnamefont {A.}~\bibnamefont {Pathak}},\
  }\href@noop {} {\bibfield  {journal} {\bibinfo  {journal} {Physica Scripta}\
  }\textbf {\bibinfo {volume} {99}},\ \bibinfo {pages} {095106} (\bibinfo
  {year} {2024})}\BibitemShut {NoStop}%
\bibitem [{\citenamefont {Dutta}\ and\ \citenamefont
  {Pathak}(2022{\natexlab{b}})}]{DP+22}%
  \BibitemOpen
  \bibfield  {author} {\bibinfo {author} {\bibfnamefont {A.}~\bibnamefont
  {Dutta}}\ and\ \bibinfo {author} {\bibfnamefont {A.}~\bibnamefont {Pathak}},\
  }\href@noop {} {\bibfield  {journal} {\bibinfo  {journal} {arXiv preprint
  arXiv:2212.13089}\ } (\bibinfo {year} {2022}{\natexlab{b}})}\BibitemShut
  {NoStop}%
\bibitem [{\citenamefont {Scheidl}\ \emph {et~al.}(2013)\citenamefont
  {Scheidl}, \citenamefont {Wille},\ and\ \citenamefont {Ursin}}]{SWU13}%
  \BibitemOpen
  \bibfield  {author} {\bibinfo {author} {\bibfnamefont {T.}~\bibnamefont
  {Scheidl}}, \bibinfo {author} {\bibfnamefont {E.}~\bibnamefont {Wille}},\
  and\ \bibinfo {author} {\bibfnamefont {R.}~\bibnamefont {Ursin}},\
  }\href@noop {} {\bibfield  {journal} {\bibinfo  {journal} {New Journal of
  Physics}\ }\textbf {\bibinfo {volume} {15}},\ \bibinfo {pages} {043008}
  (\bibinfo {year} {2013})}\BibitemShut {NoStop}%
\bibitem [{\citenamefont {Sharma}\ and\ \citenamefont {Banerjee}(2019)}]{SB19}%
  \BibitemOpen
  \bibfield  {author} {\bibinfo {author} {\bibfnamefont {V.}~\bibnamefont
  {Sharma}}\ and\ \bibinfo {author} {\bibfnamefont {S.}~\bibnamefont
  {Banerjee}},\ }\href@noop {} {\bibfield  {journal} {\bibinfo  {journal}
  {Quantum Information Processing}\ }\textbf {\bibinfo {volume} {18}},\
  \bibinfo {pages} {67} (\bibinfo {year} {2019})}\BibitemShut {NoStop}%
\bibitem [{\citenamefont {Bonato}\ \emph {et~al.}(2009)\citenamefont {Bonato},
  \citenamefont {Tomaello}, \citenamefont {Da~Deppo}, \citenamefont {Naletto},\
  and\ \citenamefont {Villoresi}}]{BTD09}%
  \BibitemOpen
  \bibfield  {author} {\bibinfo {author} {\bibfnamefont {C.}~\bibnamefont
  {Bonato}}, \bibinfo {author} {\bibfnamefont {A.}~\bibnamefont {Tomaello}},
  \bibinfo {author} {\bibfnamefont {V.}~\bibnamefont {Da~Deppo}}, \bibinfo
  {author} {\bibfnamefont {G.}~\bibnamefont {Naletto}},\ and\ \bibinfo {author}
  {\bibfnamefont {P.}~\bibnamefont {Villoresi}},\ }\href@noop {} {\bibfield
  {journal} {\bibinfo  {journal} {New Journal of Physics}\ }\textbf {\bibinfo
  {volume} {11}},\ \bibinfo {pages} {045017} (\bibinfo {year}
  {2009})}\BibitemShut {NoStop}%
\bibitem [{\citenamefont {Trinh}\ \emph {et~al.}(2022)\citenamefont {Trinh},
  \citenamefont {Carrasco-Casado}, \citenamefont {Takenaka}, \citenamefont
  {Fujiwara}, \citenamefont {Kitamura}, \citenamefont {Sasaki},\ and\
  \citenamefont {Toyoshima}}]{TCT+22}%
  \BibitemOpen
  \bibfield  {author} {\bibinfo {author} {\bibfnamefont {P.~V.}\ \bibnamefont
  {Trinh}}, \bibinfo {author} {\bibfnamefont {A.}~\bibnamefont
  {Carrasco-Casado}}, \bibinfo {author} {\bibfnamefont {H.}~\bibnamefont
  {Takenaka}}, \bibinfo {author} {\bibfnamefont {M.}~\bibnamefont {Fujiwara}},
  \bibinfo {author} {\bibfnamefont {M.}~\bibnamefont {Kitamura}}, \bibinfo
  {author} {\bibfnamefont {M.}~\bibnamefont {Sasaki}},\ and\ \bibinfo {author}
  {\bibfnamefont {M.}~\bibnamefont {Toyoshima}},\ }\href@noop {} {\bibfield
  {journal} {\bibinfo  {journal} {Communications Physics}\ }\textbf {\bibinfo
  {volume} {5}},\ \bibinfo {pages} {225} (\bibinfo {year} {2022})}\BibitemShut
  {NoStop}%
\bibitem [{\citenamefont {Vasylyev}\ \emph {et~al.}(2019)\citenamefont
  {Vasylyev}, \citenamefont {Vogel},\ and\ \citenamefont {Moll}}]{VVM19}%
  \BibitemOpen
  \bibfield  {author} {\bibinfo {author} {\bibfnamefont {D.}~\bibnamefont
  {Vasylyev}}, \bibinfo {author} {\bibfnamefont {W.}~\bibnamefont {Vogel}},\
  and\ \bibinfo {author} {\bibfnamefont {F.}~\bibnamefont {Moll}},\ }\href@noop
  {} {\bibfield  {journal} {\bibinfo  {journal} {Physical Review A}\ }\textbf
  {\bibinfo {volume} {99}},\ \bibinfo {pages} {053830} (\bibinfo {year}
  {2019})}\BibitemShut {NoStop}%
\bibitem [{\citenamefont {Bourgoin}\ \emph {et~al.}(2015)\citenamefont
  {Bourgoin}, \citenamefont {Higgins}, \citenamefont {Gigov}, \citenamefont
  {Holloway}, \citenamefont {Pugh}, \citenamefont {Kaiser}, \citenamefont
  {Cranmer},\ and\ \citenamefont {Jennewein}}]{BHG+15}%
  \BibitemOpen
  \bibfield  {author} {\bibinfo {author} {\bibfnamefont {J.-P.}\ \bibnamefont
  {Bourgoin}}, \bibinfo {author} {\bibfnamefont {B.~L.}\ \bibnamefont
  {Higgins}}, \bibinfo {author} {\bibfnamefont {N.}~\bibnamefont {Gigov}},
  \bibinfo {author} {\bibfnamefont {C.}~\bibnamefont {Holloway}}, \bibinfo
  {author} {\bibfnamefont {C.~J.}\ \bibnamefont {Pugh}}, \bibinfo {author}
  {\bibfnamefont {S.}~\bibnamefont {Kaiser}}, \bibinfo {author} {\bibfnamefont
  {M.}~\bibnamefont {Cranmer}},\ and\ \bibinfo {author} {\bibfnamefont
  {T.}~\bibnamefont {Jennewein}},\ }\href@noop {} {\bibfield  {journal}
  {\bibinfo  {journal} {Optics Express}\ }\textbf {\bibinfo {volume} {23}},\
  \bibinfo {pages} {33437} (\bibinfo {year} {2015})}\BibitemShut {NoStop}%
\bibitem [{\citenamefont {Wang}\ \emph {et~al.}(2013)\citenamefont {Wang},
  \citenamefont {Yang}, \citenamefont {Liao}, \citenamefont {Zhang},
  \citenamefont {Shen}, \citenamefont {Hu}, \citenamefont {Wu}, \citenamefont
  {Yang}, \citenamefont {Jiang}, \citenamefont {Tang} \emph {et~al.}}]{WYL+13}%
  \BibitemOpen
  \bibfield  {author} {\bibinfo {author} {\bibfnamefont {J.-Y.}\ \bibnamefont
  {Wang}}, \bibinfo {author} {\bibfnamefont {B.}~\bibnamefont {Yang}}, \bibinfo
  {author} {\bibfnamefont {S.-K.}\ \bibnamefont {Liao}}, \bibinfo {author}
  {\bibfnamefont {L.}~\bibnamefont {Zhang}}, \bibinfo {author} {\bibfnamefont
  {Q.}~\bibnamefont {Shen}}, \bibinfo {author} {\bibfnamefont {X.-F.}\
  \bibnamefont {Hu}}, \bibinfo {author} {\bibfnamefont {J.-C.}\ \bibnamefont
  {Wu}}, \bibinfo {author} {\bibfnamefont {S.-J.}\ \bibnamefont {Yang}},
  \bibinfo {author} {\bibfnamefont {H.}~\bibnamefont {Jiang}}, \bibinfo
  {author} {\bibfnamefont {Y.-L.}\ \bibnamefont {Tang}}, \emph {et~al.},\
  }\href@noop {} {\bibfield  {journal} {\bibinfo  {journal} {Nature Photonics}\
  }\textbf {\bibinfo {volume} {7}},\ \bibinfo {pages} {387} (\bibinfo {year}
  {2013})}\BibitemShut {NoStop}%
\bibitem [{\citenamefont {Nauerth}\ \emph {et~al.}(2013)\citenamefont
  {Nauerth}, \citenamefont {Moll}, \citenamefont {Rau}, \citenamefont {Fuchs},
  \citenamefont {Horwath}, \citenamefont {Frick},\ and\ \citenamefont
  {Weinfurter}}]{NMR+13}%
  \BibitemOpen
  \bibfield  {author} {\bibinfo {author} {\bibfnamefont {S.}~\bibnamefont
  {Nauerth}}, \bibinfo {author} {\bibfnamefont {F.}~\bibnamefont {Moll}},
  \bibinfo {author} {\bibfnamefont {M.}~\bibnamefont {Rau}}, \bibinfo {author}
  {\bibfnamefont {C.}~\bibnamefont {Fuchs}}, \bibinfo {author} {\bibfnamefont
  {J.}~\bibnamefont {Horwath}}, \bibinfo {author} {\bibfnamefont
  {S.}~\bibnamefont {Frick}},\ and\ \bibinfo {author} {\bibfnamefont
  {H.}~\bibnamefont {Weinfurter}},\ }\href@noop {} {\bibfield  {journal}
  {\bibinfo  {journal} {Nature Photonics}\ }\textbf {\bibinfo {volume} {7}},\
  \bibinfo {pages} {382} (\bibinfo {year} {2013})}\BibitemShut {NoStop}%
\bibitem [{\citenamefont {Pugh}\ \emph {et~al.}(2017)\citenamefont {Pugh},
  \citenamefont {Kaiser}, \citenamefont {Bourgoin}, \citenamefont {Jin},
  \citenamefont {Sultana}, \citenamefont {Agne}, \citenamefont {Anisimova},
  \citenamefont {Makarov}, \citenamefont {Choi}, \citenamefont {Higgins} \emph
  {et~al.}}]{PKB+17}%
  \BibitemOpen
  \bibfield  {author} {\bibinfo {author} {\bibfnamefont {C.~J.}\ \bibnamefont
  {Pugh}}, \bibinfo {author} {\bibfnamefont {S.}~\bibnamefont {Kaiser}},
  \bibinfo {author} {\bibfnamefont {J.-P.}\ \bibnamefont {Bourgoin}}, \bibinfo
  {author} {\bibfnamefont {J.}~\bibnamefont {Jin}}, \bibinfo {author}
  {\bibfnamefont {N.}~\bibnamefont {Sultana}}, \bibinfo {author} {\bibfnamefont
  {S.}~\bibnamefont {Agne}}, \bibinfo {author} {\bibfnamefont {E.}~\bibnamefont
  {Anisimova}}, \bibinfo {author} {\bibfnamefont {V.}~\bibnamefont {Makarov}},
  \bibinfo {author} {\bibfnamefont {E.}~\bibnamefont {Choi}}, \bibinfo {author}
  {\bibfnamefont {B.~L.}\ \bibnamefont {Higgins}}, \emph {et~al.},\ }\href@noop
  {} {\bibfield  {journal} {\bibinfo  {journal} {Quantum Science and
  Technology}\ }\textbf {\bibinfo {volume} {2}},\ \bibinfo {pages} {024009}
  (\bibinfo {year} {2017})}\BibitemShut {NoStop}%
\bibitem [{\citenamefont {Liu}\ \emph {et~al.}(2020)\citenamefont {Liu},
  \citenamefont {Tian}, \citenamefont {Gu}, \citenamefont {Fan}, \citenamefont
  {Ni}, \citenamefont {Yang}, \citenamefont {Zhang}, \citenamefont {Hu},
  \citenamefont {Guo}, \citenamefont {Cao} \emph {et~al.}}]{LTG+20}%
  \BibitemOpen
  \bibfield  {author} {\bibinfo {author} {\bibfnamefont {H.-Y.}\ \bibnamefont
  {Liu}}, \bibinfo {author} {\bibfnamefont {X.-H.}\ \bibnamefont {Tian}},
  \bibinfo {author} {\bibfnamefont {C.}~\bibnamefont {Gu}}, \bibinfo {author}
  {\bibfnamefont {P.}~\bibnamefont {Fan}}, \bibinfo {author} {\bibfnamefont
  {X.}~\bibnamefont {Ni}}, \bibinfo {author} {\bibfnamefont {R.}~\bibnamefont
  {Yang}}, \bibinfo {author} {\bibfnamefont {J.-N.}\ \bibnamefont {Zhang}},
  \bibinfo {author} {\bibfnamefont {M.}~\bibnamefont {Hu}}, \bibinfo {author}
  {\bibfnamefont {J.}~\bibnamefont {Guo}}, \bibinfo {author} {\bibfnamefont
  {X.}~\bibnamefont {Cao}}, \emph {et~al.},\ }\href@noop {} {\bibfield
  {journal} {\bibinfo  {journal} {National Science Review}\ }\textbf {\bibinfo
  {volume} {7}},\ \bibinfo {pages} {921} (\bibinfo {year} {2020})}\BibitemShut
  {NoStop}%
\bibitem [{\citenamefont {Wehner}\ \emph {et~al.}(2018)\citenamefont {Wehner},
  \citenamefont {Elkouss},\ and\ \citenamefont {Hanson}}]{WEH18}%
  \BibitemOpen
  \bibfield  {author} {\bibinfo {author} {\bibfnamefont {S.}~\bibnamefont
  {Wehner}}, \bibinfo {author} {\bibfnamefont {D.}~\bibnamefont {Elkouss}},\
  and\ \bibinfo {author} {\bibfnamefont {R.}~\bibnamefont {Hanson}},\
  }\href@noop {} {\bibfield  {journal} {\bibinfo  {journal} {Science}\ }\textbf
  {\bibinfo {volume} {362}},\ \bibinfo {pages} {eaam9288} (\bibinfo {year}
  {2018})}\BibitemShut {NoStop}%
\bibitem [{\citenamefont {Yin}\ \emph {et~al.}(2023)\citenamefont {Yin},
  \citenamefont {Fu}, \citenamefont {Li}, \citenamefont {Weng}, \citenamefont
  {Li}, \citenamefont {Gu}, \citenamefont {Lu}, \citenamefont {Huang},\ and\
  \citenamefont {Chen}}]{YFL+23}%
  \BibitemOpen
  \bibfield  {author} {\bibinfo {author} {\bibfnamefont {H.-L.}\ \bibnamefont
  {Yin}}, \bibinfo {author} {\bibfnamefont {Y.}~\bibnamefont {Fu}}, \bibinfo
  {author} {\bibfnamefont {C.-L.}\ \bibnamefont {Li}}, \bibinfo {author}
  {\bibfnamefont {C.-X.}\ \bibnamefont {Weng}}, \bibinfo {author}
  {\bibfnamefont {B.-H.}\ \bibnamefont {Li}}, \bibinfo {author} {\bibfnamefont
  {J.}~\bibnamefont {Gu}}, \bibinfo {author} {\bibfnamefont {Y.-S.}\
  \bibnamefont {Lu}}, \bibinfo {author} {\bibfnamefont {S.}~\bibnamefont
  {Huang}},\ and\ \bibinfo {author} {\bibfnamefont {Z.-B.}\ \bibnamefont
  {Chen}},\ }\href@noop {} {\bibfield  {journal} {\bibinfo  {journal} {National
  Science Review}\ }\textbf {\bibinfo {volume} {10}},\ \bibinfo {pages}
  {nwac228} (\bibinfo {year} {2023})}\BibitemShut {NoStop}%
\bibitem [{\citenamefont {Cao}\ \emph {et~al.}(2024)\citenamefont {Cao},
  \citenamefont {Li}, \citenamefont {Wang}, \citenamefont {Fu}, \citenamefont
  {Yin},\ and\ \citenamefont {Chen}}]{CLW+24}%
  \BibitemOpen
  \bibfield  {author} {\bibinfo {author} {\bibfnamefont {X.-Y.}\ \bibnamefont
  {Cao}}, \bibinfo {author} {\bibfnamefont {B.-H.}\ \bibnamefont {Li}},
  \bibinfo {author} {\bibfnamefont {Y.}~\bibnamefont {Wang}}, \bibinfo {author}
  {\bibfnamefont {Y.}~\bibnamefont {Fu}}, \bibinfo {author} {\bibfnamefont
  {H.-L.}\ \bibnamefont {Yin}},\ and\ \bibinfo {author} {\bibfnamefont {Z.-B.}\
  \bibnamefont {Chen}},\ }\href@noop {} {\bibfield  {journal} {\bibinfo
  {journal} {Science Advances}\ }\textbf {\bibinfo {volume} {10}},\ \bibinfo
  {pages} {eadk3258} (\bibinfo {year} {2024})}\BibitemShut {NoStop}%
\bibitem [{\citenamefont {Schmitt-Manderbach}\ \emph
  {et~al.}(2007)\citenamefont {Schmitt-Manderbach}, \citenamefont {Weier},
  \citenamefont {F{\"u}rst}, \citenamefont {Ursin}, \citenamefont
  {Tiefenbacher}, \citenamefont {Scheidl}, \citenamefont {Perdigues},
  \citenamefont {Sodnik}, \citenamefont {Kurtsiefer}, \citenamefont {Rarity}
  \emph {et~al.}}]{MWF+07}%
  \BibitemOpen
  \bibfield  {author} {\bibinfo {author} {\bibfnamefont {T.}~\bibnamefont
  {Schmitt-Manderbach}}, \bibinfo {author} {\bibfnamefont {H.}~\bibnamefont
  {Weier}}, \bibinfo {author} {\bibfnamefont {M.}~\bibnamefont {F{\"u}rst}},
  \bibinfo {author} {\bibfnamefont {R.}~\bibnamefont {Ursin}}, \bibinfo
  {author} {\bibfnamefont {F.}~\bibnamefont {Tiefenbacher}}, \bibinfo {author}
  {\bibfnamefont {T.}~\bibnamefont {Scheidl}}, \bibinfo {author} {\bibfnamefont
  {J.}~\bibnamefont {Perdigues}}, \bibinfo {author} {\bibfnamefont
  {Z.}~\bibnamefont {Sodnik}}, \bibinfo {author} {\bibfnamefont
  {C.}~\bibnamefont {Kurtsiefer}}, \bibinfo {author} {\bibfnamefont {J.~G.}\
  \bibnamefont {Rarity}}, \emph {et~al.},\ }\href@noop {} {\bibfield  {journal}
  {\bibinfo  {journal} {Physical Review Letters}\ }\textbf {\bibinfo {volume}
  {98}},\ \bibinfo {pages} {010504} (\bibinfo {year} {2007})}\BibitemShut
  {NoStop}%
\bibitem [{\citenamefont {Dubey}\ \emph {et~al.}(2024)\citenamefont {Dubey},
  \citenamefont {Bhole}, \citenamefont {Dutta}, \citenamefont {Behera},
  \citenamefont {Losu}, \citenamefont {Pandeeti}, \citenamefont {Metkar},
  \citenamefont {Banerjee},\ and\ \citenamefont {Pathak}}]{DBD+24}%
  \BibitemOpen
  \bibfield  {author} {\bibinfo {author} {\bibfnamefont {U.}~\bibnamefont
  {Dubey}}, \bibinfo {author} {\bibfnamefont {P.}~\bibnamefont {Bhole}},
  \bibinfo {author} {\bibfnamefont {A.}~\bibnamefont {Dutta}}, \bibinfo
  {author} {\bibfnamefont {D.~P.}\ \bibnamefont {Behera}}, \bibinfo {author}
  {\bibfnamefont {V.}~\bibnamefont {Losu}}, \bibinfo {author} {\bibfnamefont
  {G.~S.~D.}\ \bibnamefont {Pandeeti}}, \bibinfo {author} {\bibfnamefont
  {A.~R.}\ \bibnamefont {Metkar}}, \bibinfo {author} {\bibfnamefont
  {A.}~\bibnamefont {Banerjee}},\ and\ \bibinfo {author} {\bibfnamefont
  {A.}~\bibnamefont {Pathak}},\ }\href@noop {} {\bibfield  {journal} {\bibinfo
  {journal} {Physics Open}\ ,\ \bibinfo {pages} {100210}} (\bibinfo {year}
  {2024})}\BibitemShut {NoStop}%
\bibitem [{\citenamefont {Liao}\ \emph
  {et~al.}(2017{\natexlab{a}})\citenamefont {Liao}, \citenamefont {Cai},
  \citenamefont {Liu}, \citenamefont {Zhang}, \citenamefont {Li}, \citenamefont
  {Ren}, \citenamefont {Yin}, \citenamefont {Shen}, \citenamefont {Cao},
  \citenamefont {Li} \emph {et~al.}}]{LCL+17}%
  \BibitemOpen
  \bibfield  {author} {\bibinfo {author} {\bibfnamefont {S.-K.}\ \bibnamefont
  {Liao}}, \bibinfo {author} {\bibfnamefont {W.-Q.}\ \bibnamefont {Cai}},
  \bibinfo {author} {\bibfnamefont {W.-Y.}\ \bibnamefont {Liu}}, \bibinfo
  {author} {\bibfnamefont {L.}~\bibnamefont {Zhang}}, \bibinfo {author}
  {\bibfnamefont {Y.}~\bibnamefont {Li}}, \bibinfo {author} {\bibfnamefont
  {J.-G.}\ \bibnamefont {Ren}}, \bibinfo {author} {\bibfnamefont
  {J.}~\bibnamefont {Yin}}, \bibinfo {author} {\bibfnamefont {Q.}~\bibnamefont
  {Shen}}, \bibinfo {author} {\bibfnamefont {Y.}~\bibnamefont {Cao}}, \bibinfo
  {author} {\bibfnamefont {Z.-P.}\ \bibnamefont {Li}}, \emph {et~al.},\
  }\href@noop {} {\bibfield  {journal} {\bibinfo  {journal} {Nature}\ }\textbf
  {\bibinfo {volume} {549}},\ \bibinfo {pages} {43} (\bibinfo {year}
  {2017}{\natexlab{a}})}\BibitemShut {NoStop}%
\bibitem [{\citenamefont {Liao}\ \emph {et~al.}(2018)\citenamefont {Liao},
  \citenamefont {Cai}, \citenamefont {Handsteiner}, \citenamefont {Liu},
  \citenamefont {Yin}, \citenamefont {Zhang}, \citenamefont {Rauch},
  \citenamefont {Fink}, \citenamefont {Ren}, \citenamefont {Liu} \emph
  {et~al.}}]{LCH+18}%
  \BibitemOpen
  \bibfield  {author} {\bibinfo {author} {\bibfnamefont {S.-K.}\ \bibnamefont
  {Liao}}, \bibinfo {author} {\bibfnamefont {W.-Q.}\ \bibnamefont {Cai}},
  \bibinfo {author} {\bibfnamefont {J.}~\bibnamefont {Handsteiner}}, \bibinfo
  {author} {\bibfnamefont {B.}~\bibnamefont {Liu}}, \bibinfo {author}
  {\bibfnamefont {J.}~\bibnamefont {Yin}}, \bibinfo {author} {\bibfnamefont
  {L.}~\bibnamefont {Zhang}}, \bibinfo {author} {\bibfnamefont
  {D.}~\bibnamefont {Rauch}}, \bibinfo {author} {\bibfnamefont
  {M.}~\bibnamefont {Fink}}, \bibinfo {author} {\bibfnamefont {J.-G.}\
  \bibnamefont {Ren}}, \bibinfo {author} {\bibfnamefont {W.-Y.}\ \bibnamefont
  {Liu}}, \emph {et~al.},\ }\href@noop {} {\bibfield  {journal} {\bibinfo
  {journal} {Physical Review Letters}\ }\textbf {\bibinfo {volume} {120}},\
  \bibinfo {pages} {030501} (\bibinfo {year} {2018})}\BibitemShut {NoStop}%
\bibitem [{\citenamefont {Liorni}\ \emph {et~al.}(2019)\citenamefont {Liorni},
  \citenamefont {Kampermann},\ and\ \citenamefont {Bru{\ss}}}]{LKB19}%
  \BibitemOpen
  \bibfield  {author} {\bibinfo {author} {\bibfnamefont {C.}~\bibnamefont
  {Liorni}}, \bibinfo {author} {\bibfnamefont {H.}~\bibnamefont {Kampermann}},\
  and\ \bibinfo {author} {\bibfnamefont {D.}~\bibnamefont {Bru{\ss}}},\
  }\href@noop {} {\bibfield  {journal} {\bibinfo  {journal} {New Journal of
  Physics}\ }\textbf {\bibinfo {volume} {21}},\ \bibinfo {pages} {093055}
  (\bibinfo {year} {2019})}\BibitemShut {NoStop}%
\bibitem [{\citenamefont {Dutta}\ \emph {et~al.}(2024)\citenamefont {Dutta},
  \citenamefont {Muskan}, \citenamefont {Banerjee},\ and\ \citenamefont
  {Pathak}}]{DMB+24}%
  \BibitemOpen
  \bibfield  {author} {\bibinfo {author} {\bibfnamefont {A.}~\bibnamefont
  {Dutta}}, \bibinfo {author} {\bibnamefont {Muskan}}, \bibinfo {author}
  {\bibfnamefont {S.}~\bibnamefont {Banerjee}},\ and\ \bibinfo {author}
  {\bibfnamefont {A.}~\bibnamefont {Pathak}},\ }\href@noop {} {\bibfield
  {journal} {\bibinfo  {journal} {Advanced Quantum Technologies}\ ,\ \bibinfo
  {pages} {2400149}} (\bibinfo {year} {2024})}\BibitemShut {NoStop}%
\bibitem [{\citenamefont {Liang}\ and\ \citenamefont {Jiao}(2020)}]{LJ20}%
  \BibitemOpen
  \bibfield  {author} {\bibinfo {author} {\bibfnamefont {W.}~\bibnamefont
  {Liang}}\ and\ \bibinfo {author} {\bibfnamefont {R.}~\bibnamefont {Jiao}},\
  }\href@noop {} {\bibfield  {journal} {\bibinfo  {journal} {New Journal of
  Physics}\ }\textbf {\bibinfo {volume} {22}},\ \bibinfo {pages} {083074}
  (\bibinfo {year} {2020})}\BibitemShut {NoStop}%
\bibitem [{\citenamefont {Curty}\ \emph {et~al.}(2004)\citenamefont {Curty},
  \citenamefont {Lewenstein},\ and\ \citenamefont {L{\"u}tkenhaus}}]{CLL04}%
  \BibitemOpen
  \bibfield  {author} {\bibinfo {author} {\bibfnamefont {M.}~\bibnamefont
  {Curty}}, \bibinfo {author} {\bibfnamefont {M.}~\bibnamefont {Lewenstein}},\
  and\ \bibinfo {author} {\bibfnamefont {N.}~\bibnamefont {L{\"u}tkenhaus}},\
  }\href@noop {} {\bibfield  {journal} {\bibinfo  {journal} {Physical Review
  Letters}\ }\textbf {\bibinfo {volume} {92}},\ \bibinfo {pages} {217903}
  (\bibinfo {year} {2004})}\BibitemShut {NoStop}%
\bibitem [{\citenamefont {Ferenczi}\ and\ \citenamefont
  {L{\"u}tkenhaus}(2012)}]{FL12}%
  \BibitemOpen
  \bibfield  {author} {\bibinfo {author} {\bibfnamefont {A.}~\bibnamefont
  {Ferenczi}}\ and\ \bibinfo {author} {\bibfnamefont {N.}~\bibnamefont
  {L{\"u}tkenhaus}},\ }\href@noop {} {\bibfield  {journal} {\bibinfo  {journal}
  {Physical Review A}\ }\textbf {\bibinfo {volume} {85}},\ \bibinfo {pages}
  {052310} (\bibinfo {year} {2012})}\BibitemShut {NoStop}%
\bibitem [{\citenamefont {Mishra}\ \emph {et~al.}(2023)\citenamefont {Mishra},
  \citenamefont {Mondal}, \citenamefont {Arunachalam}, \citenamefont {Nayak},
  \citenamefont {Tripathy},\ and\ \citenamefont {Palai}}]{MMA+23}%
  \BibitemOpen
  \bibfield  {author} {\bibinfo {author} {\bibfnamefont {C.}~\bibnamefont
  {Mishra}}, \bibinfo {author} {\bibfnamefont {S.~R.}\ \bibnamefont {Mondal}},
  \bibinfo {author} {\bibfnamefont {R.}~\bibnamefont {Arunachalam}}, \bibinfo
  {author} {\bibfnamefont {M.}~\bibnamefont {Nayak}}, \bibinfo {author}
  {\bibfnamefont {S.}~\bibnamefont {Tripathy}},\ and\ \bibinfo {author}
  {\bibfnamefont {G.}~\bibnamefont {Palai}},\ }\href@noop {} {\bibfield
  {journal} {\bibinfo  {journal} {Optical and Quantum Electronics}\ }\textbf
  {\bibinfo {volume} {55}},\ \bibinfo {pages} {33} (\bibinfo {year}
  {2023})}\BibitemShut {NoStop}%
\bibitem [{\citenamefont {Vasylyev}\ \emph {et~al.}(2016)\citenamefont
  {Vasylyev}, \citenamefont {Semenov},\ and\ \citenamefont {Vogel}}]{VSV16}%
  \BibitemOpen
  \bibfield  {author} {\bibinfo {author} {\bibfnamefont {D.}~\bibnamefont
  {Vasylyev}}, \bibinfo {author} {\bibfnamefont {A.}~\bibnamefont {Semenov}},\
  and\ \bibinfo {author} {\bibfnamefont {W.}~\bibnamefont {Vogel}},\
  }\href@noop {} {\bibfield  {journal} {\bibinfo  {journal} {Physical Review
  Letters}\ }\textbf {\bibinfo {volume} {117}},\ \bibinfo {pages} {090501}
  (\bibinfo {year} {2016})}\BibitemShut {NoStop}%
\bibitem [{\citenamefont {Vasylyev}\ \emph {et~al.}(2017)\citenamefont
  {Vasylyev}, \citenamefont {Semenov}, \citenamefont {Vogel}, \citenamefont
  {G{\"u}nthner}, \citenamefont {Thurn}, \citenamefont {Bayraktar},\ and\
  \citenamefont {Marquardt}}]{VSV+17}%
  \BibitemOpen
  \bibfield  {author} {\bibinfo {author} {\bibfnamefont {D.}~\bibnamefont
  {Vasylyev}}, \bibinfo {author} {\bibfnamefont {A.}~\bibnamefont {Semenov}},
  \bibinfo {author} {\bibfnamefont {W.}~\bibnamefont {Vogel}}, \bibinfo
  {author} {\bibfnamefont {K.}~\bibnamefont {G{\"u}nthner}}, \bibinfo {author}
  {\bibfnamefont {A.}~\bibnamefont {Thurn}}, \bibinfo {author} {\bibfnamefont
  {{\"O}.}~\bibnamefont {Bayraktar}},\ and\ \bibinfo {author} {\bibfnamefont
  {C.}~\bibnamefont {Marquardt}},\ }\href@noop {} {\bibfield  {journal}
  {\bibinfo  {journal} {Physical Review A}\ }\textbf {\bibinfo {volume} {96}},\
  \bibinfo {pages} {043856} (\bibinfo {year} {2017})}\BibitemShut {NoStop}%
\bibitem [{\citenamefont {Vasylyev}\ \emph {et~al.}(2012)\citenamefont
  {Vasylyev}, \citenamefont {Semenov},\ and\ \citenamefont {Vogel}}]{VSV12}%
  \BibitemOpen
  \bibfield  {author} {\bibinfo {author} {\bibfnamefont {D.~Y.}\ \bibnamefont
  {Vasylyev}}, \bibinfo {author} {\bibfnamefont {A.}~\bibnamefont {Semenov}},\
  and\ \bibinfo {author} {\bibfnamefont {W.}~\bibnamefont {Vogel}},\
  }\href@noop {} {\bibfield  {journal} {\bibinfo  {journal} {Physical Review
  Letters}\ }\textbf {\bibinfo {volume} {108}},\ \bibinfo {pages} {220501}
  (\bibinfo {year} {2012})}\BibitemShut {NoStop}%
\bibitem [{\citenamefont {Wang}\ \emph {et~al.}(2018)\citenamefont {Wang},
  \citenamefont {Huang}, \citenamefont {Wang},\ and\ \citenamefont
  {Zeng}}]{WHW+18}%
  \BibitemOpen
  \bibfield  {author} {\bibinfo {author} {\bibfnamefont {S.}~\bibnamefont
  {Wang}}, \bibinfo {author} {\bibfnamefont {P.}~\bibnamefont {Huang}},
  \bibinfo {author} {\bibfnamefont {T.}~\bibnamefont {Wang}},\ and\ \bibinfo
  {author} {\bibfnamefont {G.}~\bibnamefont {Zeng}},\ }\href@noop {} {\bibfield
   {journal} {\bibinfo  {journal} {New Journal of Physics}\ }\textbf {\bibinfo
  {volume} {20}},\ \bibinfo {pages} {083037} (\bibinfo {year}
  {2018})}\BibitemShut {NoStop}%
\bibitem [{\citenamefont {Bourgoin}\ \emph {et~al.}(2013)\citenamefont
  {Bourgoin}, \citenamefont {Meyer-Scott}, \citenamefont {Higgins},
  \citenamefont {Helou}, \citenamefont {Erven}, \citenamefont {Huebel},
  \citenamefont {Kumar}, \citenamefont {Hudson}, \citenamefont {D'Souza},
  \citenamefont {Girard} \emph {et~al.}}]{BSH+13}%
  \BibitemOpen
  \bibfield  {author} {\bibinfo {author} {\bibfnamefont {J.}~\bibnamefont
  {Bourgoin}}, \bibinfo {author} {\bibfnamefont {E.}~\bibnamefont
  {Meyer-Scott}}, \bibinfo {author} {\bibfnamefont {B.~L.}\ \bibnamefont
  {Higgins}}, \bibinfo {author} {\bibfnamefont {B.}~\bibnamefont {Helou}},
  \bibinfo {author} {\bibfnamefont {C.}~\bibnamefont {Erven}}, \bibinfo
  {author} {\bibfnamefont {H.}~\bibnamefont {Huebel}}, \bibinfo {author}
  {\bibfnamefont {B.}~\bibnamefont {Kumar}}, \bibinfo {author} {\bibfnamefont
  {D.}~\bibnamefont {Hudson}}, \bibinfo {author} {\bibfnamefont
  {I.}~\bibnamefont {D'Souza}}, \bibinfo {author} {\bibfnamefont
  {R.}~\bibnamefont {Girard}}, \emph {et~al.},\ }\href@noop {} {\bibfield
  {journal} {\bibinfo  {journal} {New Journal of Physics}\ }\textbf {\bibinfo
  {volume} {15}},\ \bibinfo {pages} {023006} (\bibinfo {year}
  {2013})}\BibitemShut {NoStop}%
\bibitem [{\citenamefont {Vargas}\ \emph {et~al.}(2000)\citenamefont {Vargas},
  \citenamefont {Ben{\'\i}tez},\ and\ \citenamefont {Bajo}}]{VBB2000}%
  \BibitemOpen
  \bibfield  {author} {\bibinfo {author} {\bibfnamefont {M.~J.}\ \bibnamefont
  {Vargas}}, \bibinfo {author} {\bibfnamefont {P.~M.}\ \bibnamefont
  {Ben{\'\i}tez}},\ and\ \bibinfo {author} {\bibfnamefont {F.~S.}\ \bibnamefont
  {Bajo}},\ }\href@noop {} {\bibfield  {journal} {\bibinfo  {journal} {European
  Journal of Physics}\ }\textbf {\bibinfo {volume} {21}},\ \bibinfo {pages}
  {245} (\bibinfo {year} {2000})}\BibitemShut {NoStop}%
\bibitem [{\citenamefont {Yin}\ \emph {et~al.}(2017{\natexlab{a}})\citenamefont
  {Yin}, \citenamefont {Cao}, \citenamefont {Li}, \citenamefont {Ren},
  \citenamefont {Liao}, \citenamefont {Zhang}, \citenamefont {Cai},
  \citenamefont {Liu}, \citenamefont {Li}, \citenamefont {Dai} \emph
  {et~al.}}]{YCL+17}%
  \BibitemOpen
  \bibfield  {author} {\bibinfo {author} {\bibfnamefont {J.}~\bibnamefont
  {Yin}}, \bibinfo {author} {\bibfnamefont {Y.}~\bibnamefont {Cao}}, \bibinfo
  {author} {\bibfnamefont {Y.-H.}\ \bibnamefont {Li}}, \bibinfo {author}
  {\bibfnamefont {J.-G.}\ \bibnamefont {Ren}}, \bibinfo {author} {\bibfnamefont
  {S.-K.}\ \bibnamefont {Liao}}, \bibinfo {author} {\bibfnamefont
  {L.}~\bibnamefont {Zhang}}, \bibinfo {author} {\bibfnamefont {W.-Q.}\
  \bibnamefont {Cai}}, \bibinfo {author} {\bibfnamefont {W.-Y.}\ \bibnamefont
  {Liu}}, \bibinfo {author} {\bibfnamefont {B.}~\bibnamefont {Li}}, \bibinfo
  {author} {\bibfnamefont {H.}~\bibnamefont {Dai}}, \emph {et~al.},\
  }\href@noop {} {\bibfield  {journal} {\bibinfo  {journal} {Physical Review
  Letters}\ }\textbf {\bibinfo {volume} {119}},\ \bibinfo {pages} {200501}
  (\bibinfo {year} {2017}{\natexlab{a}})}\BibitemShut {NoStop}%
\bibitem [{\citenamefont {Ren}\ \emph {et~al.}(2017)\citenamefont {Ren},
  \citenamefont {Xu}, \citenamefont {Yong}, \citenamefont {Zhang},
  \citenamefont {Liao}, \citenamefont {Yin}, \citenamefont {Liu}, \citenamefont
  {Cai}, \citenamefont {Yang}, \citenamefont {Li} \emph {et~al.}}]{RXY+17}%
  \BibitemOpen
  \bibfield  {author} {\bibinfo {author} {\bibfnamefont {J.-G.}\ \bibnamefont
  {Ren}}, \bibinfo {author} {\bibfnamefont {P.}~\bibnamefont {Xu}}, \bibinfo
  {author} {\bibfnamefont {H.-L.}\ \bibnamefont {Yong}}, \bibinfo {author}
  {\bibfnamefont {L.}~\bibnamefont {Zhang}}, \bibinfo {author} {\bibfnamefont
  {S.-K.}\ \bibnamefont {Liao}}, \bibinfo {author} {\bibfnamefont
  {J.}~\bibnamefont {Yin}}, \bibinfo {author} {\bibfnamefont {W.-Y.}\
  \bibnamefont {Liu}}, \bibinfo {author} {\bibfnamefont {W.-Q.}\ \bibnamefont
  {Cai}}, \bibinfo {author} {\bibfnamefont {M.}~\bibnamefont {Yang}}, \bibinfo
  {author} {\bibfnamefont {L.}~\bibnamefont {Li}}, \emph {et~al.},\ }\href@noop
  {} {\bibfield  {journal} {\bibinfo  {journal} {Nature}\ }\textbf {\bibinfo
  {volume} {549}},\ \bibinfo {pages} {70} (\bibinfo {year} {2017})}\BibitemShut
  {NoStop}%
\bibitem [{\citenamefont {Yin}\ \emph {et~al.}(2017{\natexlab{b}})\citenamefont
  {Yin}, \citenamefont {Cao}, \citenamefont {Li}, \citenamefont {Liao},
  \citenamefont {Zhang}, \citenamefont {Ren}, \citenamefont {Cai},
  \citenamefont {Liu}, \citenamefont {Li}, \citenamefont {Dai} \emph
  {et~al.}}]{YCLL+17}%
  \BibitemOpen
  \bibfield  {author} {\bibinfo {author} {\bibfnamefont {J.}~\bibnamefont
  {Yin}}, \bibinfo {author} {\bibfnamefont {Y.}~\bibnamefont {Cao}}, \bibinfo
  {author} {\bibfnamefont {Y.-H.}\ \bibnamefont {Li}}, \bibinfo {author}
  {\bibfnamefont {S.-K.}\ \bibnamefont {Liao}}, \bibinfo {author}
  {\bibfnamefont {L.}~\bibnamefont {Zhang}}, \bibinfo {author} {\bibfnamefont
  {J.-G.}\ \bibnamefont {Ren}}, \bibinfo {author} {\bibfnamefont {W.-Q.}\
  \bibnamefont {Cai}}, \bibinfo {author} {\bibfnamefont {W.-Y.}\ \bibnamefont
  {Liu}}, \bibinfo {author} {\bibfnamefont {B.}~\bibnamefont {Li}}, \bibinfo
  {author} {\bibfnamefont {H.}~\bibnamefont {Dai}}, \emph {et~al.},\
  }\href@noop {} {\bibfield  {journal} {\bibinfo  {journal} {Science}\ }\textbf
  {\bibinfo {volume} {356}},\ \bibinfo {pages} {1140} (\bibinfo {year}
  {2017}{\natexlab{b}})}\BibitemShut {NoStop}%
\bibitem [{\citenamefont {Liao}\ \emph
  {et~al.}(2017{\natexlab{b}})\citenamefont {Liao}, \citenamefont {Yong},
  \citenamefont {Liu}, \citenamefont {Shentu}, \citenamefont {Li},
  \citenamefont {Lin}, \citenamefont {Dai}, \citenamefont {Zhao}, \citenamefont
  {Li}, \citenamefont {Guan} \emph {et~al.}}]{LYL+17}%
  \BibitemOpen
  \bibfield  {author} {\bibinfo {author} {\bibfnamefont {S.-K.}\ \bibnamefont
  {Liao}}, \bibinfo {author} {\bibfnamefont {H.-L.}\ \bibnamefont {Yong}},
  \bibinfo {author} {\bibfnamefont {C.}~\bibnamefont {Liu}}, \bibinfo {author}
  {\bibfnamefont {G.-L.}\ \bibnamefont {Shentu}}, \bibinfo {author}
  {\bibfnamefont {D.-D.}\ \bibnamefont {Li}}, \bibinfo {author} {\bibfnamefont
  {J.}~\bibnamefont {Lin}}, \bibinfo {author} {\bibfnamefont {H.}~\bibnamefont
  {Dai}}, \bibinfo {author} {\bibfnamefont {S.-Q.}\ \bibnamefont {Zhao}},
  \bibinfo {author} {\bibfnamefont {B.}~\bibnamefont {Li}}, \bibinfo {author}
  {\bibfnamefont {J.-Y.}\ \bibnamefont {Guan}}, \emph {et~al.},\ }\href@noop {}
  {\bibfield  {journal} {\bibinfo  {journal} {Nature Photonics}\ }\textbf
  {\bibinfo {volume} {11}},\ \bibinfo {pages} {509} (\bibinfo {year}
  {2017}{\natexlab{b}})}\BibitemShut {NoStop}%
\end{thebibliography}%

\appendix
%dummy comment inserted by tex2lyx to ensure that this paragraph is not empty

\section*{Appendix A}

\subsection*{Loss-only scenario}

Here, we demonstrate how to obtain $\langle\boldsymbol{\mathbf{\alpha}}|{\rm E}^{\gamma}|\boldsymbol{\beta}\rangle$,
where ${\rm E}^{\gamma}$ represents Eve's POVM associated with the
loss-only scenario and $\boldsymbol{\mathbf{\alpha}},\boldsymbol{\mathbf{\beta}}\in\mathcal{S}$.
First, Alice (Bob) prepares coherent state $|\alpha_{{\rm A}}\rangle_{{\rm A^{\prime}}}$
$\left(|\alpha_{{\rm B}}\rangle_{{\rm B^{\prime}}}\right)$ in the
registers ${\rm A}^{'}$ $\left({\rm B}^{'}\right)$, and send it
to Charlie. After passing through the lossy channel, the state becomes
$\left|\eta^{\frac{1}{4}}\alpha_{{\rm A}},\eta^{\frac{1}{4}}\alpha_{{\rm B}}\right\rangle _{I_{{\rm A}}I_{{\rm B}}}$,
while Eve's state is $\left|\sqrt{1-\eta^{\frac{1}{2}}}\alpha_{{\rm A}},\sqrt{1-\eta^{\frac{1}{2}}}\alpha_{{\rm B}}\right\rangle _{E_{{\rm A}}E_{{\rm B}}}$.
After the beam splitter, the state transforms to $\left|\frac{\eta^{\frac{1}{4}}\left(\alpha_{{\rm A}}+\alpha_{{\rm B}}\right)}{\sqrt{2}},\frac{\eta^{\frac{1}{4}}\left(\alpha_{{\rm A}}-\alpha_{{\rm B}}\right)}{\sqrt{2}}\right\rangle _{{\rm O_{A}}{\rm O_{B}}}$.
In this setup, the input modes of the beam splitter at the central
node (Charlie) are $I_{{\rm A}}$ and $I_{{\rm B}}$, and the output
modes are ${\rm O_{A}}$ and ${\rm O_{B}}$, which reach detectors
${\rm D_{+}}$ and ${\rm D_{-}}$, respectively. We now apply the
POVM of the detectors to this state. The ideal detectors used by Charlie
are characterized by the following POVM,

\begin{equation}
\begin{array}{lcl}
\Omega_{{\rm ideal}}^{+} & = & \left(\mathds{1}_{{\rm O_{A}}}-|0\rangle\langle0|_{{\rm O_{A}}}\right)\otimes|0\rangle\langle0|_{{\rm O_{B}}},\\
\\\Omega_{{\rm ideal}}^{-} & = & |0\rangle\langle0|_{{\rm O_{A}}}\otimes\left(\mathds{1}_{{\rm O_{B}}}-|0\rangle\langle0|_{{\rm O_{B}}}\right),\\
\\\Omega_{{\rm ideal}}^{?} & = & |0\rangle\langle0|_{{\rm O_{A}}}\otimes|0\rangle\langle0|_{{\rm O_{B}}},\\
\\\Omega_{{\rm ideal}}^{{\rm d}} & = & \left(\mathds{1}_{{\rm O_{A}}}-|0\rangle\langle0|_{{\rm O_{A}}}\right)\otimes\left(\mathds{1}_{{\rm O_{B}}}-|0\rangle\langle0|_{{\rm O_{B}}}\right),
\end{array}\label{eq:Detectors'=000020POVM=000020Loss-Only}
\end{equation}
where $\mathds{1}$ denotes the identity operator, while $|0\rangle$
represents the vacuum state. Considering $|\boldsymbol{\alpha}\rangle=|\alpha_{{\rm A}},\alpha_{{\rm B}}\rangle$
and $|\boldsymbol{\beta}\rangle=|\beta_{{\rm A}},\beta_{{\rm B}}\rangle$,
the general expression for the final state resulting from the output
of Charlie's beam splitter and Eve's disposal can be expressed, 

\begin{equation}
\begin{array}{lcl}
\langle\boldsymbol{\mathbf{\alpha}}|{\rm E}^{\gamma}|\boldsymbol{\beta}\rangle & = & \left\langle \frac{\eta^{\frac{1}{4}}\left(\alpha_{{\rm A}}+\alpha_{{\rm B}}\right)}{\sqrt{2}},\frac{\eta^{\frac{1}{4}}\left(\alpha_{{\rm A}}+\alpha_{{\rm B}}\right)}{\sqrt{2}}\right|\Omega_{{\rm ideal}}^{\gamma}\left|\frac{\eta^{\frac{1}{4}}\left(\beta_{{\rm A}}+\beta_{{\rm B}}\right)}{\sqrt{2}},\frac{\eta^{\frac{1}{4}}\left(\beta_{{\rm A}}+\beta_{{\rm B}}\right)}{\sqrt{2}}\right\rangle _{{\rm O_{A}O_{B}}}\left\langle \sqrt{1-\eta^{\frac{1}{2}}}\alpha_{{\rm A}}|\sqrt{1-\eta^{\frac{1}{2}}}\beta_{{\rm A}}\right\rangle _{E_{{\rm A}}}\\
 & \times & \left\langle \sqrt{1-\eta^{\frac{1}{2}}}\alpha_{{\rm B}}|\sqrt{1-\eta^{\frac{1}{2}}}\beta_{{\rm B}}\right\rangle _{E_{{\rm B}}}.
\end{array}\label{eq:Genral=000020Probability=000020with=000020General=000020POVM}
\end{equation}
From Eqs. (\ref{eq:Genral=000020Probability=000020with=000020General=000020POVM})
and (\ref{eq:Detectors'=000020POVM=000020Loss-Only}), we derive the
following relationships,

\begin{equation}
\begin{array}{lcl}
\langle\boldsymbol{\mathbf{\alpha}}|{\rm E}^{+}|\boldsymbol{\beta}\rangle & = & \left[e^{\frac{\sqrt{\eta}}{2}\left(-\frac{\left|\alpha_{{\rm A}}+\alpha_{{\rm B}}\right|^{2}}{2}-\frac{\left|\beta_{{\rm A}}+\beta_{{\rm B}}\right|^{2}}{2}+\left(\alpha_{{\rm A}}+\alpha_{{\rm B}}\right)^{*}\left(\beta_{{\rm A}}+\beta_{{\rm B}}\right)\right)}-e^{-\frac{\sqrt{\eta}}{4}\left(\left|\alpha_{{\rm A}}+\alpha_{{\rm B}}\right|^{2}+\left|\beta_{{\rm A}}+\beta_{{\rm B}}\right|^{2}\right)}\right]\\
 & \times & e^{-\frac{\sqrt{\eta}}{4}\left(\left|\alpha_{{\rm A}}-\alpha_{{\rm B}}\right|^{2}+\left|\beta_{{\rm A}}-\beta_{{\rm B}}\right|^{2}\right)}e^{\left(1-\sqrt{\eta}\right)\left(-\frac{\left|\alpha_{{\rm A}}\right|^{2}}{2}-\frac{\left|\beta_{{\rm A}}\right|^{2}}{2}+\alpha_{{\rm A}}^{*}\beta_{{\rm A}}\right)}e^{\left(1-\sqrt{\eta}\right)\left(-\frac{\left|\alpha_{{\rm B}}\right|^{2}}{2}-\frac{\left|\beta_{{\rm B}}\right|^{2}}{2}+\alpha_{{\rm B}}^{*}\beta_{{\rm B}}\right),}\\
\\\langle\boldsymbol{\mathbf{\alpha}}|{\rm E}^{-}|\boldsymbol{\beta}\rangle & = & e^{-\frac{\sqrt{\eta}}{4}\left(\left|\alpha_{{\rm A}}+\alpha_{{\rm B}}\right|^{2}+\left|\beta_{{\rm A}}+\beta_{{\rm B}}\right|^{2}\right)}\left[e^{\frac{\sqrt{\eta}}{2}\left(-\frac{\left|\alpha_{{\rm A}}-\alpha_{{\rm B}}\right|^{2}}{2}-\frac{\left|\beta_{{\rm A}}-\beta_{{\rm B}}\right|^{2}}{2}+\left(\alpha_{{\rm A}}-\alpha_{{\rm B}}\right)^{*}\left(\beta_{{\rm A}}-\beta_{{\rm B}}\right)\right)}\right.\\
 & - & \left.e^{-\frac{\sqrt{\eta}}{4}\left(\left|\alpha_{{\rm A}}-\alpha_{{\rm B}}\right|^{2}+\left|\beta_{{\rm A}}-\beta_{{\rm B}}\right|^{2}\right)}\right]e^{\left(1-\sqrt{\eta}\right)\left(-\frac{\left|\alpha_{{\rm A}}\right|^{2}}{2}-\frac{\left|\beta_{{\rm A}}\right|^{2}}{2}+\alpha_{{\rm A}}^{*}\beta_{{\rm A}}\right)}e^{\left(1-\sqrt{\eta}\right)\left(-\frac{\left|\alpha_{{\rm B}}\right|^{2}}{2}-\frac{\left|\beta_{{\rm B}}\right|^{2}}{2}+\alpha_{{\rm B}}^{*}\beta_{{\rm B}}\right),}\\
\\\langle\boldsymbol{\mathbf{\alpha}}|{\rm E}^{?}|\boldsymbol{\beta}\rangle & = & e^{-\frac{\sqrt{\eta}}{4}\left(\left|\alpha_{{\rm A}}+\alpha_{{\rm B}}\right|^{2}+\left|\beta_{{\rm A}}+\beta_{{\rm B}}\right|^{2}\right)}e^{-\frac{\sqrt{\eta}}{4}\left(\left|\alpha_{{\rm A}}-\alpha_{{\rm B}}\right|^{2}+\left|\beta_{{\rm A}}-\beta_{{\rm B}}\right|^{2}\right)}\\
 & \times & e^{\left(1-\sqrt{\eta}\right)\left(-\frac{\left|\alpha_{{\rm A}}\right|^{2}}{2}-\frac{\left|\beta_{{\rm A}}\right|^{2}}{2}+\alpha_{{\rm A}}^{*}\beta_{{\rm A}}\right)}e^{\left(1-\sqrt{\eta}\right)\left(-\frac{\left|\alpha_{{\rm B}}\right|^{2}}{2}-\frac{\left|\beta_{{\rm B}}\right|^{2}}{2}+\alpha_{{\rm B}}^{*}\beta_{{\rm B}}\right),}\\
\\\langle\boldsymbol{\mathbf{\alpha}}|{\rm E}^{{\rm d}}|\boldsymbol{\beta}\rangle & = & \mathds{1-}\langle\boldsymbol{\mathbf{\alpha}}|{\rm E}^{+}|\boldsymbol{\beta}\rangle-\langle\boldsymbol{\mathbf{\alpha}}|{\rm E}^{-}|\boldsymbol{\beta}\rangle-\langle\boldsymbol{\mathbf{\alpha}}|{\rm E}^{?}|\boldsymbol{\beta}\rangle-\langle\boldsymbol{\mathbf{\alpha}}|{\rm E}^{{\rm d}}|\boldsymbol{\beta}\rangle.
\end{array}\label{eq:General=000020Probability=000020with=000020Specific=000020POVM}
\end{equation}
Now, we have the expression for ${\rm E}^{\gamma}$ in basis $\mathds{B}$,
allowing us to determine the values of $\langle\boldsymbol{\mathbf{\alpha}}|{\rm E}^{\gamma}|\boldsymbol{\beta}\rangle$
using Eq. (\ref{eq:General=000020Probability=000020with=000020Specific=000020POVM})
(also see Eq. (A6) in \cite{LL18}),

\begin{equation}
\begin{array}{lcl}
{\rm E}^{+} & = & \left(1-\pounds^{2}\right)\left(\begin{array}{cccc}
\frac{1-\pounds^{2}\zeta^{2}}{8c_{0}^{4}} & \frac{1-\pounds^{2}\zeta^{2}}{8c_{0}^{2}c_{1}^{2}} & 0 & 0\\
\\\frac{1-\pounds^{2}\zeta^{2}}{8c_{0}^{2}c_{1}^{2}} & \frac{1-\pounds^{2}\zeta^{2}}{8c_{1}^{4}} & 0 & 0\\
\\0 & 0 & \frac{1+\pounds^{2}\zeta^{2}}{8c_{0}^{2}c_{1}^{2}} & \frac{1+\pounds^{2}\zeta^{2}}{8c_{0}^{2}c_{1}^{2}}\\
\\0 & 0 & \frac{1+\pounds^{2}\zeta^{2}}{8c_{0}^{2}c_{1}^{2}} & \frac{1+\pounds^{2}\zeta^{2}}{8c_{0}^{2}c_{1}^{2}}
\end{array}\right),\\
\\{\rm E}^{-} & = & \left(1-\pounds^{2}\right)\left(\begin{array}{cccc}
\frac{1-\pounds^{2}\zeta^{2}}{8c_{0}^{4}} & \frac{-1+\pounds^{2}\zeta^{2}}{8c_{0}^{2}c_{1}^{2}} & 0 & 0\\
\\\frac{-1+\pounds^{2}\Omega^{2}}{8c_{0}^{2}c_{1}^{2}} & \frac{1-\pounds^{2}\zeta^{2}}{8c_{1}^{4}} & 0 & 0\\
\\0 & 0 & \frac{1+\pounds^{2}\zeta^{2}}{8c_{0}^{2}c_{1}^{2}} & \frac{-1-\pounds^{2}\zeta^{2}}{8c_{0}^{2}c_{1}^{2}}\\
\\0 & 0 & \frac{-1-\pounds^{2}\zeta^{2}}{8c_{0}^{2}c_{1}^{2}} & \frac{1+\pounds^{2}\zeta^{2}}{8c_{0}^{2}c_{1}^{2}}
\end{array}\right),\\
\\{\rm E}^{?} & = & \pounds^{2}\left(\begin{array}{cccc}
\frac{\left(1+\zeta\right)^{2}}{4c_{0}^{4}} & 0 & 0 & 0\\
\\0 & \frac{\left(1-\zeta\right)^{2}}{4c_{1}^{4}} & 0 & 0\\
\\0 & 0 & \frac{1-\zeta^{2}}{4c_{0}^{2}c_{1}^{2}} & 0\\
\\0 & 0 & 0 & \frac{1-\zeta^{2}}{4c_{0}^{2}c_{1}^{2}}
\end{array}\right),\\
\\{\rm E}^{{\rm d}} & = & 0.
\end{array}\label{eq:POVM=000020F=000020for=000020Loss-Only=000020Scenario}
\end{equation}
When Alice and Bob transmit coherent states $|\alpha_{{\rm A}}\rangle$
and $|\alpha_{{\rm B}}\rangle$ in the same optical mode, respectively,
the combined state becomes $|\eta^{\frac{1}{4}}\alpha_{{\rm A}},\eta^{\frac{1}{4}}\alpha_{{\rm B}}\rangle$
after passing through the lossy channel. Charlie measures that state,
the probability for each announcement outcome $\gamma$, represented
as $\langle\boldsymbol{\mathbf{\alpha}}|{\rm E}^{\gamma}|\boldsymbol{\mathbf{\alpha}}\rangle$,
can be computed using Eq. (\ref{eq:General=000020Probability=000020with=000020Specific=000020POVM})
as follows:

\begin{equation}
\begin{array}{lcl}
\left\langle \alpha_{{\rm A}},\alpha_{{\rm B}}\right|{\rm E}^{+}\left|\alpha_{{\rm A}},\alpha_{{\rm B}}\right\rangle  & = & \left(1-e^{-\frac{\sqrt{\eta}\left|\alpha_{{\rm A}}+\alpha_{{\rm B}}\right|^{2}}{2}}\right)e^{-\frac{\sqrt{\eta}\left|\alpha_{{\rm A}}-\alpha_{{\rm B}}\right|^{2}}{2}},\\
\\\left\langle \alpha_{{\rm A}},\alpha_{{\rm B}}\right|{\rm E}^{-}\left|\alpha_{{\rm A}},\alpha_{{\rm B}}\right\rangle  & = & e^{-\frac{\sqrt{\eta}\left|\alpha_{{\rm A}}+\alpha_{{\rm B}}\right|^{2}}{2}}\left(1-e^{-\frac{\sqrt{\eta}\left|\alpha_{{\rm A}}-\alpha_{{\rm B}}\right|^{2}}{2}}\right),\\
\\\left\langle \alpha_{{\rm A}},\alpha_{{\rm B}}\right|{\rm E}^{?}\left|\alpha_{{\rm A}},\alpha_{{\rm B}}\right\rangle  & = & e^{-\frac{\sqrt{\eta}\left|\alpha_{{\rm A}}+\alpha_{{\rm B}}\right|^{2}}{2}}e^{-\frac{\sqrt{\eta}\left|\alpha_{{\rm A}}-\alpha_{{\rm B}}\right|^{2}}{2}},\\
\\\left\langle \alpha_{{\rm A}},\alpha_{{\rm B}}\right|{\rm E}^{{\rm d}}\left|\alpha_{{\rm A}},\alpha_{{\rm B}}\right\rangle  & = & \left(1-e^{-\frac{\sqrt{\eta}\left|\alpha_{{\rm A}}+\alpha_{{\rm B}}\right|^{2}}{2}}\right)\left(1-e^{-\frac{\sqrt{\eta}\left|\alpha_{{\rm A}}-\alpha_{{\rm B}}\right|^{2}}{2}}\right).
\end{array}\label{eq:Probabilities=000020of=000020Charlie's=000020Announcement}
\end{equation}
To determine the final secret key rate achievable in the loss-only
scenario, we require the conditional probabilities of Alice and Bob's
initial coherent states (denoted by $|\alpha_{{\rm A}}\rangle$ and
$|\alpha_{{\rm B}}\rangle$) given Charlie's measurement outcome (denoted
by $\gamma$). Table \ref{tab:Conditional-Probability-Loss-Only-Scenario}
summarizes these conditional probabilities for each announcement outcome
($\gamma$) across all states in the set $\mathcal{S}$. We analyze
the mutual information between Alice and Bob's registers conditioned
on Charlie's different announcements. The mutual information, denoted
by $I\left(A:B\right)_{\rho_{ABE}^{\gamma}}$, is found to be $1$
for successful announcements ($\gamma=+\,{\rm or}\,-$) and $0$ for
inconclusive announcements ($\gamma=?\,{\rm or}\,{\rm d}$). These
results indicate that no secret key can be established from inconclusive
announcements, and also $\delta_{{\rm EC}}^{+}=\delta_{{\rm EC}}^{-}=0$.
Given $S\left(\rho^{k,+}\right)=S\left(\rho^{k,-}\right)=0$, the
key rate contribution arises from the von Neumann entropy of $\rho^{+}$
and $\rho^{-}$, expressed as $S\left(\rho^{+}\right)=S\left(\rho^{-}\right)=h\left(\frac{1-e^{-4\mu\left(1-\sqrt{\eta}\right)}e^{-2\mu\sqrt{\eta}}}{2}\right)$.
Finally, the formula for the key generation rate, expressed as a function
of $\eta$ and the intensity $\mu$ in this loss-only scenario, is
detailed in Section IV.A of Ref. \cite{LL18}.

\begin{equation}
R_{{\rm loss}}^{\infty}=\left(1-e^{-2\mu\sqrt{\eta}}\right)\left[1-h\left(\frac{1-e^{-4\mu\left(1-\sqrt{\eta}\right)}e^{-2\mu\sqrt{\eta}}}{2}\right)\right]\label{eq:Key-rate=000020of=000020Loss-Only=000020Scenario}
\end{equation}

\begin{table}
\begin{centering}
\begin{tabular}{|c|c|c|c|c|}
\hline 
\backslashbox{$\langle\boldsymbol{\mathbf{\alpha}}|{\rm E}^{\gamma}|\boldsymbol{\alpha}\rangle$}{$\boldsymbol{\mathbf{\alpha}}=|\alpha_{{\rm A}},\alpha_{{\rm B}}\rangle$} & $\left|+\sqrt{\mu},+\sqrt{\mu}\right\rangle $ & $\left|-\sqrt{\mu},-\sqrt{\mu}\right\rangle $ & $\left|+\sqrt{\mu},-\sqrt{\mu}\right\rangle $ & $\left|-\sqrt{\mu},+\sqrt{\mu}\right\rangle $\tabularnewline
\hline 
$\langle\boldsymbol{\mathbf{\alpha}}|{\rm E}^{+}|\boldsymbol{\mathbf{\alpha}}\rangle$ & $1-e^{-2\sqrt{\eta}\mu}$ & $1-e^{-2\sqrt{\eta}\mu}$ & 0 & 0\tabularnewline
$\langle\boldsymbol{\mathbf{\alpha}}|{\rm E}^{-}|\boldsymbol{\mathbf{\alpha}}\rangle$ & 0 & 0 & $1-e^{-2\sqrt{\eta}\mu}$ & $1-e^{-2\sqrt{\eta}\mu}$\tabularnewline
$\langle\boldsymbol{\mathbf{\alpha}}|{\rm E}^{?}|\boldsymbol{\mathbf{\alpha}}\rangle$ & $e^{-2\sqrt{\eta}\mu}$ & $e^{-2\sqrt{\eta}\mu}$ & $e^{-2\sqrt{\eta}\mu}$ & $e^{-2\sqrt{\eta}\mu}$\tabularnewline
$\langle\boldsymbol{\mathbf{\alpha}}|{\rm E}^{{\rm d}}|\boldsymbol{\mathbf{\alpha}}\rangle$ & 0 & 0 & 0 & 0\tabularnewline
\hline 
\end{tabular}
\par\end{centering}
\caption{ The conditional probability distribution of announcement outcomes,
given the states from $\mathcal{S}$, under the loss-only scenario.
Here, $\eta$ denotes the single-photon transmission efficiency between
Alice and Bob, while $\mu$ represents the intensity of coherent states
used in the key-generation mode.}\label{tab:Conditional-Probability-Loss-Only-Scenario}
\end{table}

\section*{Appendix B}

\subsection*{Realistic Imperfection}

In the practical implementation of a protocol, various realistic imperfections
associated with experimental devices can arise. These include dark
counts of detectors, mode mismatch, phase mismatch, detector inefficiency
and error correction inefficiency, as discussed in Ref. \cite{LL18}.
Here, we provide a brief description of these imperfections and present
an analytical solution for the key rate equation, taking these realistic
imperfections into account.

For the sake of simplicity, we assume that both detectors have identical
efficiency, denoted as $\eta_{d}$, and the same dark count probability,
$p_{d}$. Ideally, Alice and Bob should prepare coherent states in
the same optical mode, sharing identical spectral and temporal profiles,
as well as the same polarization, to achieve single-photon interference
at the beam splitter. However, in practice, their states may originate
from different lasers and traverse distinct optical components before
reaching the central node, leading to potential mode mismatches. We
account for this by introducing a simulation parameter ${\rm M}$,
which represents the relative mode mismatch. In the simulation, if
there is no mode mismatch, the state arriving at the central node
from Alice and Bob would be $\left|\alpha_{{\rm A}},\alpha_{{\rm B}}\right\rangle $.
With mode mismatch, the state becomes $\left|\alpha_{{\rm A}},\sqrt{{\rm M}}\,\alpha_{{\rm B}}\right\rangle $
in the original mode, and $\left|0,\sqrt{1-{\rm M}}\,\alpha_{{\rm B}}\right\rangle $
in a secondary mode, labeled with subscripts $1$ and $2$, respectively.
Both modes then enter Charlie\textquoteright s devices independently.
Another imperfection considered in the simulation model is phase mismatch.
In key-generation mode, Alice and Bob are expected to prepare states
from the set $\mathcal{S}$, which are coherent states with a uniform
global phase and encoding information in the relative phases. In reality,
the global phase may not remain consistent when the states reach the
detectors. Thus, we consider the case of a relative phase mismatch
between Alice's and Bob's signal states. Without phase mismatch, the
state would be $\left|\alpha_{{\rm A}},\alpha_{{\rm B}}\right\rangle $.
Due to phase mismatch, the state changes to $\left|\alpha_{{\rm A}},\alpha_{{\rm B}}e^{i\delta}\right\rangle $,
with $\delta$ being the phase mismatch simulation parameter\footnote{We use the simulation parameters provided in Table II of Ref. \cite{LL18}.}.

We now present the foundational formulations required to derive the
analytical key rate equation under realistic imperfection scenarios.
Initially, we define Eve\textquoteright s POVM ${\rm E}_{{\rm mismatch}}^{\gamma}$,
which accounts for both mode and phase mismatches. Subsequently, we
derive Eve\textquoteright s POVM ${\rm E}_{{\rm model}}^{\gamma}$
by incorporating detector dark counts. Finally, we consider the effects
of detector efficiency through a redefinition of $\eta$. For an input
coherent state $|\alpha_{{\rm A}}\alpha_{{\rm B}}\rangle_{{\rm A^{\prime}{\rm B^{\prime}}}}$,
the state after passing through lossy channels and accounting for
mode and phase mismatches becomes $\left|\eta^{\frac{1}{4}}\alpha_{{\rm A}},\eta^{\frac{1}{4}}\sqrt{{\rm M}}\,\alpha_{{\rm B}}e^{i\delta}\right\rangle _{I_{{\rm A1}}I_{{\rm B1}}}\otimes\left|0,\eta^{\frac{1}{4}}\sqrt{1-{\rm M}}\,\alpha_{{\rm B}}e^{i\delta}\right\rangle _{I_{{\rm A2}}I_{{\rm B2}}}$,
while Eve has $\left|\sqrt{1-\eta^{\frac{1}{2}}}\alpha_{{\rm A}},\sqrt{1-\eta^{\frac{1}{2}}}\alpha_{{\rm B}}\right\rangle _{E_{{\rm A}}E_{{\rm B}}}$.
The state arriving at the detectors is $\left|\frac{\eta^{\frac{1}{4}}\left(\alpha_{{\rm A}}+\sqrt{{\rm M}}\,\alpha_{{\rm B}}e^{i\delta}\right)}{\sqrt{2}},\frac{\eta^{\frac{1}{4}}\left(\alpha_{{\rm A}}-\sqrt{{\rm M}}\alpha_{{\rm B}}e^{i\delta}\right)}{\sqrt{2}}\right\rangle _{{\rm O_{A1}}{\rm O_{B1}}}\otimes\left|\frac{\eta^{\frac{1}{4}}\left(\alpha_{{\rm A}}+\sqrt{1-{\rm M}}\,\alpha_{{\rm B}}e^{i\delta}\right)}{\sqrt{2}},\frac{\eta^{\frac{1}{4}}\left(\alpha_{{\rm A}}-\sqrt{1-{\rm M}}\,\alpha_{{\rm B}}e^{i\delta}\right)}{\sqrt{2}}\right\rangle _{{\rm O_{A2}}{\rm O_{B2}}}$.
Next, we define the POVM of ideal detectors when two independent modes
enter the detectors due to mode mismatch.

\begin{equation}
\begin{array}{lcl}
\Omega_{{\rm mismatch}}^{+} & = & \left(\mathds{1}_{{\rm O_{A1}O_{A2}}}-|00\rangle\langle00|_{{\rm O_{A1}O_{A2}}}\right)\otimes|00\rangle\langle00|_{{\rm O_{B1}O_{B2}}},\\
\\\Omega_{{\rm mismatch}}^{-} & = & |00\rangle\langle00|_{{\rm O_{A1}O_{A2}}}\otimes\left(\mathds{1}_{{\rm O_{B1}O_{B2}}}-|00\rangle\langle00|_{{\rm O_{B1}O_{B2}}}\right),\\
\\\Omega_{{\rm mismatch}}^{?} & = & |00\rangle\langle00|_{{\rm O_{A1}O_{A2}}}\otimes|00\rangle\langle00|_{{\rm O_{B1}O_{B2}}},\\
\\\Omega_{{\rm mismatch}}^{{\rm d}} & = & \left(\mathds{1}_{{\rm O_{A1}O_{A2}}}-|00\rangle\langle00|_{{\rm O_{A1}O_{A2}}}\right)\otimes\left(\mathds{1}_{{\rm O_{B1}O_{B2}}}-|00\rangle\langle00|_{{\rm O_{B1}O_{B2}}}\right).
\end{array}\label{eq:Detectors'=000020POVM=000020Realistic=000020Imperfections}
\end{equation}
When the two-mode coherent states are $|\boldsymbol{\alpha}\rangle=|\alpha_{{\rm A}},\alpha_{{\rm B}}\rangle$
and $|\boldsymbol{\beta}\rangle=|\beta_{{\rm A}},\beta_{{\rm B}}\rangle$,
they transform into the following states after accounting for realistic
imperfections:

\begin{equation}
\begin{array}{lcl}
|\alpha_{{\rm final}}\rangle & = & \left|\frac{\eta^{\frac{1}{4}}\left(\alpha_{{\rm A}}+\sqrt{{\rm M}}\,\alpha_{{\rm B}}e^{i\delta}\right)}{\sqrt{2}},\frac{\eta^{\frac{1}{4}}\sqrt{1-{\rm M}}\,\alpha_{{\rm B}}e^{i\delta}}{\sqrt{2}},\frac{\eta^{\frac{1}{4}}\left(\alpha_{{\rm A}}-\sqrt{{\rm M}}\,\alpha_{{\rm B}}e^{i\delta}\right)}{\sqrt{2}},-\frac{\eta^{\frac{1}{4}}\sqrt{1-{\rm M}}\,\alpha_{{\rm B}}e^{i\delta}}{\sqrt{2}}\right\rangle _{{\rm O_{A1}O_{A2}}{\rm O_{B1}O_{B2}}},\\
|\beta_{{\rm final}}\rangle & = & \left|\frac{\eta^{\frac{1}{4}}\left(\beta_{{\rm A}}+\sqrt{{\rm M}}\,\beta_{{\rm B}}e^{i\delta}\right)}{\sqrt{2}},\frac{\eta^{\frac{1}{4}}\sqrt{1-{\rm M}}\,\beta_{{\rm B}}e^{i\delta}}{\sqrt{2}},\frac{\eta^{\frac{1}{4}}\left(\beta_{{\rm A}}-\sqrt{{\rm M}}\,\beta_{{\rm B}}e^{i\delta}\right)}{\sqrt{2}},-\frac{\eta^{\frac{1}{4}}\sqrt{1-{\rm M}}\,\beta_{{\rm B}}e^{i\delta}}{\sqrt{2}}\right\rangle _{{\rm O_{A1}O_{A2}}{\rm O_{B1}O_{B2}}},
\end{array}\label{eq:Final_State=000020after=000020realistic=000020imperfection}
\end{equation}
then we have,

\begin{equation}
\begin{array}{lcl}
\left\langle \boldsymbol{\alpha}\right|{\rm E}_{{\rm mismatch}}^{\gamma}\left|\boldsymbol{\beta}\right\rangle  & = & \left\langle \alpha_{{\rm final}}\right|\Omega_{{\rm mismatch}}^{\gamma}\left|\beta_{{\rm final}}\right\rangle \left\langle \sqrt{1-\eta^{\frac{1}{2}}}\alpha_{{\rm A}}|\sqrt{1-\eta^{\frac{1}{2}}}\beta_{{\rm A}}\right\rangle _{E_{{\rm A}}}\left\langle \sqrt{1-\eta^{\frac{1}{2}}}\alpha_{{\rm B}}|\sqrt{1-\eta^{\frac{1}{2}}}\beta_{{\rm B}}\right\rangle _{E_{{\rm B}}},\end{array}\label{eq:Probability_Realistic=000020Imperfection}
\end{equation}
where the total transmittance is given by $\eta=\eta_{s}\eta_{t}\eta_{d}^{2}$,
with $\eta_{s}$ representing the transmittance in satellite communication,
and $\eta_{t}=10^{-\frac{0.2L}{10}}$ for a distance $L$ in km. We
define a few variables to express ${\rm E}_{{\rm mismatch}}^{\gamma}$
in the $\mathds{B}$ basis: $\pounds=e^{-\sqrt{\eta}\mu}$ and $\zeta=e^{-2\left(1-\sqrt{\eta}\right)\mu}$.
Using these two variables, we can define additional variables as follows:\\
$\begin{array}{lcl}
p & = & \left(1-\pounds^{\left(1+\sqrt{{\rm M}}\,{\rm cos}\delta\right)}\right)\pounds^{\left(1-\sqrt{{\rm M}}\,{\rm cos}\delta\right)}\end{array},$ $\begin{array}{lcl}
q & = & \left(\pounds^{2\left(1+\sqrt{{\rm M}}\,{\rm cos}\delta\right)}-\pounds^{\left(1+\sqrt{{\rm M}}\,{\rm cos}\delta\right)}\right)\pounds^{\left(1-\sqrt{{\rm M}}\,{\rm cos}\delta\right)}\zeta^{2}\end{array},$ $\begin{array}{lcl}
r & = & \left(\pounds^{1+i\sqrt{{\rm M}}\,{\rm sin}\delta}-\pounds\right)\pounds\zeta\end{array},$ $\begin{array}{lcl}
s & = & \left(\pounds^{1-i\sqrt{{\rm M}}\,{\rm sin}\delta}-\pounds\right)\pounds\zeta\end{array},$ $\begin{array}{lcl}
a & = & \left(1-\pounds^{\left(1-\sqrt{{\rm M}}\,{\rm cos}\delta\right)}\right)\pounds^{\left(1+\sqrt{{\rm M}}\,{\rm cos}\delta\right)},\end{array}$ $\begin{array}{lcl}
b & = & \left(\pounds^{2\left(1-\sqrt{{\rm M}}\,{\rm cos}\delta\right)}-\pounds^{\left(1-\sqrt{{\rm M}}\,{\rm cos}\delta\right)}\right)\pounds^{\left(1+\sqrt{{\rm M}}\,{\rm cos}\delta\right)}\zeta^{2},\end{array}$\\
$\begin{array}{lcl}
c & = & \left(\pounds^{1+i\sqrt{{\rm M}}\,{\rm sin}\delta}-\pounds\right)\left(\pounds^{1-i\sqrt{{\rm M}}\,{\rm sin}\delta}-\pounds\right)\zeta\end{array}$, $\begin{array}{lcl}
d & = & \left(1-\pounds^{\left(1+\sqrt{{\rm M}}\,{\rm cos}\delta\right)}\right)\left(1-\pounds^{\left(1-\sqrt{{\rm M}}\,{\rm cos}\delta\right)}\right),\end{array}$ and \\
$\begin{array}{lcl}
e & = & \left(\pounds^{2\left(1+\sqrt{{\rm M}}\,{\rm cos}\delta\right)}-\pounds^{\left(1+\sqrt{{\rm M}}\,{\rm cos}\delta\right)}\right)\left(\pounds^{2\left(1-\sqrt{{\rm M}}\,{\rm cos}\delta\right)}-\pounds^{\left(1-\sqrt{{\rm M}}\,{\rm cos}\delta\right)}\right)\pounds^{2}.\end{array}$

Now, we have ${\rm E}_{{\rm mismatch}}^{\gamma}$ in the basis $\mathds{B}$,

\begin{equation}
\begin{array}{lcl}
{\rm E}_{{\rm mismatch}}^{+} & = & \left(\begin{array}{cccc}
\frac{p+q+2r+2s+a+b}{8c_{0}^{4}} & \frac{p+q-a-b}{8c_{0}^{2}c_{1}^{2}} & 0 & 0\\
\\\frac{p+q-a-b}{8c_{0}^{2}c_{1}^{2}} & \frac{p+q-2r-2s+a+b}{8c_{1}^{4}} & 0 & 0\\
\\0 & 0 & \frac{p-q+a-b}{8c_{0}^{2}c_{1}^{2}} & \frac{p-q+2r-2s-a+b}{8c_{0}^{2}c_{1}^{2}}\\
\\0 & 0 & \frac{p-q-2r+2s-a+b}{8c_{0}^{2}c_{1}^{2}} & \frac{p-q+a-b}{8c_{0}^{2}c_{1}^{2}}
\end{array}\right),\\
\\{\rm E}_{{\rm mismatch}}^{-} & = & \left(\begin{array}{cccc}
\frac{p+q+2r+2s+a+b}{8c_{0}^{4}} & -\frac{p+q-a-b}{8c_{0}^{2}c_{1}^{2}} & 0 & 0\\
\\-\frac{p+q-a-b}{8c_{0}^{2}c_{1}^{2}} & \frac{p+q-2r-2s+a+b}{8c_{1}^{4}} & 0 & 0\\
\\0 & 0 & \frac{p-q+a-b}{8c_{0}^{2}c_{1}^{2}} & -\frac{p-q+2r-2s-a+b}{8c_{0}^{2}c_{1}^{2}}\\
\\0 & 0 & -\frac{p-q-2r+2s-a+b}{8c_{0}^{2}c_{1}^{2}} & \frac{p-q+a-b}{8c_{0}^{2}c_{1}^{2}}
\end{array}\right),\\
\\{\rm E}_{{\rm mismatch}}^{?} & = & \pounds^{2}\left(\begin{array}{cccc}
\frac{\left(1+\zeta\right)^{2}}{4c_{0}^{4}} & 0 & 0 & 0\\
\\0 & \frac{\left(1-\zeta\right)^{2}}{4c_{1}^{4}} & 0 & 0\\
\\0 & 0 & \frac{1-\zeta^{2}}{4c_{0}^{2}c_{1}^{2}} & 0\\
\\0 & 0 & 0 & \frac{1-\zeta^{2}}{4c_{0}^{2}c_{1}^{2}}
\end{array}\right),\\
\\{\rm E}_{{\rm mismatch}}^{{\rm d}} & = & \left(\begin{array}{cccc}
\frac{d+e+2c}{4c_{0}^{4}} & 0 & 0 & 0\\
\\0 & \frac{d+e-2c}{4c_{1}^{4}} & 0 & 0\\
\\0 & 0 & \frac{d-e}{4c_{0}^{2}c_{1}^{2}} & 0\\
\\0 & 0 & 0 & \frac{d-e}{4c_{0}^{2}c_{1}^{2}}
\end{array}\right).
\end{array}\label{eq:POVM=000020corresponding=000020to=000020mode=000020mismatch}
\end{equation}
Finally, Eve's effective POVM, accounting for mode mismatch, phase
mismatch and detector dark counts, is given as follows:

\begin{equation}
\begin{array}{lcl}
{\rm E}_{{\rm model}}^{+} & = & \left(1-p_{d}\right){\rm E}_{{\rm mismatch}}^{+}+\left(1-p_{d}\right)p_{d}{\rm E}_{{\rm mismatch}}^{?},\\
\\{\rm E}_{{\rm model}}^{-} & = & \left(1-p_{d}\right){\rm E}_{{\rm mismatch}}^{-}+\left(1-p_{d}\right)p_{d}{\rm E}_{{\rm mismatch}}^{?},\\
\\{\rm E}_{{\rm model}}^{?} & = & \left(1-p_{d}\right)^{2}{\rm E}_{{\rm mismatch}}^{?},\\
\\{\rm E}_{{\rm model}}^{{\rm d}} & = & p_{d}\,{\rm E}_{{\rm mismatch}}^{+}+p_{d}\,{\rm E}_{{\rm mismatch}}^{-}+p_{d}^{2}\,{\rm E}_{{\rm mismatch}}^{?}+{\rm E}_{{\rm mismatch}}^{{\rm d}}.
\end{array}\label{eq:POVM=000020for=000020realistic=000020model}
\end{equation}
When Alice transmits a coherent state $|\alpha_{{\rm A}}\rangle$
and Bob sends a coherent state $|\alpha_{{\rm B}}\rangle$ with mode
mismatch parameter ${\rm M}$, the state becomes $|\alpha_{{\rm final}},\beta_{{\rm final}}\rangle$
after passing through the lossy channel. When Charlie conducts a measurement
on that state, the probability for each announcement outcome as $\gamma$,
represented by $\langle\boldsymbol{\mathbf{\alpha}}|{\rm E}^{\gamma}|\boldsymbol{\mathbf{\alpha}}\rangle$,
can be determined using Eqs. (\ref{eq:Detectors'=000020POVM=000020Realistic=000020Imperfections})
- (\ref{eq:Probability_Realistic=000020Imperfection}) as follows:

\begin{equation}
\begin{array}{lcl}
\left\langle \alpha_{{\rm A}},\alpha_{{\rm B}}\right|{\rm E}^{+}\left|\alpha_{{\rm A}},\alpha_{{\rm B}}\right\rangle  & = & \left(1-p_{d}\right)\left(1-\omega_{1}\omega_{2}\right)\omega_{2}\omega_{3}+\left(1-p_{d}\right)p_{d}\,\omega_{1}\omega_{2}^{2}\omega_{3},\\
\\\left\langle \alpha_{{\rm A}},\alpha_{{\rm B}}\right|{\rm E}^{-}\left|\alpha_{{\rm A}},\alpha_{{\rm B}}\right\rangle  & = & \left(1-p_{d}\right)\omega_{1}\omega_{2}\left(1-\omega_{2}\omega_{3}\right)+\left(1-p_{d}\right)p_{d}\,\omega_{1}\omega_{2}^{2}\omega_{3},\\
\\\left\langle \alpha_{{\rm A}},\alpha_{{\rm B}}\right|{\rm E}^{?}\left|\alpha_{{\rm A}},\alpha_{{\rm B}}\right\rangle  & = & \left(1-p_{d}\right)^{2}\omega_{1}\omega_{2}^{2}\omega_{3},\\
\\\left\langle \alpha_{{\rm A}},\alpha_{{\rm B}}\right|{\rm E}^{{\rm d}}\left|\alpha_{{\rm A}},\alpha_{{\rm B}}\right\rangle  & = & p_{d}\left(1-\omega_{1}\omega_{2}\right)\omega_{2}\omega_{3}+p_{d}\,\omega_{1}\omega_{2}\left(1-\omega_{2}\omega_{3}\right)+p_{d}^{2}\,\omega_{1}\omega_{2}^{2}\omega_{3}+\left(1-\omega_{1}\omega_{2}\right)\left(1-\omega_{2}\omega_{3}\right),
\end{array}\label{eq:Probabilities=000020of=000020Charlie's=000020Announcement-Realistic=000020Imperfection}
\end{equation}
For the sake of simplicity, we introduce these definitions, $\begin{array}{lcl}
\omega_{1} & = & e^{-\frac{1}{2}\sqrt{\eta}\left|\alpha_{{\rm A}}+\sqrt{{\rm M}}\alpha_{{\rm B}}e^{i\delta}\right|^{2}},\end{array}$ $\begin{array}{lcl}
\omega_{2} & = & e^{-\frac{1}{2}\sqrt{\eta}\left(1-{\rm M}\right)\left|\alpha_{{\rm B}}\right|^{2}},\end{array}$ $\begin{array}{lcl}
\omega_{3} & = & e^{-\frac{1}{2}\sqrt{\eta}\left|\alpha_{{\rm A}}-\sqrt{{\rm M}}\alpha_{{\rm B}}e^{i\delta}\right|^{2}}.\end{array}$ Now, we proceed to compute each term of the key rate equation (refer
to Eq. (\ref{eq:Key-Rate=000020Equation})). Subsequently, we evaluate
the Holevo information $\chi\left(A:E\right)_{\rho_{ABE}^{\gamma}}$.
The expressions for the first and second terms of Eq. (\ref{eq:Holevo=000020Information})
given Charlie's announcement $\gamma$ are as follows:

\begin{equation}
\begin{array}{lcl}
S\left(\rho_{E}^{\gamma}\right) & = & S\left(\underset{a}{\sum}\,p\left(a|\gamma\right)\,\rho_{E}^{a,\gamma}\right)\\
 & = & S\left(\underset{a,y}{\sum}\,p\left(a,b|\gamma\right)\,\rho_{E}^{a,b,\gamma}\right)\\
 & = & S\left(p\left(0,0|\gamma\right)\,\rho_{E}^{0,0,\gamma}+p\left(1,1|\gamma\right)\,\rho_{E}^{1,1,\gamma}+p\left(0,1|\gamma\right)\,\rho_{E}^{0,1,\gamma}+p\left(1,0|\gamma\right)\,\rho_{E}^{1,0,\gamma}\right)
\end{array},\label{eq:S_Rho_Gamma}
\end{equation}
and

\begin{equation}
\begin{array}{lcl}
\underset{a}{\sum}p\left(a|\gamma\right)S\left(\rho_{E}^{a,\gamma}\right) & = & \underset{a}{\sum}\,p\left(a|\gamma\right)S\left(\underset{b}{\sum}\,p\left(b|a,\gamma\right)\,\rho_{E}^{a,b,\gamma}\right)\\
 & = & \underset{a}{\sum}\,p\left(a|\gamma\right)S\left(\underset{b}{\sum}\,\frac{p\left(a,b|\gamma\right)}{p\left(a|\gamma\right)}\,\rho_{E}^{a,b,\gamma}\right)\\
 & = & \underset{a}{\sum}\,p\left(a|\gamma\right)S\left(\frac{p\left(a,0|\gamma\right)}{p\left(a|\gamma\right)}\,\rho_{E}^{a,0,\gamma}+\frac{p\left(a,1|\gamma\right)}{p\left(a|\gamma\right)}\,\rho_{E}^{a,1,\gamma}\right)\\
 & = & p\left(0|\gamma\right)S\left(\frac{p\left(0,0|\gamma\right)}{p\left(0|\gamma\right)}\,\rho_{E}^{0,0,\gamma}+\frac{p\left(0,1|\gamma\right)}{p\left(0|\gamma\right)}\,\rho_{E}^{0,1,\gamma}\right)\\
 & + & p\left(1|\gamma\right)S\left(\frac{p\left(1,0|\gamma\right)}{p\left(1|\gamma\right)}\,\rho_{E}^{1,0,\gamma}+\frac{p\left(1,1|\gamma\right)}{p\left(1|\gamma\right)}\,\rho_{E}^{1,1,\gamma}\right)
\end{array},\label{eq:S_Rho_A_Gamma}
\end{equation}
where $p\left(\gamma\right)$ represents the marginal probability
of the joint probability distribution $p\left(a,b,\gamma\right)$.
The conditional probability$p\left(a,b|\gamma\right)=\frac{p\left(a,b,\gamma\right)}{p\left(\gamma\right)}$
describes the scenario where Alice holds $a$ in her register $A$,
Bob holds $b$ in his register $B$, and the central node (Charlie)
announces $\gamma$. Similarly, $p\left(a|\gamma\right)$ is the conditional
probability of Alice having $a$ given Charlie's announcement $\gamma$,
where, $a,b\in\left\{ 0,1\right\} $ and $\gamma\in\left\{ +,-\right\} $.
The conditional probabilities for each announcement outcome for each
state in the set $\mathcal{S}$ are summarized in Table \ref{tab:Conditional-Probability-Noisy-Scenario},
utilizing Eq. (\ref{eq:Probabilities=000020of=000020Charlie's=000020Announcement-Realistic=000020Imperfection}).
For simplicity, we introduce new variables: $x=\omega_{1}\omega_{2}=e^{-\sqrt{\eta}\mu\left(1+\sqrt{{\rm M}}{\rm cos}\delta\right)},$
$y=\omega_{2}\omega_{3}=e^{-\sqrt{\eta}\mu\left(1-\sqrt{{\rm M}}{\rm cos}\delta\right)},$
and $z=\omega_{1}\omega_{2}^{2}\omega_{3}=e^{-2\sqrt{\eta}\mu}$.

\begin{table}[h]
\begin{centering}
\begin{tabular}{|c|c|c|c|c|}
\hline 
\backslashbox{$\langle\boldsymbol{\mathbf{\alpha}}|{\rm E}^{\gamma}|\boldsymbol{\alpha}\rangle$}{$\boldsymbol{\mathbf{\alpha}}=|\alpha_{{\rm A}},\alpha_{{\rm B}}\rangle$} & $\left|+\sqrt{\mu},+\sqrt{\mu}\right\rangle $ & $\left|-\sqrt{\mu},-\sqrt{\mu}\right\rangle $ & $\left|+\sqrt{\mu},-\sqrt{\mu}\right\rangle $ & $\left|-\sqrt{\mu},+\sqrt{\mu}\right\rangle $\tabularnewline
\hline 
\multirow{2}{*}{$\langle\boldsymbol{\mathbf{\alpha}}|{\rm E}^{+}|\boldsymbol{\mathbf{\alpha}}\rangle$} & $\left(1-p_{d}\right)\left(1-x\right)y$ & $\left(1-p_{d}\right)\left(1-x\right)y$ & $\left(1-p_{d}\right)\left(1-y\right)x$ & $\left(1-p_{d}\right)\left(1-y\right)x$\tabularnewline
 & $+\left(1-p_{d}\right)p_{d}z$ & $+\left(1-p_{d}\right)p_{d}z$ & $+\left(1-p_{d}\right)p_{d}z$ & $+\left(1-p_{d}\right)p_{d}z$\tabularnewline
\hline 
\multirow{2}{*}{$\langle\boldsymbol{\mathbf{\alpha}}|{\rm E}^{-}|\boldsymbol{\mathbf{\alpha}}\rangle$} & $\left(1-p_{d}\right)x\left(1-y\right)$ & $\left(1-p_{d}\right)x\left(1-y\right)$ & $\left(1-p_{d}\right)y\left(1-x\right)$ & $\left(1-p_{d}\right)y\left(1-x\right)$\tabularnewline
 & $+\left(1-p_{d}\right)p_{d}z$ & $+\left(1-p_{d}\right)p_{d}z$ & $+\left(1-p_{d}\right)p_{d}z$ & $+\left(1-p_{d}\right)p_{d}z$\tabularnewline
\hline 
$\langle\boldsymbol{\mathbf{\alpha}}|{\rm E}^{?}|\boldsymbol{\mathbf{\alpha}}\rangle$ & $\left(1-p_{d}\right)^{2}z$ & $\left(1-p_{d}\right)^{2}z$ & $\left(1-p_{d}\right)^{2}z$ & $\left(1-p_{d}\right)^{2}z$\tabularnewline
\hline 
\multirow{2}{*}{$\langle\boldsymbol{\mathbf{\alpha}}|{\rm E}^{{\rm d}}|\boldsymbol{\mathbf{\alpha}}\rangle$} & $p_{d}\left(1-x\right)y+p_{d}x\left(1-y\right)$ & $p_{d}\left(1-x\right)y+p_{d}x\left(1-y\right)$ & $p_{d}\left(1-y\right)x+p_{d}y\left(1-x\right)$ & $p_{d}\left(1-y\right)x+p_{d}y\left(1-x\right)$\tabularnewline
 & $+p_{d}^{2}z+\left(1-x\right)\left(1-y\right)$ & $+p_{d}^{2}z+\left(1-x\right)\left(1-y\right)$ & $+p_{d}^{2}z+\left(1-y\right)\left(1-x\right)$ & $+p_{d}^{2}z+\left(1-y\right)\left(1-x\right)$\tabularnewline
\hline 
\end{tabular}
\par\end{centering}
\caption{Conditional probability distribution of the announcement outcomes
given the states from $\mathcal{S}$ in the presence of noise. Here,
$\eta$ represents the single-photon transmissivity between Alice
and Bob and $\mu$ denotes the intensity of coherent states in the
key-generation mode.}\label{tab:Conditional-Probability-Noisy-Scenario}
\end{table}
We now have the probabilities to calculate the Holevo information
$\chi\left(A:E\right)_{\rho_{AYE}^{\gamma}}$. The probabilities are
$p\left(0,0,+\right)=p\left(1,1,+\right)=p\left(0,1,-\right)=p\left(1,0,-\right)=\frac{\left(1-p_{d}\right)\left(\left(1-x\right)y+p_{d}z\right)}{4\left(1-xy+z\right)},$
$p\left(0,1,+\right)=p\left(1,0,+\right)=p\left(0,1,-\right)=p\left(1,1,-\right)=\frac{\left(1-p_{d}\right)\left(\left(1-y\right)x+p_{d}z\right)}{4\left(1-xy+z\right)},$
$p\left(0,+\right)=p\left(1,+\right)=p\left(0,-\right)=p\left(1,-\right)=\frac{\left(1-p_{d}\right)\left(\left(1-x\right)y+\left(1-y\right)x+2p_{d}z\right)}{4\left(1-xy+z\right)}$
and $p\left(+\right)=p\left(-\right)=\frac{\left(1-p_{d}\right)\left(\left(1-x\right)y+\left(1-y\right)x+2p_{d}z\right)}{2\left(1-xy+z\right)}$.
To obtain the conditional probability, use the relations $p\left(a,b|\gamma\right)=\frac{p\left(a,b,\gamma\right)}{p\left(\gamma\right)}$
and $p\left(a|\gamma\right)=\frac{p\left(a,\gamma\right)}{p\left(\gamma\right)}$.
From the Table \ref{tab:Conditional-Probability-Noisy-Scenario},
we can directly evaluate the classical mutual information between
bit values in Alice's and Bob's registers given Charlie's announcement
$\gamma$, $I\left(A|B\right)_{\rho_{ABE}^{+}}=I\left(A|B\right)_{\rho_{ABE}^{-}}=2{\rm H}\left(\frac{\left(1-p_{d}\right)\left(\left(1-x\right)y+\left(1-y\right)x+2p_{d}z\right)}{4\left(1-xy+z\right)}\right)-\frac{\left(1-p_{d}\right)\left(\left(1-x\right)y+\left(1-y\right)x+2p_{d}z\right)}{\left(1-xy+z\right)}\left({\rm H}\left(\frac{\left(1-x\right)y+p_{d}z}{\left(1-x\right)y+\left(1-y\right)x+2p_{d}z}\right)+{\rm H}\left(\frac{\left(1-y\right)x+p_{d}z}{\left(1-x\right)y+\left(1-y\right)x+2p_{d}z}\right)\right)$
and $I\left(A|B\right)_{\rho_{ABE}^{?}}=I\left(A|B\right)_{\rho_{ABE}^{{\rm d}}}=0,$
where ${\rm H}\left(x\right)=-x\log_{2}x$ is the Shannon entropy.
Clearly, we obtain the key rate from $\gamma=+$ and $\gamma=-$.
The mutual information for $\gamma=?$ and $\gamma=d$ is zero because
the probability of these announcements is independent of the signal
states sent from Alice and Bob in that simulation \cite{LL18}. To
evaluate Eqs. (\ref{eq:S_Rho_Gamma}) and (\ref{eq:S_Rho_A_Gamma}),
we derive the expression $\rho_{E}^{a,b,\gamma}$ in the basis set
$\mathcal{S}$,

\begin{equation}
\begin{array}{lcl}
\rho_{E}^{0,0,\gamma} & = & \frac{\sqrt{{\rm E}_{{\rm model}}^{\gamma}}\left|+\sqrt{\mu},+\sqrt{\mu}\right\rangle \left\langle +\sqrt{\mu},+\sqrt{\mu}\right|\sqrt{\left({\rm E}_{{\rm model}}^{\gamma}\right)^{\dagger}}}{{\rm \boldsymbol{Tr}\left(\sqrt{{\rm E}_{{\rm model}}^{\gamma}}\left|+\sqrt{\mu},+\sqrt{\mu}\right\rangle \left\langle +\sqrt{\mu},+\sqrt{\mu}\right|\sqrt{\left({\rm E}_{{\rm model}}^{\gamma}\right)^{\dagger}}\right)}},\\
\\\rho_{E}^{1,1,\gamma} & = & \frac{\sqrt{{\rm E}_{{\rm model}}^{\gamma}}\left|-\sqrt{\mu},-\sqrt{\mu}\right\rangle \left\langle -\sqrt{\mu},-\sqrt{\mu}\right|\sqrt{\left({\rm E}_{{\rm model}}^{\gamma}\right)^{\dagger}}}{{\rm \boldsymbol{Tr}\left(\sqrt{{\rm E}_{{\rm model}}^{\gamma}}\left|-\sqrt{\mu},-\sqrt{\mu}\right\rangle \left\langle -\sqrt{\mu},-\sqrt{\mu}\right|\sqrt{\left({\rm E}_{{\rm model}}^{\gamma}\right)^{\dagger}}\right)}},\\
\\\rho_{E}^{0,1,\gamma} & = & \frac{\sqrt{{\rm E}_{{\rm model}}^{\gamma}}\left|+\sqrt{\mu},-\sqrt{\mu}\right\rangle \left\langle +\sqrt{\mu},-\sqrt{\mu}\right|\sqrt{\left({\rm E}_{{\rm model}}^{\gamma}\right)^{\dagger}}}{{\rm \boldsymbol{Tr}\left(\sqrt{{\rm E}_{{\rm model}}^{\gamma}}\left|+\sqrt{\mu},-\sqrt{\mu}\right\rangle \left\langle +\sqrt{\mu},-\sqrt{\mu}\right|\sqrt{\left({\rm E}_{{\rm model}}^{\gamma}\right)^{\dagger}}\right)}},\\
\\\rho_{E}^{1,0,\gamma} & = & \frac{\sqrt{{\rm E}_{{\rm model}}^{\gamma}}\left|-\sqrt{\mu},+\sqrt{\mu}\right\rangle \left\langle -\sqrt{\mu},+\sqrt{\mu}\right|\sqrt{\left({\rm E}_{{\rm model}}^{\gamma}\right)^{\dagger}}}{{\rm \boldsymbol{Tr}\left(\sqrt{{\rm E}_{{\rm model}}^{\gamma}}\left|-\sqrt{\mu},+\sqrt{\mu}\right\rangle \left\langle -\sqrt{\mu},+\sqrt{\mu}\right|\sqrt{\left({\rm E}_{{\rm model}}^{\gamma}\right)^{\dagger}}\right)}}.
\end{array}\label{eq:Eve's_Sate_Noisy_Scenario}
\end{equation}
The preceding derivation provides an analytical evaluation of $\chi\left(A:E\right)_{\rho_{ABE}^{\gamma}}$.
This can be expressed as follows: 

\begin{equation}
\begin{array}{lcl}
\chi\left(A:E\right)_{\rho_{ABE}^{+}} & = & S\left(p\left(0,0|+\right)\,\rho_{E}^{0,0,+}+p\left(1,1|+\right)\,\rho_{E}^{1,1,+}+p\left(0,1|+\right)\,\rho_{E}^{0,1,+}+p\left(1,0|+\right)\,\rho_{E}^{1,0,+}\right)\\
 & - & \left(p\left(0|+\right)S\left(\frac{p\left(0,0|+\right)}{p\left(0|+\right)}\,\rho_{E}^{0,0,+}+\frac{p\left(0,1|+\right)}{p\left(0|+\right)}\,\rho_{E}^{0,1,+}\right)+p\left(1|+\right)S\left(\frac{p\left(1,0|+\right)}{p\left(1|+\right)}\,\rho_{E}^{1,0,+}+\frac{p\left(1,1|+\right)}{p\left(1|+\right)}\,\rho_{E}^{1,1,+}\right)\right),\\
\\\chi\left(A:E\right)_{\rho_{ABE}^{-}} & = & S\left(p\left(0,0|-\right)\,\rho_{E}^{0,0,-}+p\left(1,1|-\right)\,\rho_{E}^{1,1,-}+p\left(0,1|-\right)\,\rho_{E}^{0,1,-}+p\left(1,0|-\right)\,\rho_{E}^{1,0,-}\right)\\
 & - & \left(p\left(0|-\right)S\left(\frac{p\left(0,0|-\right)}{p\left(0|-\right)}\,\rho_{E}^{0,0,-}+\frac{p\left(0,1|-\right)}{p\left(0|-\right)}\,\rho_{E}^{0,1,-}\right)+p\left(1|-\right)S\left(\frac{p\left(1,0|-\right)}{p\left(1|-\right)}\,\rho_{E}^{1,0,-}+\frac{p\left(1,1|-\right)}{p\left(1|-\right)}\,\rho_{E}^{1,1,-}\right)\right).
\end{array}\label{eq:Holevo_Information_Noisy_Scenario}
\end{equation}
Now, we evaluate $\delta_{{\rm EC}}^{\gamma}$ to obtain the final
expression of Eq. (\ref{eq:Key-Rate=000020Equation}) in a realistic
imperfection scenario. When the states prepared by Alice and Bob are
either in the same phase or differ by $\pi$, Charlie's (Eve's) detectors
$D_{+}$ and $D_{-}$ will click. Consequently, we define the error
rates $\epsilon_{+}$ and $\epsilon_{-}$ corresponding to the announcement
outcomes $\gamma=+$ and $\gamma=-$, respectively,

\begin{equation}
\begin{array}{lcl}
\epsilon_{+} & = & p\left(0,1|+\right)+p\left(1,0|+\right)\\
 & = & \frac{x-\left(1-p_{d}\right)xy}{x+y-2\left(1-p_{d}\right)xy},\\
\\\epsilon_{-} & = & p\left(0,0|-\right)+p\left(1,1|-\right)=\epsilon_{+}.
\end{array}\label{eq:Error_Rate_Noisy_Scenario}
\end{equation}
To account for the inefficiency of error correction, we use the following
values for $\delta_{{\rm EC}}^{+}$ and $\delta_{{\rm EC}}^{-}$,

\begin{equation}
\begin{array}{lcl}
\delta_{{\rm EC}}^{+} & = & f_{{\rm EC}}h\left(\epsilon_{+}\right),\\
\\\delta_{{\rm EC}}^{-} & = & f_{{\rm EC}}h\left(\epsilon_{-}\right)=\delta_{{\rm EC}}^{+},
\end{array}\label{eq:Error_Correction_Noisy_Scenario}
\end{equation}
where $f_{{\rm EC}}$ represents the efficiency of error correction,
and $h\left(x\right)=-x\log_{2}x-\left(1-x\right)\log_{2}\left(1-x\right)$
denotes the binary entropy function. The key rate achieved for $\gamma=+$
and $\gamma=-$ is given by the Eqs. (\ref{eq:Holevo_Information_Noisy_Scenario})
and (\ref{eq:Error_Correction_Noisy_Scenario}), respectively,

\begin{equation}
\begin{array}{lcl}
r\left(\rho_{ABE}^{+}\right) & = & {\rm max}\left[1-\delta_{{\rm EC}}^{+}-\chi\left(A:E\right)_{\rho_{ABE}^{+}},0\right],\\
\\r\left(\rho_{ABE}^{-}\right) & = & {\rm max}\left[1-\delta_{{\rm EC}}^{-}-\chi\left(A:E\right)_{\rho_{ABE}^{-}},0\right].
\end{array}\label{eq:Key_Rate_Plus_Minus_Noisy}
\end{equation}
Finally, we derive the expression for the secret key generation rate
under realistic imperfections as follows:

\begin{equation}
\begin{array}{lcl}
R_{{\rm noisy}}^{\infty} & = & \underset{\gamma}{\sum}p\left(\gamma\right)r\left(\rho_{ABE}^{\gamma}\right)\\
 & = & p\left(+\right)r\left(\rho_{ABE}^{+}\right)+p\left(-\right)r\left(\rho_{ABE}^{-}\right),
\end{array}\label{eq:Final_Key_Rate_Noisy_Scenario}
\end{equation}
where the probability of the corresponding announcement outcomes $\gamma=+$
and $\gamma=-$ is given by $p\left(+\right)=p\left(-\right)=\frac{\left(1-p_{d}\right)\left(\left(1-x\right)y+\left(1-y\right)x+2p_{d}z\right)}{2\left(1-xy+z\right)}.$
\end{document}